\documentclass[12pt,twoside]{cunythesis}
\usepackage[centertags]{amsmath}
\usepackage{amssymb}
\usepackage[top=1in, bottom=1in, left=1in, right=1in]{geometry}
\usepackage{graphicx}
\usepackage{epstopdf}
\usepackage{subcaption}
\usepackage[usenames,dvipsnames]{xcolor}
\usepackage{listings}
\usepackage[numbers,sort&compress]{natbib}
\usepackage[hidelinks]{hyperref}
\usepackage{rotating}
\usepackage{capt-of}
\usepackage{tabularx}
\usepackage{array}
\usepackage{tabu}

\begin{document}

\begin{frontmatter}
\title{Measuring the transmission matrix for microwave radiation propagating through random waveguides: fundamentals and applications}
\author{Zhou Shi}
\maketitle{}
\copyrightpage
\abstract{
\sloppy
This thesis describes the measurement and analysis of the transmission matrix (TM) for microwave radiation propagating through multichannel random waveguides in the crossover to Anderson localization. Eigenvalues of the transmission matrix and the associated eigenchannels are obtained via a singular value decomposition of the TM. The sum of the transmission eigenvalues yields the transmittance {\it T}, which is the classical analog of the dimensionless conductance {\textsl g}. The dimensionless conductance {\textsl g} is the electronic conductance in units of the quantum conductance, $G/(e^2/h)$.

For diffusive waves ${\textsl g}>1$, approximately {\textsl g} transmission eigenchannels contribute appreciably to the transmittance {\it T}. In contrast, for localized waves with ${\textsl g}<1$, {\it T} is dominated by the highest transmission eigenvalue, $\tau_1$. For localized waves, the inverse of the localization lengths of different eigenchannels are found to be equally spaced. 

Measurement of the TM allows us to explore the statistics of the transmittance {\it T}. A one-sided log-normal distribution of {\it T} is found for a random ensemble with ${\textsl g}=0.37$ and explained using an intuitive Coulomb gas model for the transmission eigenvalues. Single parameter scaling (SPS) predicted for one dimension random system is approached in multichannel systems once {\it T} is dominated by a single transmission eigenchannel. 

In addition to the statistics of the TM for ensembles of random samples, we investigated the statistics of a single TM. The statistics within a large single TM are found to depend upon a single parameter, the eigenchannel participation number, $M\equiv (\sum_{n}^N \tau_n)^2/\sum_{n}^N \tau_n^2$. The variance of the total transmission normalized by its averaging in the TM is equal to $M^{-1}$. We found universal fluctuation of $M$, reminiscent of the well known universal conductance fluctuations for diffusive waves.

We demonstrate focusing of steady state and pulse transmission through a random medium via phase conjugation of the TM. The contrast between the focus and the background is determined by $M$ and the size of the transmission matrix $N$. The spatio-temporal profile of focused radiation in the diffusive limit is shown to be the square of the field-field correlation function in space and time. 

We determine the density of states (DOS) of a disordered medium from the dynamics of transmission eigenchannels and from the quasi-normal modes of the medium for localized samples. The intensity profile of each eigenchannel within the random media is closely linked to the dynamics of transmission eigenchannels and an analytical expression for intensity profile of each of the eigenchannel based on numerical simulation was provided. 

}
\begin{acknowledgements}
First and foremost, I would like to thank my advisor Dr. Azriel Z. Genack for his great guidance, encouragement and support throughout the research and the writing of this thesis. I am deeply touched by his enthusiasm for physics and extraordinary insights about many complex problems. In addition, his optimistic attitude about research and life has greatly inspired me and is invaluable for my future life. I would also like to thank Prof. A. Douglas Stone, Prof. Li Ge, Prof. Vinod M. Menon, and Prof. Vadim Oganesyan for being on my supervisory committee and Dr. Sajan Saini for being on my second exam committee.

Many thanks also go to the former and current students and postdoctoral fellows in this group. In particular, I acknowledge great help from Dr. Jing Wang and Dr. Matthieu Davy. I have benefited from many discussions with Jing on the field of mesoscopic physic and I have learnt a lot regarding the technique of time reversal in acoustics from Matthieu. I also want to thank Professor Jerome Klosner for his continued interest in our work and for reading my thesis. I am also happy to thank Howard Rose for his technical assistance on machinery.

I would like to thank all my friends at Queens College for their valuable friendship, especially Bidisha and Harish from the batch of 2007. I would also like to thank Dr. Huan Zhan and Dr. Yi He, who put a roof over my head when I came to US for the first time, for their friendship over the many years of my stay in New York. I will never forget all the happy time we spent drinking beers and playing world of warcraft together. Special thanks go to Ms. Duo Yang for her encouragements. 

Finally, I would like to express my gratitude to my family for their unconditional support and love. Many thanks go to my dad, for his advice about life, my mom for many hours on the telephone trying to cheer me up when I was blue or homesick and to my sister for all the help whenever I need it. Without all your love, I cannot imagine I would make it this far. 

\vfill
\end{acknowledgements}
\tableofcontents
\listoftables
\listoffigures
\end{frontmatter}
\chapter {Introduction}
\section{General introduction}
Waves play a central role in our daily lives. Classical waves, such as acoustic, ultra sound and electromagnetic waves are the means by which we probe and image our living environment and communicate with each other. Quantum waves, such as electron waves and matter waves, are responsible for information storage and stability of solids. Therefore, the study of wave propagation through diverse media, natural and man-made, simple and complex, has always been an important topic in the physics realm because of its fundamental and applied interest. In this thesis, we study classical waves propagation in complex media, in which the waves undergo multiple scattering before escaping from the media.
 
Multiple scattering of waves is a common phenomenon occurring in our surroundings. For example, sunlight is multiply scattered by clouds on a rainy day \cite{1}. Another case is the appearance of an intensity pattern consisting of bright and dark spots when a laser pointer is passed through a scotch tape. The intensity pattern is due to the interference between scattered waves and is referred as speckle pattern \cite{2}. Because of the fluctuation of intensity within a speckle pattern, a single measurement of the intensity at a speckle spot is rather accidental and only minimum information regarding the media could be obtained. For this reason, the behavior of nature of scattering in a complex medium is usually characterized in a statistical manner in which measurements are made over a random ensemble of sample realizations.

In the weak scattering limit in which the interference effects are negligible, transport of wave through random media is well described by the classical particle diffusion theory. The average of the spatial and temporal intensity over a random ensemble therefore follows the diffusion equation \cite{3,4,5},
\begin{equation} 
\frac{\partial I(r,t)}{\partial t}-D\nabla^2I(r,t)+\frac{1}{\tau_a}I(r,t)=Q(r,t)
\end{equation}
where $I(r,t)$ is the intensity, $\tau_a$ is the absorption time, $D$ is the diffusion coefficient and $Q(r,t)$ is the source term. When the scattering becomes stronger, the probability of a wave returning to the same coherence volume after a random sequence of scattering events and form a closed loop increases. With the presence of time reversal symmetry, there exists an identical closed loop in the ``time-reversed" sense. This is sketched in Fig. 1-1.
\begin{figure}[htbp]
\centering
\includegraphics[width=4in]{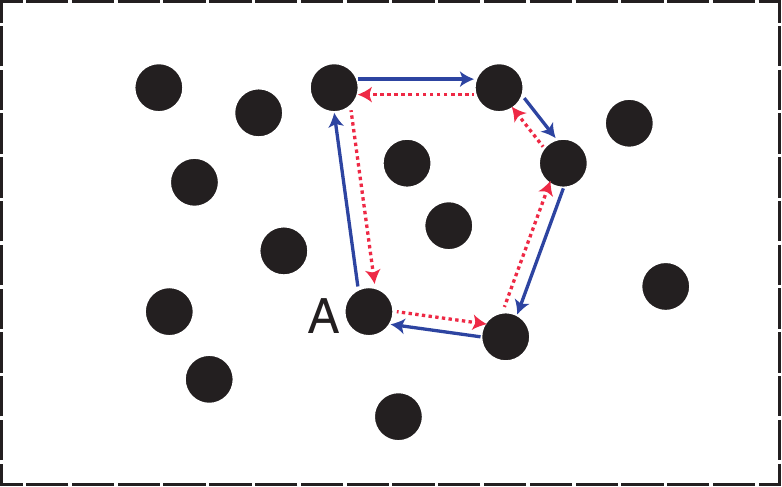}
\caption{Weak localization effect. Interference between a closed wave path and its time-reversed path leads to a higher probability of wave returning to position A.} \label{Fig1}
\end{figure}
Due to constructive interference, the probability of a wave returning to the same coherent volume is twice of the value when the interference is absent. This could be readily observed in the enhanced backscattering of light from the surface of random media, where the peak in the ensemble average of retroreflection is twice of the background \cite{6,7,8}. Coherent backscattering is also known as weak localization effect and results in suppression of the transport through random media below the level of incoherent diffusion. As the strength of scattering increases, the return probability to the coherent volume grows. When the return probability is close to unity, wave becomes exponentially localized within the media and diffusion ceases. This is known as Anderson localization \cite{9}, which we will discuss in the following section.  

\section{Anderson localization}
\subsection{Localization of electron in disordered conductors}
In 1958, Anderson \cite{9} proposed that beyond a certain threshold in disorder the electron wave function within a material becomes exponentially peaked and a good conductor becomes an insulator. Subsequently, Thouless \cite{10,11} showed that the ratio of the average energy level width to the spacing of energy levels of the sample, $\delta=\frac{\delta E}{\Delta E}$, appears to be a natural parameter to characterize localization of electrons in disordered lattices. The energy levels are quasi-normal modes in open systems and are often referred to as modes. The Thouless number $\delta$ is a measure of the sensitivity of energy level to the changes at the sample boundaries. When an electron eigenstate is localized exponentially within the sample, it is insensitive to the changes happening at the boundary since it is only weakly coupled to the surroundings. The lifetime of that eigenstate is therefore long and its corresponding level width is narrow, $\delta E<\Delta E$. On the other hand, for an extended eigenstate, the coupling to surroundings is strong and electron readily escapes from the sample and its lifetime is short and the associated level width is large, $\delta E>\Delta E$. Thus, the metal-insulator transition occurs at $\delta=1$. 

A further step was taken by Thouless, when he argued that $\delta$ is linearly related to the dimensionless conductance $\textsl{g}=\frac{G}{e^2/\hbar}$ \cite{11}. $G$ is the electronic conductance; $e$ is the electron charge and $\hbar$ is the Plank constant. Therefore, the threshold for Anderson localization lies at $\textsl{g}=1$. This is important since in many circumstances, $\delta$ is not experimentally available while $G$ can be readily measured. Shortly afterwards, a scaling approach to Anderson localization was developed by Abrahams, Anderson, Licciardello, and Ramakrishnan, in which the dimensionless conductance $\textsl{g}$ is considered as the single parameter that controls the behavior of $\textsl{g}$ as a function of system size $L$ \cite{12}. For diffusive conductors, an Ohmic scaling law is valid, $\textsl{g}=\sigma L^{d-2}$, where $\sigma$ is the conductivity and $d$ is the dimensionality of the sample. When localization occurs, the dimensionless conductance $\textsl{g}$ decreases exponentially with $L$, $\textsl{g}(L)\sim e^{-L/\xi}$, where $\xi$ is the localization length. The scaling theory shows that in disordered 2D systems, all electron eigenstates are weakly localized. While in 1D disordered systems, it is rigorously proved that all eigenstates are localized no matter how weak the disorder is. It is also shown that in 3D disordered systems,a metal-insulator transition exists. 

\subsection{Localization of classical waves in random media}
It was later realized that the Anderson localization is ultimately a wave phenomenon, in which wave interference is the key ingredient. Therefore, Anderson localization can be realized for classical waves in random media. In 1987, John \cite{13} proposed that photons can be localized by introducing disorder in a structure with periodic alternating refractive index. Compared with localization of electrons in disordered conductors, the advantage of studying Anderson localization of photons is the fact that temporal phase coherence is preserved when photons are scattered by static disorder. In electronic systems, to suppress the electron-phonon scattering, the sample has to be cooled down below 1K so that the electron wave is temporally coherent throughout the entire sample \cite{14,15}. In addition, since photons are bosons, there is no mutual interaction, while the Coulomb interaction between electrons is inevitable. This makes the transport of photons in random media an ideal model for observing pure Anderson localization without any ambiguity.  

By Landauer-Fisher-Lee relation \cite{16,17}, average of the transmittance $T$ over random ensemble collections of samples is equal to the dimensionless conductance $\textsl{g}$, $\langle T\rangle=\textsl{g}$. The optical transmittance $T$ is obtained by summing over all possible pairs of transmission coefficients from incoming channels $a$ to outgoing channels $b$, $T=\sum_{a,b=1}^{N} |t_{ba}^2|$, where $N$ is the number of channels allowed in the systems. Squaring the field transmission coefficient  $t_{ba}$ will give the intensity transmitted from incident channel $a$ to outgoing channel $b$. These channels can be independent points on the front and back of the sample. In optics, it can also refer to different illumination and detection angles of light radiation. Besides the optical transmittance $T$, we could also measure and study the intensity $T_{ba}=|t_{ba}^2|$ and the total transmission $T_a=\sum_{b=1}^{N}T_{ba}$ for transport of classical wave through random media. A general picture of measurable quantity for wave scattering in random media is shown in Fig. 1-2.
\vspace{1em}
\begin{figure}[htc]
\centering
\includegraphics[width=4in]{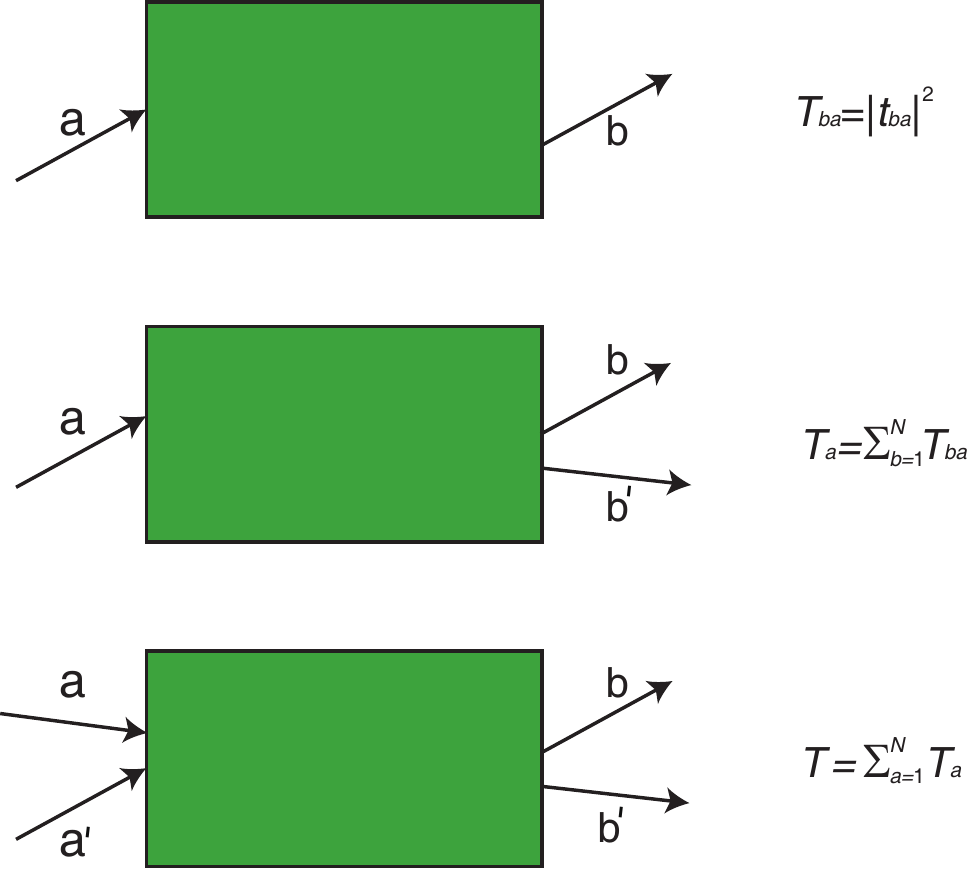}
\caption{Sketch of three transmission quantities: transmission $T_{ba}$, total transmission $T_a$ and transmittance $T$.} \label{Fig1}
\end{figure}   

The direct way to claim photon localization will be the observation of the exponentially decay of the intensity profile inside the random media \cite{18}. Nevertheless, this is in general not possible in experiments for high dimension disordered media, especially for 3D systems. An indirect evidence for photon localization is the exponentially decay of transmittance $T$ with respect to the increase of sample length $L$, $T(L)\sim e^{-L/\xi}$, which is based on the scaling theory of Anderson localization. This approach is compromised when there is loss inside the random media, since absorption will also cause the exponential decay of transmission \cite{19,20}. Furthermore, impact of absorption itself actually suppresses the Anderson localization. This is easy to understand since the partial waves with longer trajectories inside the random medium will have better chances to visit the same coherence volume and interfere with themselves which lead to localization. But the probability of obtaining partial waves with trajectories greater than the absorption length $L_a$ is greatly reduced by the absorption. To overcome the ambiguity of photon localization raised by absorption, a statistical approach to photon localization in random media was developed, in which the large fluctuation of transmission was considered as a signature of approaching photon localization \cite{21}. For localized waves, the mode is spectrally isolated, so the transmission is largely due to coupling to a single mode or few modes with similar output speckle patterns. Thus, the transmission varies dramatically when tuning the energy on and off resonance with certain mode. On the other hand, because of the spectrally overlap of modes for diffusive waves, the wave is always on resonance with several modes. Therefore, the transmission spectrum will be smooth and the fluctuation low. Hence, though the average of transmission which is severely attenuated by absorption, the fluctuation of transmission is less sensitive to absorption.
\\ \indent The probability distribution of the total transmission normalized by its ensemble average, $s_a=T_a/\langle T_a\rangle$, for a diffusive non-absorbing medium has been calculated by Nieuwenhuizen and van Romssum \cite{23} with diagrammatic method and by Kogan and Kaveh \cite{24} via random matrix theory. Assuming the density of eigenvalues of transmission matrix follows a bimodal distribution, they showed that the probability distribution of $s_a$ is:
\begin{eqnarray}
P(s_a)=\frac{1}{2\pi i}\int\limits_{-i\infty}^{i\infty}\exp(qs_a)F(q/\textsl{g})dq  \nonumber \\
F(q) = \exp(-\textsl{g}\log^2(\sqrt{1+q}+\sqrt{q}))
\end{eqnarray}
This yields the relation between the fluctuation of normalized total transmission and the dimensionless conductance $\textsl{g}$, $\text{var}(s_a)=\frac{2}{3\textsl{g}}$. We may use a new parameter $\textsl{g}^\prime=\frac{2}{3\text{var}(s_a)}$ as an indicator for approaching Anderson localization \cite{21}, where the threshold is at $\textsl{g}^\prime=1$, corresponding to $\text{var}(s_a)=\frac{2}{3}$. We will refer to $\textsl{g}^\prime$ as the statistical conductance. Once the field within the speckle pattern is normalized by the square root of the average value of the intensity within the speckle pattern, it is a complex Gaussian random variable over the entire random ensemble. This gives the Rayleigh distribution for the normalized intensity in a given speckle pattern \cite{2,22,23}. Combined with Equation 1.2, the probability distribution of transmission relative to its ensemble average, $s_{ba}=T_{ba}/\langle T_{ba}\rangle$ can be given as: 
\begin{equation}
P(s_{ba})=\int\limits_{0}^{\infty}P(s_a)\frac{\exp(-s_{ba}/s_a)}{s_a}ds_a
\end{equation}
This corresponds to the relationship between the moments of $s_a$ and $s_{ba}$,
\begin{equation}
\langle s^n_{ba}\rangle = n!\langle s^n_a\rangle
\end{equation}
and allows us to expressed the statistical conductance in terms of the fluctuation of intensity over the random ensemble, $\textsl{g}^\prime=\frac{3}{4}[\text{var}(s_{ba})-1]$. The key point here is that in some experiments, total transmission cannot be measured but transmission could always be determined. 

Localization of classical wave has been observed in 1D, 2D and Quasi-1D random media for microwave and light radiation \cite{21,24,25a,25,26}. It has been a long route to experimentally realize localization of classical waves in 3D random media. In an unbounded 3D system, the probability of returning to the same coherent volume is low. To achieve localization, the scattering has to be strong so that the scattering mean free path is comparable with the wavelength, which is the Ioffe-Regel criterion for 3D localization, $k\ell=1$ \cite{28}. Recently, Hu {\it et. al.,} \cite{27} reported localization of ultrasound in 3D random networks of aluminum beads brazed together, where the statistical conductance is found to be 0.8, just beyond the localization threshold. 

\section{Random matrix theory of wave transport in random media}
Localization of wave in disordered media is the consequence of interference of different partial waves. Even for diffusive waves $\textsl{g}\gg 1$, the wave interference leads to many striking phenomena, one of which is the universal conductance fluctuation (UCF) in disordered conductors \cite{29,30,31}. Naively speaking, one would expect that the variance of conductance increases with its mean value. However, measurements made in small metal wires and rings in low temperature uncovered an unusual fluctuations of conductance as a function of external magnetic field. It was observed that the variance of conductance is of order of unity, independent of the sample length or the strength of applied magnetic field \cite{29}. The experimental finding was first explained by Altshuler \cite{30}, Lee and Stone \cite{31} with a diagrammatic theory. Subsequently, Imry \cite{32} argued that wave transport through random media can be modeled by a large random transmission matrix and the rigidity of the eigenvalues of the random transmission matrix is the source of UCF. 

Random matrix theory (RMT) was developed into a powerful mathematical tool in the 1960’s, notably by Wigner, Dyson and Mehta \cite{33,34,35}. The original motivation was to understand the neutron scattering cross section from heavy nuclei which was related to the energy levels and level spacing of these nuclei. It is conjectured that the statistics of complex system are determined by the statistics of the eigenvalues of large random matrices. RMT is concerned with the following question: with a given large matrix whose elements are random variables with given probability laws, what is the probability distribution of its eigenvalues. The correlation functions of eigenvalues are derived from the probability distribution of the matrices. From the correlation functions one then computes the physical properties of the system. RMT was able to describe the characteristic feature of energy level statistics measured in experiments. In 1984, Bohigas \cite{36} conjectured that RMT can be applied to explain the universal behavior of level statistics in chaotic quantum spectra, in which the chaotic system can be considered as a large random Hamiltonian matrix. In open systems, the Hamiltonian is replaced with a scattering matrix. The universality of wave propagation through disordered media may therefore as well be explained by statistics of a random scattering matrix as proposed by Imry. 
\\ \indent The scattering matrix represents the solution of the Schrodinger equation for a sample that is connected to semi-infinite leads. In Fig. 1-3, a disordered sample in which the wave is temporally coherent is connected to two ideal leads.\vspace{1em}
\begin{figure}[htc]
\centering
\includegraphics[width=4in]{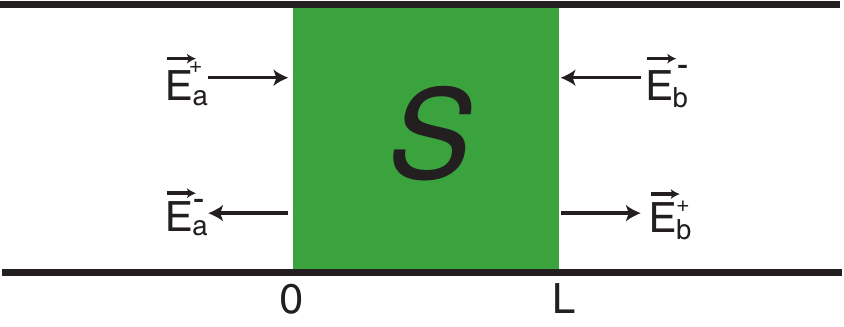}
\caption{Disordered media can be considered as a large random scattering matrix that connects the electric field in the two sample leads.} \label{Fig1}
\end{figure} 
Inside the leads, the wave can be written as a sum of incoming and outgoing propagating waves and evanescent waves. Far from the sample, only the propagating waves survive. The solution of Schrodinger’s equation provides a linear relation between incoming and outgoing waves, which we write as:
\vspace{1em}
\begin{equation}
\begin{pmatrix}
E^{-}_a\\
E^{+}_b
\end{pmatrix}
=S
\begin{pmatrix}
E^{+}_a\\
E^{-}_b
\end{pmatrix}
\end{equation}
\\where symbols {\it a} and {\it b} label the channels in the left and right leads, respectively. The plus (minus) sign identifies incoming (outgoing) waves as indicated in the sketch. The scattering matrix can be written in terms of four sub-matrices as,
\vspace{1em}
\begin{equation}
S=
\begin{pmatrix}
r \hspace{2 em} & t^\prime \\
t \hspace{2 em} & r^\prime
\end{pmatrix}
\end{equation}
\\in which each sub-matrix is a $N\times N$ random matrix. Flux conservation requires the $S$ is a unitary matrix, $SS^\dagger=I$. A direct consequence of unitarity of $S$ is that matrices $tt^\dagger, t^\prime t^{\prime\dagger},I-rr^\prime$ and $I-r^\prime r^{\prime \dagger}$ share the same set of eigenvalues $\tau_n$. The scattering matrix may be expressed in terms of the transmission eigenvalues $\tau_n$ via a polar decomposition \cite{37},
\vspace{1em}
\begin{equation}
S=
\begin{pmatrix}
U \hspace{2 em} & 0 \\
0 \hspace{2 em} & V
\end{pmatrix}
\begin{pmatrix}
-\sqrt{1-\tau} \hspace{0.5 em} & \sqrt{\tau} \\
\sqrt{\tau} \hspace{0.5 em} & \sqrt{1-\tau}
\end{pmatrix}
\begin{pmatrix}
U^\prime \hspace{2 em} & 0 \\
0 \hspace{2 em} & V^\prime
\end{pmatrix}
\end{equation}
\\Here, $U, U^\prime, V$ and $V^\prime$ are $N\times N$ unitary matrices and $\tau$ is a diagonal matrix with transmission eigenvalues along its main diagonal. Depending on the entries of the matrices, the random scattering matrices fall into three different ensembles. If the elements are real, the random ensemble is called Gaussian orthogonal ensemble. In case of complex elements or real quaternion, it is called Gaussian unitary ensemble and Gaussian symplectic ensemble, respectively. The joint probability distribution of the transmission eigenvalues $\tau_n$ can be found from the Jacobian between the volume elements $dS$ on one hand and $dU, dV, dU^\prime, dV^\prime$ and $\prod_{n=1}^N d\tau_n$ on the other hand,
\begin{equation}
dS=dUdVdU^\prime dV^\prime\prod_{m<n=1}^N|\tau_m-\tau_n|^\beta\prod_{n=1}^N\tau_n^{-1+\beta/2}d\tau_n
\end{equation}
where the index $\beta$ takes the value of 1,2 or 4 for Gaussian orthogonal, unitary and symplectic ensembles. Here, it is worth noting that the probability of having two eigenvalues close to each other is low, mimicking the repulsion between energy levels. The dimensionless conductance $\textsl{g}$ can be expressed in terms of the transmission eigenvalues, $\textsl{g}=\sum_{n=1}^N\tau_n$. It is precisely that the repulsion between transmission eigenvalues for diffusive waves which gives rise to UCF \cite{32}.   
\section{Outline}
In this thesis, we will discuss the measurement of transmission matrix for microwave propagation through random waveguides and explore the statistics of transmission on the Anderson localization from the perspective of transmission eigenvalues. In addition, the application of transmission matrix, such as focusing monochromatic and pulse transmission through multiply scattering samples will be considered. This thesis consists of 8 chapters. In the introductory chapter, it has been shown that interference plays a central role in understanding of transport of wave through disordered media. We also briefly reviewed the theory of random matrices, which is the approach we will take to describe wave scattering in random systems. In Chapter 2, we will discuss the measurement of transmission matrix for microwave radiation propagation through random waveguides in great details. Then the statistics of transmission eigenvalues on the Anderson localization transition will be discussed in Chapter 3. We will show that there exists some eigenchannel through which nearly 100\% radiation energy could be transmitted. This is yet another novel consequence of wave interference. Once taking the wave interaction at the surface of the random media, we could determine the value of bare conductance for localized waves which is the value of $\textsl{g}$ as if the interference were turned off. In Chapter 4, the transmission statistics within single random samples instead of random ensembles will be discussed and we will introduce a new statistical parameter $M$, which is the participation number of the transmission eigenvalues. The statistics of transmission within a large transmission matrix depends only on the value of $M$ and the value of $T$ serves as an overall normalization. After that, in Chapter 5, we present the distribution of transmittance, known as ``optical" conductance, on the localization transition. A highly asymmetric distribution of logarithm of transmittance observed just beyond the localization threshold as well as the universal fluctuation of transmittance found for diffusive samples will be explained in terms of an intuitive Coulomb charge model for the transmission eigenvalues. In Chapter 6, a review of  the method developed in the field of focusing transmission and imaging through random media will be presented. The focusing parameters such as the resolution and contrast will be related  to the property of the random systems. The determination of density of states of a disordered medium and the intensity profiles for each eigenchannel will be discussed in Chapter 7. We will summarize our findings in Chapter 8. 

\chapter{Measuring the transmission matrix for microwave propagating through random waveguides}
\section{Introduction}
No matter how complicated it is, a closed system can be well characterized by its Hamiltonian, giving its energy levels and associated wave functions. Open systems are described analogously by their scattering matrix, which provides the system response to the incoming excitation. For instance, the scattering matrix gives the transmission coefficient $t_{ba}$ and reflection coefficients $r_{ba}$ for excitation in ``channel" {\it a} and detection in ``channel" {\it b}. In this chapter, we will introduce an experiment for measuring the microwave transmission matrix {\it t} of random waveguides. The transmission matrix is part of the scattering matrix, which encompasses all transmission information regarding the random system. 
\section{Measurement of transmission matrix}
\subsection{Samples and experimental setup}
\begin{figure}[htc]
\centering
\includegraphics[width=4in]{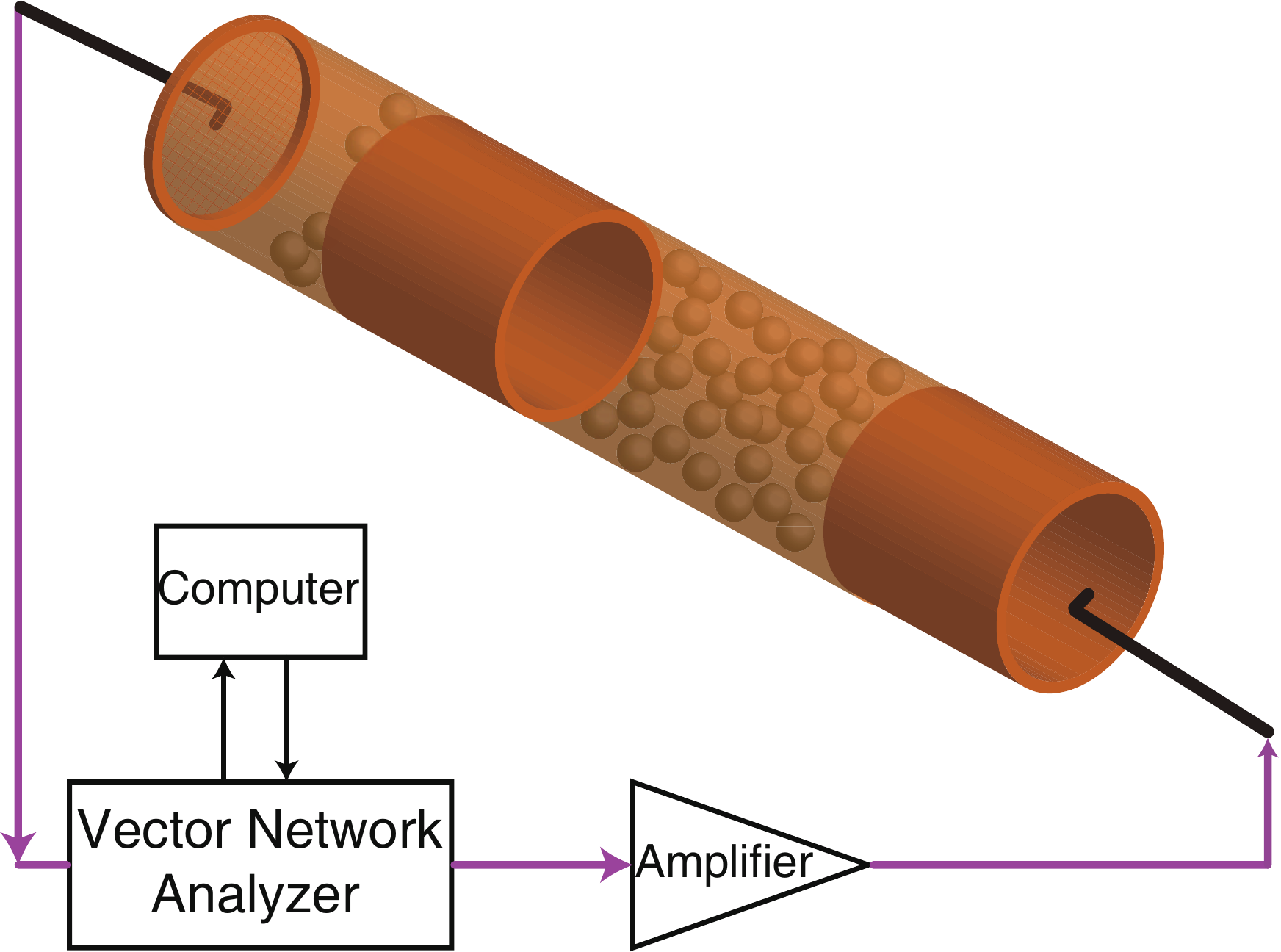}
\caption{Sketch of experimental setup. Collections of alumina spheres embedded in Styrofoam spheres are randomly positioned inside a copper waveguide.} \label{Fig1}
\end{figure} 
The random sample is a copper tube of diameter 7.3 cm filled with randomly positioned 99.95\% alumina spheres of diameter 0.95 cm and refractive index 3.14 embedded in Styrofoam spheres of diameter 1.9 cm and refractive index 1.04 to achieve an alumina filling fraction of 0.068. The tube was 99.999\% copper to reduce losses in reflection. The dimensionality of the sample can be considered as quasi-one-dimension since the sample length is usually much greater than its transverse dimension, $L\gg W$. The reflecting boundary prevents wave escaping from the sides of sample and therefore greatly increases the probability of wave returning to its coherent volume. Meanwhile, the number of allowed propagation modes is reduced to the number of waveguides modes supported in the transverse dimension of the sample. This allows us to obtain samples with low values of {\textsl g} and study the phenomenon of Anderson localization. We now refer the direction of axis of the sample tube as z-direction and input and output surface as x-y plane. Polarized microwave radiation is produced by a wired dipole antenna connected to a vector network analyzer where the polarization is determined by the orientation of the antenna. The transmitted electric field coefficient is picked up by a 4-mm long wire antenna. Both antennas are mounted on a two-dimension translation state so that they can move freely on the input and output surface. The sketch of experimental setup is shown in Fig. 2.1. Spectra of transmission are measured in two frequency ranges 10-10.24 GHz and 14.7-14.94 GHz in which the wave is localized and diffusive, respectively \cite{38}. The number of propagation waveguide modes {\it N} can be estimated as $N=\frac{Ak^2}{2\pi}$, where {\it A} is the area of the transverse dimension and {\it k} is the wave vector in vacuum. This gives $N\sim 30$ and 66 in the low and high frequency ranges, in agreement with direct calculation of number of propagation waveguide modes. 
\begin{figure}[htc]
\centering
\includegraphics[width=3in]{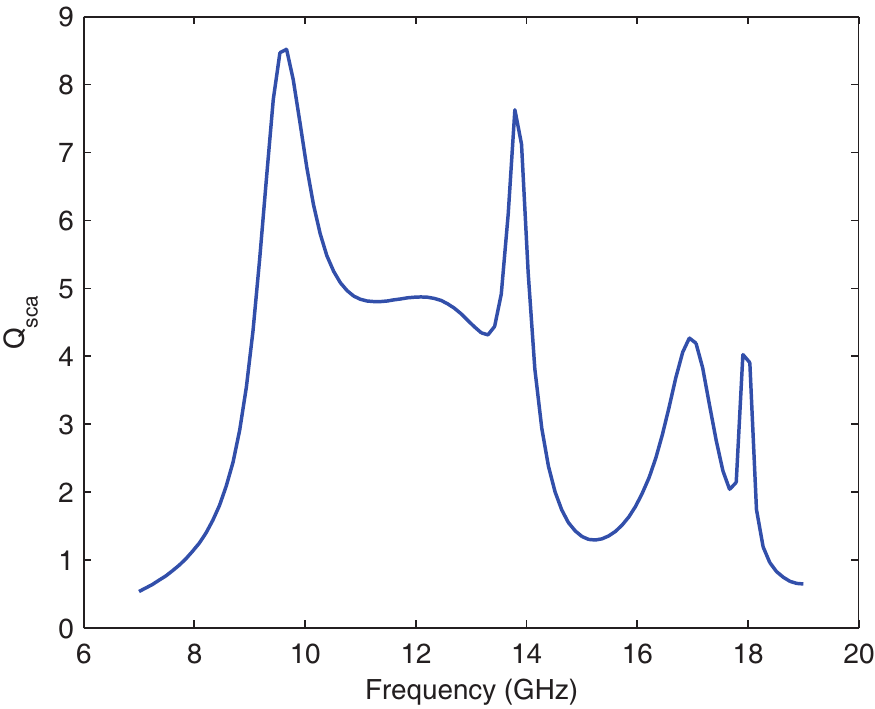}
\caption{Scattering efficient for a sphere with diameter 0.95 cm and refractive index $n=3.14$ over a broad frequency range.} \label{Fig2}
\end{figure} 
The wave is localized in the lower frequency range because scattering is strong since the frequency range is close to the first Mie resonance of the individual alumina sphere \cite{39}. In Fig. 2.2, a spectrum of scattering efficients for a single sphere over a broad frequency range is presented. Since photons are scattered by collections of alumina spheres instead of individual spheres, the window of photon localization is shifted from the first Mie resonance of the alumina sphere.

\subsection{Removing the impact of absorption from the measured spectrum}
In order to reveal the true statistics of transmission through random systems without absorption, we compensate for the affect of the loss due to absorption. Here, we will employ two different approaches for removing the impact of absorption and show that they are in excellent agreement. In the first approach, the measured spectrum of field transmission coefficient $E_{ba}(\nu)$ is multiplied by a Gaussian field, $E(\nu)=\exp(-\frac{(\nu-\nu_0)^2}{2\sigma^2})$ centered at $\nu_0$ and Fourier transformed into the time domain to give response to an incident Gaussian pulse, $E_{ba}(t)$ \cite{21}. Then the time-dependent electric field is multiplied by an exponential function $\exp(t/2\tau_a)$ to compensate for the loss due to absorption in the sample, where $\tau_a$ is the absorption rate. $\tau_a$ can be found by fitting the field-field correlation function with respect to frequency shift or to the photon time of flight distribution \cite{42,40,41}. The compensated $E_{ba}(t)$ is then Fourier transformed back to frequency domain and divided by the incident Gaussian field to give the electric field at the frequency point $E_{ba}(\nu_0)$. By sweeping the carrier central frequency of the incident pulse, we can obtain the spectrum of $E_{ba}(\nu)$. In order to avoid the edge of the measured spectrum, we only take values of $\nu_0$ from $\nu_i+3\sigma$ to $\nu_f-3\sigma$, in which $\nu_i$ and $\nu_f$ are the starting and ending frequency points in the measured spectrum, respectively. We found that the correction process is not sensitive to the width of the incident Gaussian field, provided that it is greater than the field-field correlation length in frequency. This method is tested in 1D simulation for wave propagation through a periodic structure of binary elements with two defects placed with equal distance from the center of the otherwise periodic structure. The sample is composed of periodically distributed 100-nm thick binary elements with refractive indices of $n=1$ and 1.55 and defects with thickness of one period of 200 nm and refractive index 1. The absorption level is controlled by adding an imaginary part to the refractive index. 
\begin{figure}[htc]
\centering
\includegraphics[width=2.8in]{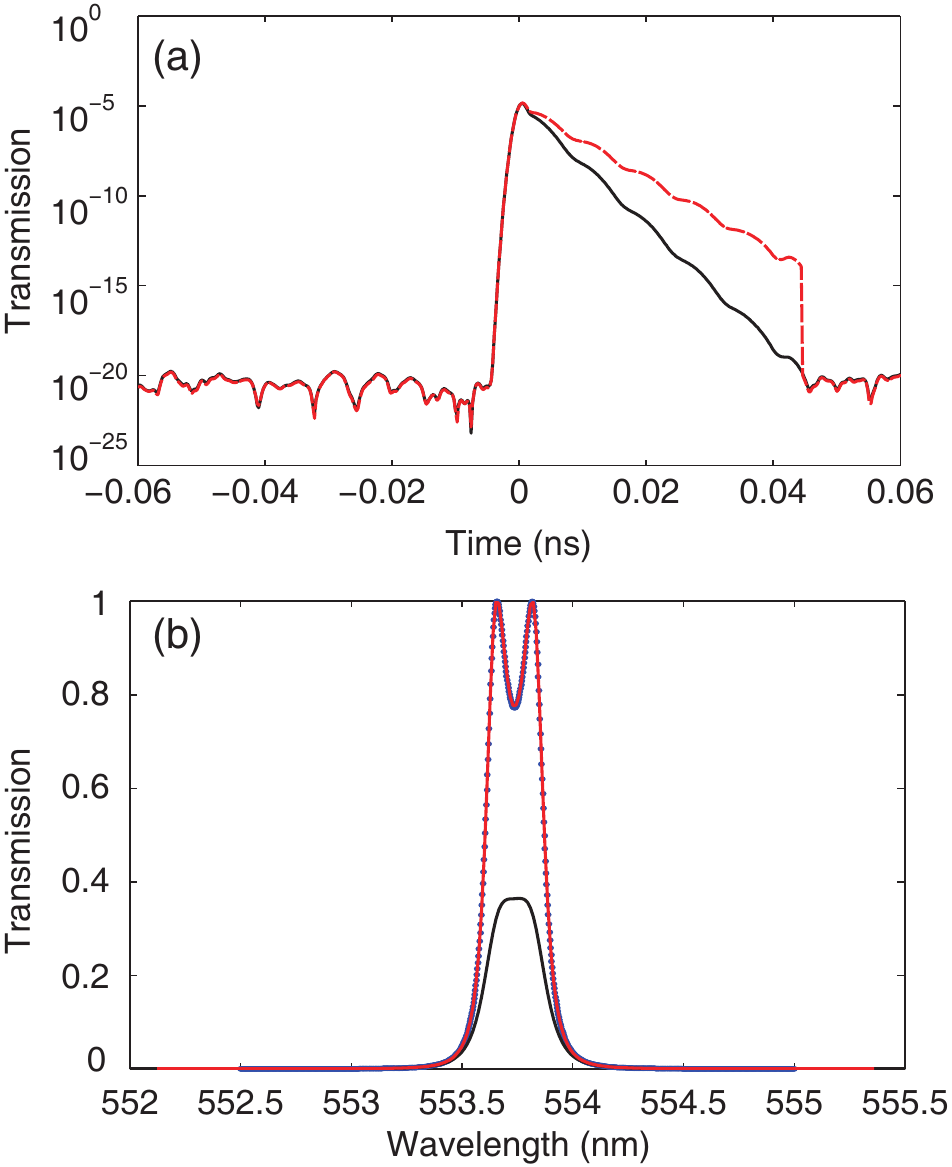}
\caption{Transmitted field spectrum before and after the impact of absorption is removed. (a) The absorption is compensated in time domain by multiplying the field by $\exp(t/2\tau_a)$. (b) The intensity spectrum when compensated for loss is the same as the intensity spectrum without absorption. The spectrum with absorption is shown in black, while the red curve shows the spectrum compensated for the loss. This curve overlaps the transmission spectrum for the sample without absorption shown in blue dots.} \label{Fig3}
\end{figure} 
Figure 2.3 shows a comparison between the spectrum of transmission without absorption and the spectrum corrected for absorption in an absorbing sample.

The second way of eliminating effect of absorption is to find the quasi-normal modes in the system \cite{43,43a,43b,44}. The measured spectrum can be decomposed into a sum of contributions from quasi-normal modes,
\begin{equation} 
E_{ba}(x,y,\nu) = \sum_{n} a_n(x,y)\frac{\Gamma_n/2}{\Gamma_n/2+i(\nu-\nu_n)}.
\end{equation}
in which $\nu_n$ is the central frequency of the $n^{th}$ mode, $\Gamma_n$ is the width of the $n^{th}$ mode and $a_n$ is the amplitude associated with the $n^{th}$ mode. The width of the mode is a sum of leakage from the open boundary as well as absorption rate inside the system $\Gamma_n=\Gamma^{leak}_n+\Gamma^{as}_n$. The impact of absorption is removed by subtracting a constant value from the width of the each mode found by modal decomposition and construct spectrum without absorption,
\begin{equation} 
E^0_{ba}(\nu) = \sum_{n} a^0_n\frac{\Gamma^0_n/2}{\Gamma^0_n/2+i(\nu-\nu_n)}.
\end{equation}
where the superscript 0 indicates the quantity without absorption and $\Gamma_n^0=\Gamma_n-\Gamma_n^{abs}$, $a_n^0=a_n\frac{\Gamma_n}{\Gamma_n^0}$. $\Gamma_n^{abs}$ is equal to the inverse of absorption rate $\tau_a$. Two ways of accounting for absorption are shown in Fig. 2.4 and an excellent agreement is found.
\begin{figure}[htc]
\centering
\includegraphics[width=3in]{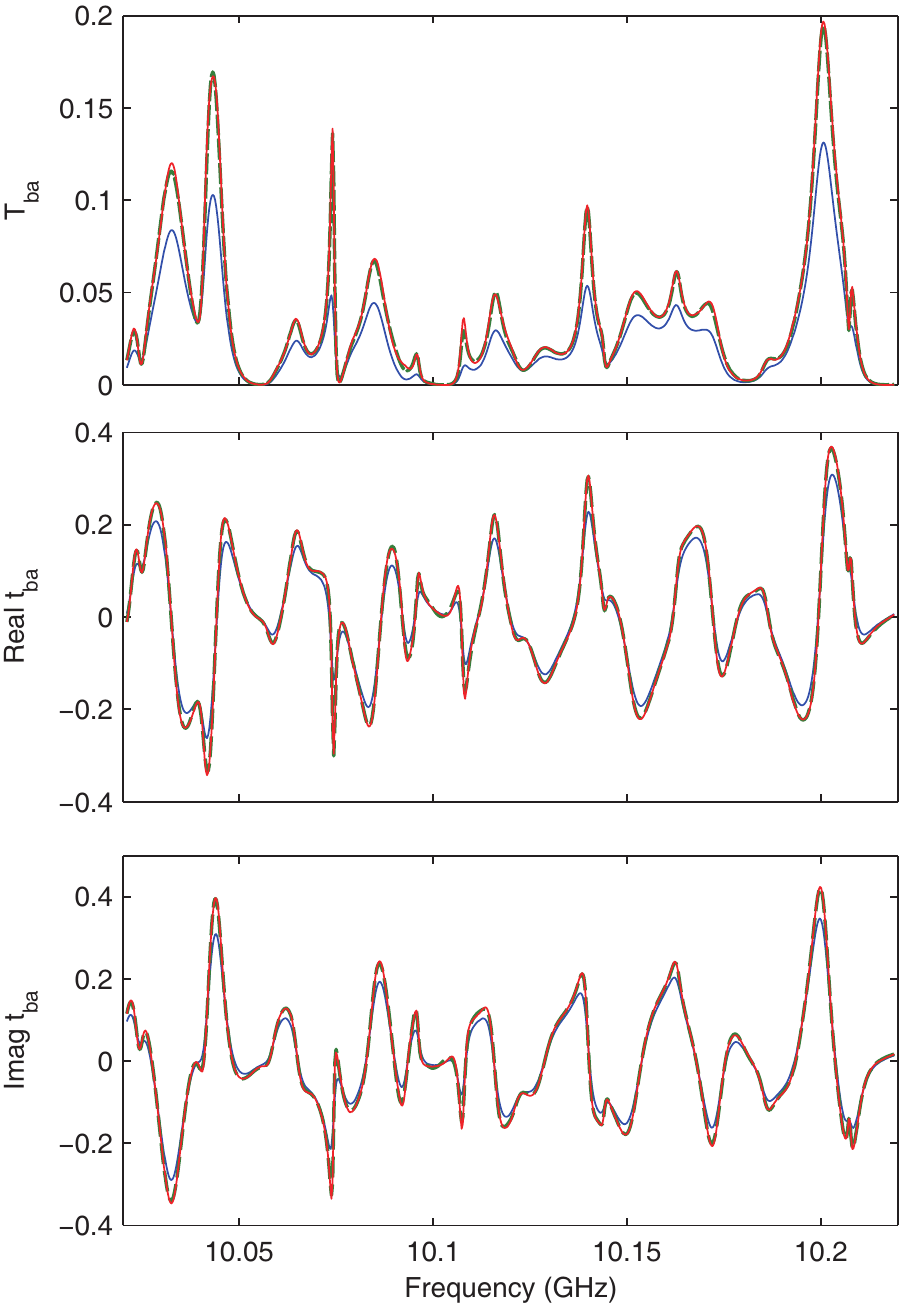}
\caption{Comparison of two methods to remove the impact of absorption. The blue line is the measured spectrum, the dashed green line is the spectrum compensated for loss by subtracting a constant from the width of the modes inside the medium and the red line is obtained by compensating the loss in time domain. } \label{Fig4}
\end{figure} 

\subsection{Measuring the transmission matrix}
The transmission matrix for microwave propagation through random waveguides is recorded on arrays of points on the input and output surface by translating the antennas \cite{45}. Two orthogonal polarized excitation along {\it x} and y {\it i}s produced by rotating the source antenna between $x$ and $y$ direction. The transmitted electric field coefficient is as well measured for two orthogonal polarization, namely $E_x$ and $E_y$. The complete transmission matrix is then,
\begin{equation}
S=
\begin{pmatrix}
t^{xx} \hspace{2 em} & t^{xy} \\
t^{yx} \hspace{2 em} & t^{yy}
\end{pmatrix}
\end{equation}
where the superscript $x$ and $y$ indicates the polarization on the input and output surfaces.

The real part of the field-field correlation with displacement of the source or detection antenna on the output plane is:
\begin{equation}
F_E(\Delta r)=\frac{\sin k\Delta r}{k \Delta r}\exp(-\frac{\Delta r}{2\ell_s}).
\end{equation} 
in which $\Delta r$ is the displacement, $k$ is the wave-vector in vacuum and $\ell_s$ is the scattering mean free path. The imaginary part of the field-field correlation function in space is predicted to be zero \cite{41}. Thus, the correlation function vanishes at a series of spacings, the first of which at $\delta r=\lambda/2$, where $\lambda$ is the wavelength of the incident wave in vacuum. Therefore, the number of independent measured transmission coefficient on the output or number of independent excitation position on the incident plane for single polarization is $\sim \frac{A}{(\lambda/2)^2}$. There are extra degrees of freedom due to the polarization of a photon. The real part of the field-field correlation of the transmitted field for different polarization excitation is $F_E(\Delta \theta)=\cos(\Delta \theta)$ \cite{46}. Without losing any generality, we will refer the polarization along-x-direction as $\Delta\theta=0$. The imaginary part of the field-correlation function with respect to polarization will be zero. Including the spatial freedom, the size of the transmission matrix will be $\sim\frac{2A}{(\lambda/2)^2}$. To construct the transmission matrix, $N/2$ points are selected from each of the polarizations.
 
\subsection{Mesoscopic correlation and enhanced transmission fluctuations}
Though field-field correlation vanishes beyond certain characteristic length, intensity is correlated well beyond that length. This non-local intensity correlation give rises to the enhanced mesoscopic fluctuations of transmission \cite{47,48,49,50,51,52,53,54,55,56,57}. For instance, one bright spot in the transmitted speckle pattern could result in high transmission in all the speckle spots because of this non-local correlation, which leads to enhanced fluctuation of total transmission $T_a$. Microwave measurement suggests, and diagrammatic calculations confirm, that the cumulant correlation function of the normalized intensity can be expressed as the sum of three terms in Q1D samples. Each term involves only the product or sum of the square of the field-field correlation function normalized to the square root of the average intensity \cite{46,57}. This gives,
\begin{equation}
C_I=F_{in}F_{out}+A_2(F_{in}+F_{out})+A_3(F_{in}F_{out}+F_{in}+F_{out}+1).
\end{equation}
where $F_{in}$ and $F_{out}$ are the squares of the field correlation function with respect to change of position or polarization of the source and detector, respectively. The cumulant correlation function can further be expressed as the sum of multiplicative, additive and constant terms,
\begin{equation}
C_I=(1+A_3)F_{in}F_{out}+(1+A_2)(F_{in}+F_{out})+A_3.
\end{equation}
For diffusive waves, the multiplicative, additive, and constant terms, which correspond to short-, long-, and infinite-range contributions to $C_I$, dominating fluctuations of intensity, total transmission and transmittance \cite{4}. $A_2$ is of order 2/3{\textsl g} and A3 is of order 2/15$\textsl{g}^2$. In Fig. 2.5, we show a spectrum of normalized intensity, total transmission and transmittance for both localized and diffusive waves. The lack of self-averaging for localized waves is clearly seen.
\begin{figure}[htc]
\centering
\includegraphics[width=4in]{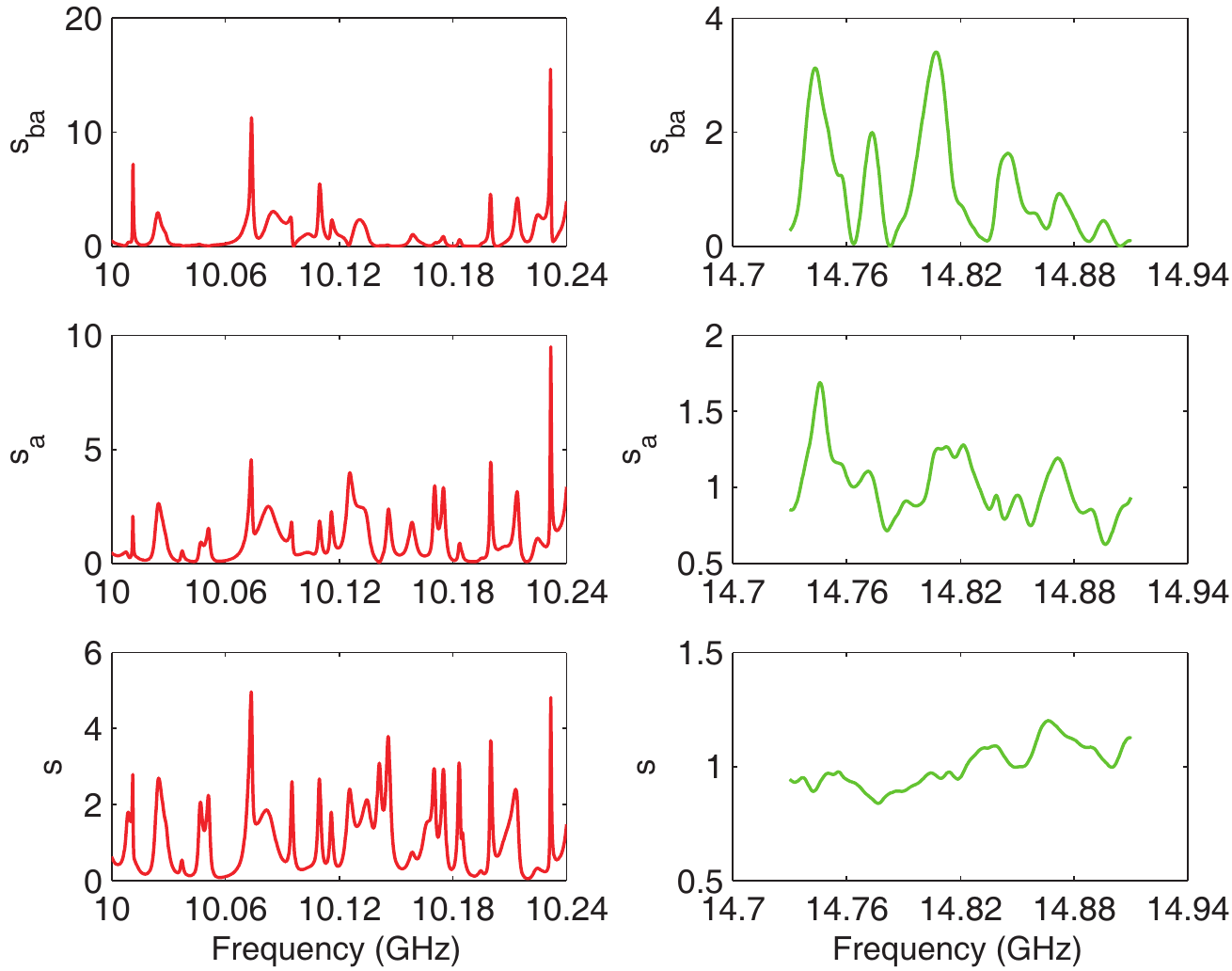}
\caption{Spectra of normalized intensity $s_{ba}=T_{ba}/\langle T_{ba}\rangle$, total transmission $s_a=T_a/\langle T_a\rangle$ and transmittance $s=T/\langle T\rangle$ for one sample realization for diffusive wave with {\textsl g=6.9} and localized wave with {\textsl g}=0.37.} \label{Fig6}
\end{figure} 

\subsection{Generic speckle pattern}
We have shown that non-local intensity in the transmitted speckle pattern leads to large fluctuation in the speckle pattern, from which we could separate diffusive wave from localized waves. Nonetheless, a single measurement of speckle pattern is generic in that their structure and statistics are robust under perturbation and are governed by Gaussian field statistics. The impact of mesoscopic correlation also vanishes when the field measured at every point in the speckle pattern is normalized to the square root of the average transmission within the speckle pattern. In this sense, sets of normalized speckle patterns are also generic since it only reflects the Gaussian statistics \cite{2,58}. This is seen in Fig. 2.6, in which probability distribution of $T_{ba}/{T_a}$ for a localized ensemble is plotted and seen to follow an exponential function. 
\begin{figure}[htc]
\centering
\includegraphics[width=3in]{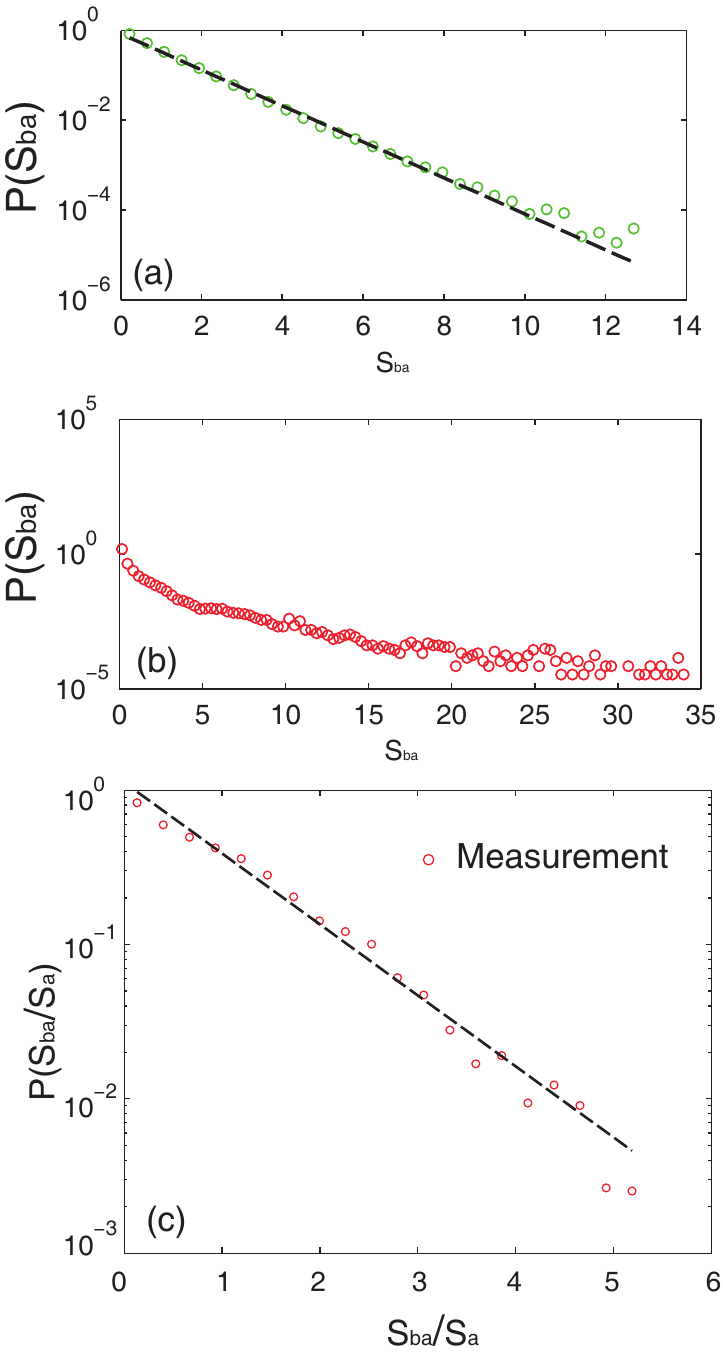}
\caption{Generic speckle pattern. The probability distribution of the intensity $T_{ba}$ normalized by the total transmission within the speckle pattern $T_a$ is a universal negative exponential. Probability distribution of $s_{ba}$ for a diffusive ensemble (a) and a localized ensembel (b). $P(s_{ba})$ is fitted with a negative exponential function for the diffusive sample. When normalized by the total transmission, $P(s_{ba}/s_a)$ is seen to follow an exponential function for the localized ensemble.} \label{Fig8}
\end{figure} 

\subsection{Some experimental details}
In order to improve the statistics, the transmission matrix is always measured over a range of frequencies. Thus, we have to remove the frequency response due to the experimental apparatus so that the statistics over the measured spectrum can be considered as equivalent. To do this, we usually measure thousands of spectra of $t_{ba}$ for antennas centered at both the input and output surface and we obtain the average intensity over the spectrum, $\langle I_{ba}(\nu)\rangle$. The measured spectrum is then divided by the square root of $\langle I_{ba}(\nu)\rangle$ to get rid of the frequency response. We also find that due to the limit number of waveguide modes in the measured frequency ranges and coupling efficiency of the antenna to the empty waveguides, the ensemble averaged intensity patterns of the transmission matrix is not uniform. Near the boundary of the waveguides, the coupling into the system is poor and the transmitted intensity is low. Therefore, when constructing the transmission matrix, we should select points where the average transmission is about the same. Because of the finite size of the antenna, what we actually measured is the electric field integrated along the length of the wire antenna and we found that there is some residual correlation between {\it x} and {\it y} polarized excitation which should be zero in the ideal case. 

\section{Summary}
In this chapter, we have discussed measurements of microwave transmission matrix on a gird, in which the number of statistically independent points in the grid is about the same as the number of free propagating waveguide modes allowed in the empty waveguide. In this manner, the subtle mesoscopic intensity correlation is preserved in the measurement and therefore the transmission matrix reflects the nature of wave propagation through random systems. We have also introduced two ways of removing the impact of absorption on the measured transmission matrix, which is crucial since absorption can dramatically change the structure of the eigenvalues of the transmission matrix. 

\chapter{Transmission eigenvalues in the Anderson localization transition}
\section{Introduction}
An experimentally important difference between classical and quantum transport is that coherent propagation is the rule for classical waves such as sound, light and microwave radiation in granular or imperfectly fabricated structures, whereas the electron waves are only coherent at ultralow temperatures in micron-sized samples. For classical waves in “static" samples, the wavelength is typically long compared to the scale of thermal fluctuations so that the wave remains temporally coherent within the sample even as its phase is random in space. In contrast, mesoscopic features of transport in random systems are achieved only in samples with dimensions of several microns at ultralow temperatures \cite{14,15}. Electrons are typically multiply scattered within conducting samples so their dimensions are larger than the electron mean free path, which is on the scale of or larger than the microscopic atomic spacing and electron wavelength. At the same time, electronic samples are typically smaller than the macroscopic scale on which the electrons are inelastically scattered so that the wave function is no longer coherent. Thus mesoscopic electronic samples are intermediate in size between the microscopic atomic scale and the macroscopic scale. In contrast, monochromatic classical waves are generally temporally coherent over the average dwell time of the wave within human-sized samples. It is therefore possible to explore the statistics of mesoscopic phenomena with classical waves. Such studies may also be instructive regarding the statistics of transport in electronic mesoscopic samples.

It was conjectured that statistics of transmission through random systems are determined by the eigenvalues of large transmission matrices \cite{32}. The field transmission matrix $t$ connects the transmitted wave in channel $b$ to the incident wave in channel $a$, $E_b=\sum_{a}t_{ba}E_a$ \cite{37,59,60}. Via a singular value decomposition, the field transmission $t$ can be expressed as $t=U\Lambda V^\dagger$. Here, $U$ and $V$ are unitary matrices, of which the elements are complex Gaussian random variables and $\Lambda$ is a diagonal matrix with the singular value $\lambda$ of the matrix $t$ along the diagonal \cite{61}. Summing all the elements in the transmission matrix $t$ gives the transmittance $T$, $T=\sum_{a,b}|t_{ba}|^2$. The ensemble average of the transmittance is equal to the dimensionless conductance {\textsl g}, $\langle T\rangle={\textsl g}$ \cite{16,17}. The transmittance may as well be expressed in terms of the eigenvalues $\tau_n$ of the matrix product $tt^\dagger$, $T=\sum_{n}\tau_n$. Random matrix theory predicts that the transmission eigenvalues follow the bimodal distribution, $\rho (\tau)=\frac{{\textsl g}}{2\tau \sqrt{1-\tau}}$ \cite{101,61,62,63}. Most of the contribution to $T$ comes from approximately {\textsl g} eigenvalues that are larger than $1/e$ \cite{32,37,60,61,62}, while most of the eigenvalues are close to zero. The characteristic of these ``open" and``closed" channels were first discussed by Dorokhov \cite{60} where he considered the conduction of electrons in disordered conductors. For classical waves, the existence of the ``open" channels indicates that even when the sample length $L$ is much greater than the mean free path $\ell$, near 100\% transmission is possible when the incident wave corresponds to the eigenvector associated with the highest transmission eigenvalues, which is crucial in biomedical imaging where the samples are usually optically opaque.   

\begin{figure}[htc!]
\centering
\includegraphics[width=4.5in]{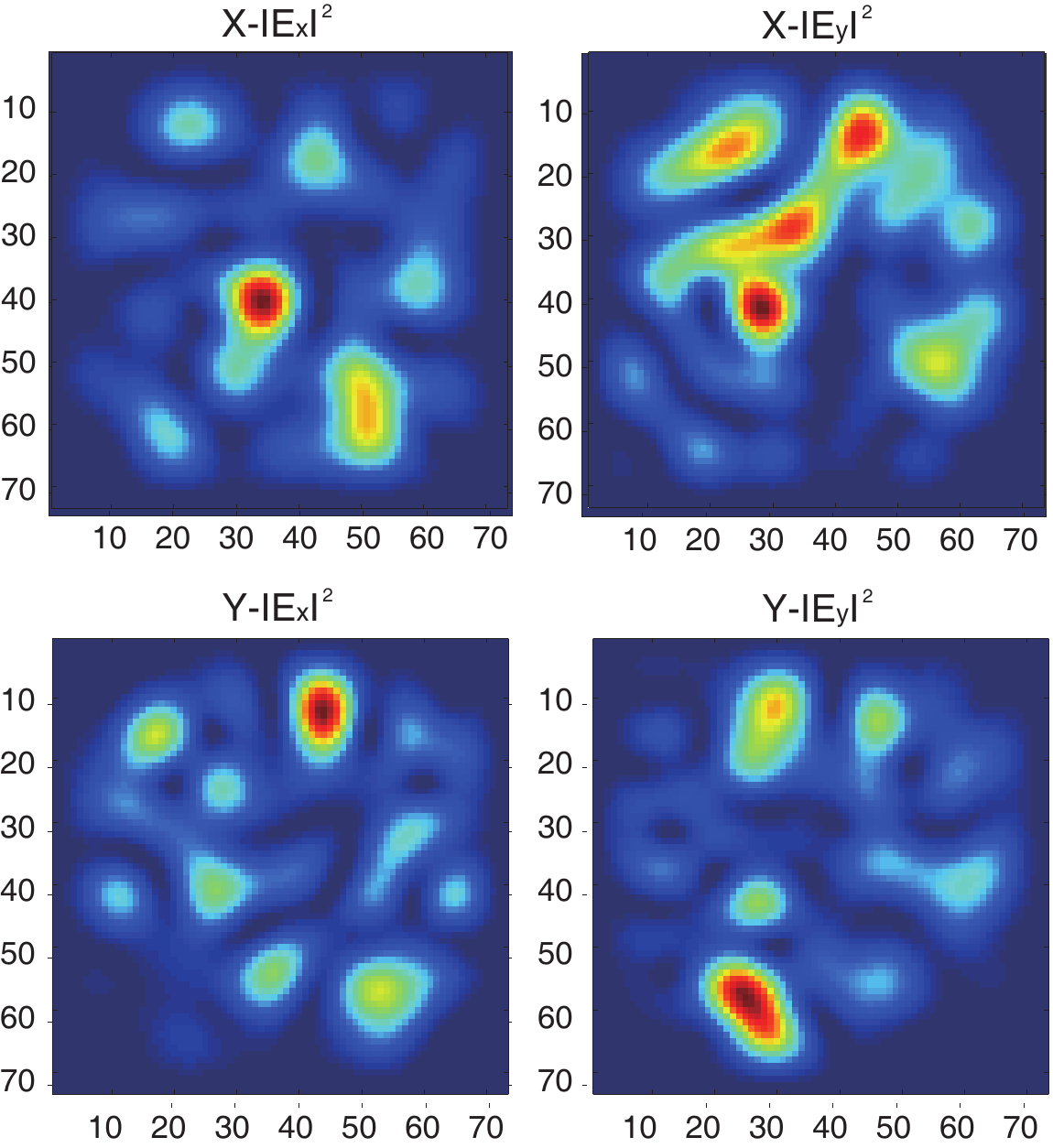}
\caption{Speckle patterns of $|E_x|^2$ and $|E_y|^2$ measured for two perpendicular polarized excitation along X and Y direction for one random diffusive sample of 61 cm.} \label{Fig3.1}
\end{figure} 
We have discussed in Chapter 2 that the microwave transmission matrix is built upon measurements of transmitted electric field, $E_x$ and $E_y$ for two perpendicular polarized excitation. In Fig. 3.1 and 3.2, we present speckle patterns of $|E_x|^2$ and $|E_y|^2$ for a source antenna placed at the center of the incident surface for two polarizations at a single frequency for both diffusive and localized waves with sample length of 61 cm. The Whittaker-Shannon 2-D sampling theorem is applied to obtain the speckle patterns.
\begin{figure}[htc!]
\centering
\includegraphics[width=4.5in]{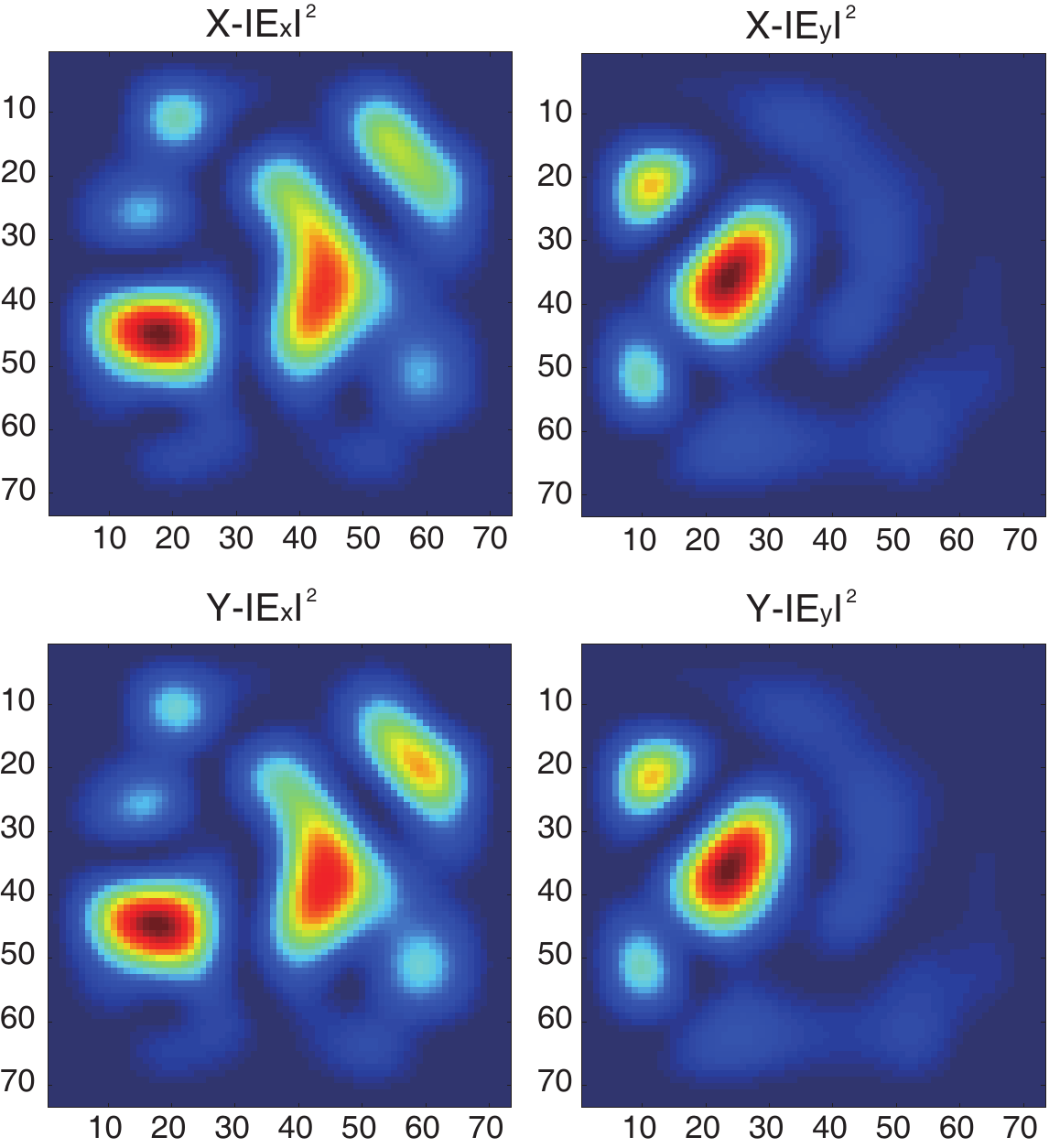}
\caption{Speckle patterns of $|E_x|^2$ and $|E_y|^2$ measured for two perpendicular polarized excitation X and Y for one random localized sample of 61 cm. For two orthogonal excitation of the random sample, the transmitted intensity speckle patterns are similar.} \label{Fig3.2}
\end{figure}  
The transmission matrix is constructed by selecting $N/2$ points from each of the polarization and images of the transmission matrix corresponding to the configurations in Figs. 3.1 and 3.2 are shown in Fig. 3.3. 
\begin{figure}[htc!]
\centering
\includegraphics[width=4.5in]{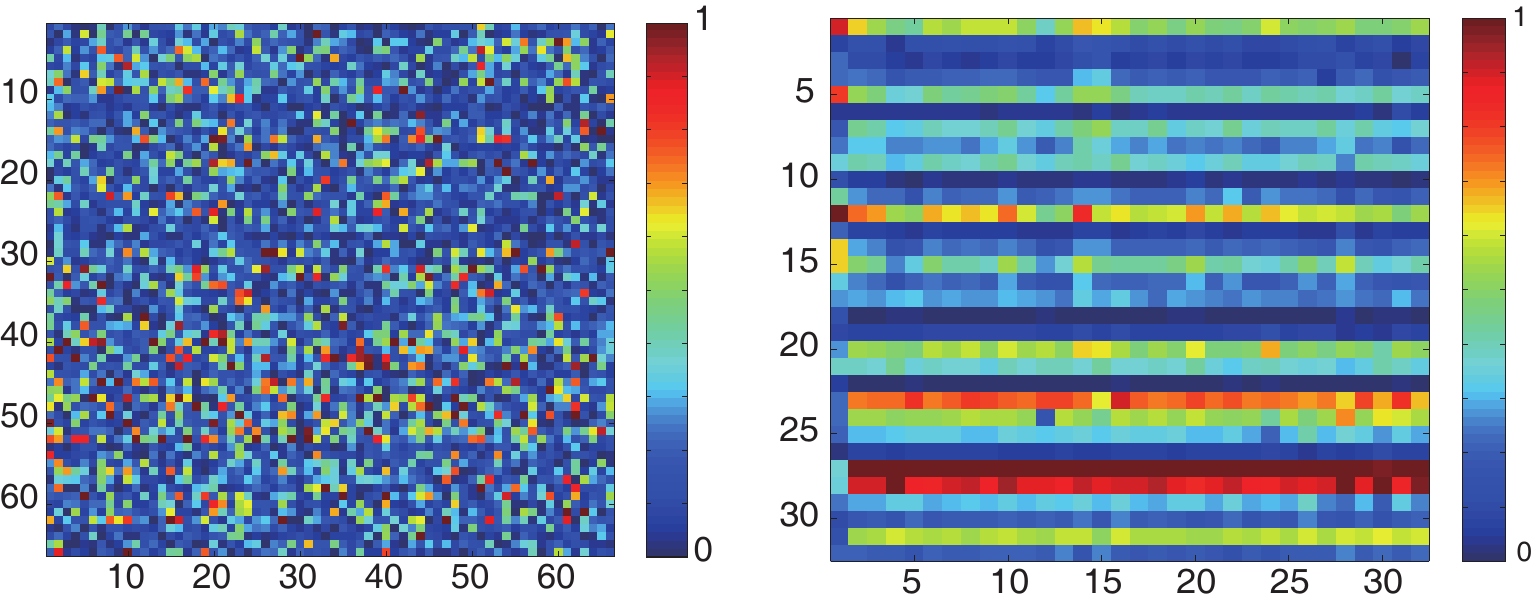}
\caption{Images of intensity within a single transmission matrix. The intensity in each column represents the transmission of incident waves from one channel {\it a} on the input surface with a specified polarization to all possible $N$ output channels {\it b} including two orthogonal polarization. The intensity within each column is normalized by its maximum value within the column. The intensity patterns from different input channels {\it a} are similar for localized waves.} \label{Fig3.3}
\end{figure}  

\section{Localization lengths of eigenchannels}
In 1982, Dorokhov \cite{59} calculated the conduction of electrons through $N$ parallel disordered chains with weak transverse coupling, where he showed that even for conductors, the current in most channels would be exponentially small so that the conduction is dominated by a number of high conducting channels with $\tau_n\ge 1/e$. Thus the number of such channels will be close to the conductance {\textsl g}. These conducting channels were later termed as active eigenchannels by Imry \cite{32}. For localized samples , Dorokhov \cite{59} proposed that there exist $N$ different localization lengths  associated with each of the transmission eigenchannel with the inverse of the localization length given by:
\begin{equation}
\frac{1}{\xi_n}=\frac{2n-1}{2N\ell}.
\end{equation}
in which $\ell$ is the mean free path. The localization length of the isolated chain is of order $\ell$. This suggests the spacing of the inverse localization lengths between neighboring eigenchannels is equal to $1/N\ell$. In analogy with the localization length for the disordered systems, $\langle \ln T\rangle=-L/\xi$, we define the channel localization length via the relation $\langle \ln \tau_n \rangle=-L/\xi_n$. The spacing between neighboring values of $\langle \ln \tau_n \rangle$ will then be $L/N\ell$, which is equal to the inverse of the bare conductance $\textsl{g}_0$. The bare conductance is the conductance one would get if wave interference were turned off. It is inversely proportional to the sample length even for localized waves in which interference substantially suppresses transmission through random media.
\begin{figure}[htc!]
\centering
\includegraphics[width=3in]{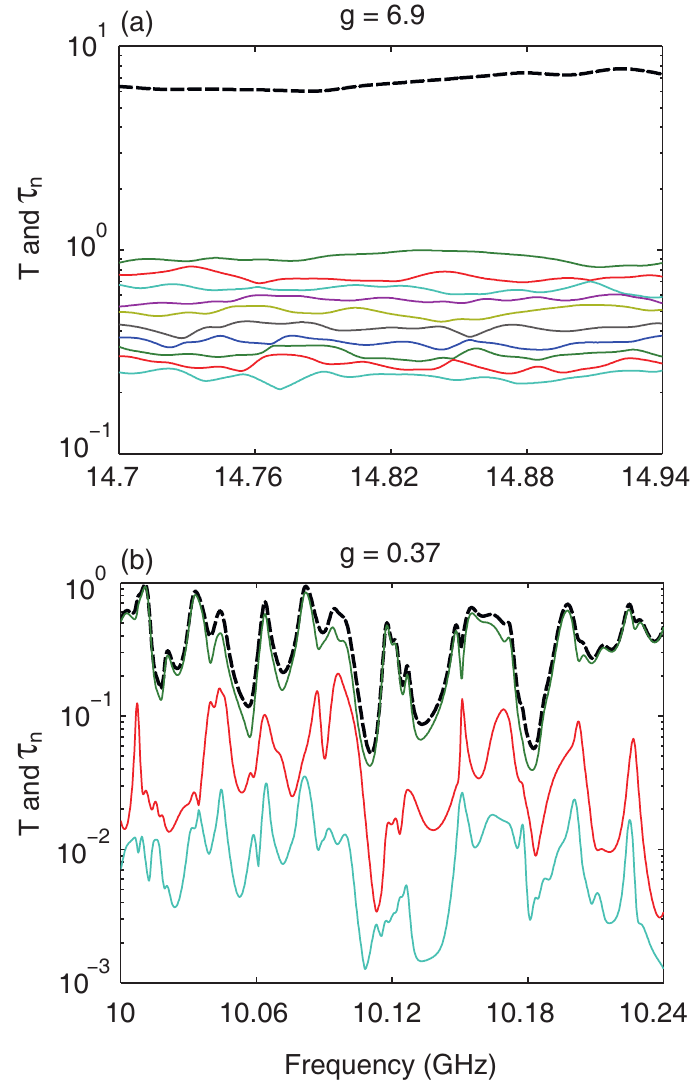}
\caption{Spectra of the transmission eigenvalues and the transmittance for one sample realization drawn from ensemble with {\textsl g}=6.9 and 0.37, respectively.} \label{Fig3.4}
\end{figure} 
In Fig. 3.4, we present the spectra of the optical conductance and underlying transmission eigenvalues for random realizations drawn from random ensembles with values of {\textsl g}=6.9 and 0.37. For diffusive waves, a number of eigenvalues contribute appreciably to the conductance {\textsl g} and the highest transmission eigenvalues is of order of unity. It is therefore possible to transmit nearly 100\% of the radiation energy through a diffusive system when coupling to the high transmitting eigenchannel. For localized waves, the highest transmission eigenvalue is seen to dominate the conductance and it is typically much lower than unity. 
\begin{figure}[htc!]
\centering
\includegraphics[width=3in]{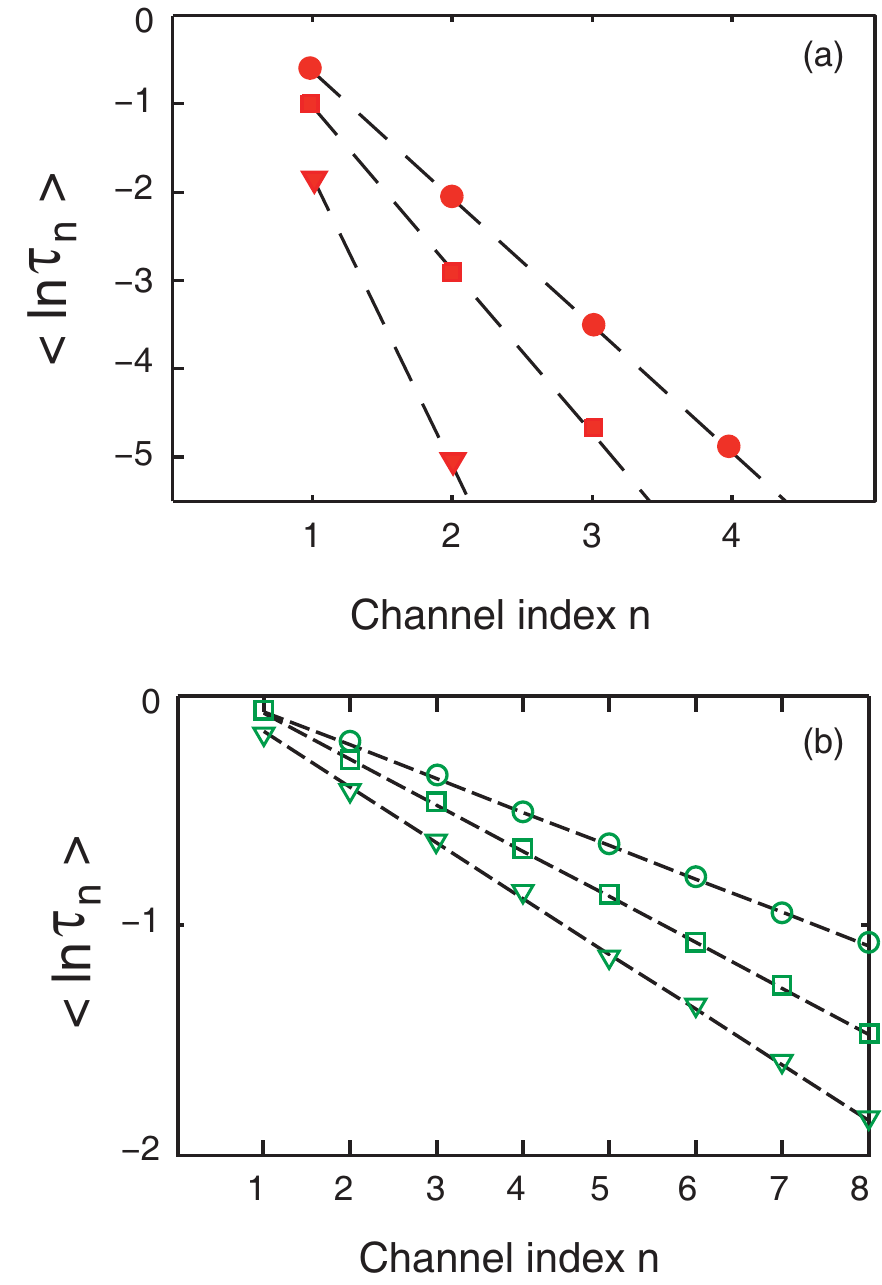}
\caption{Variation of $\langle \ln \tau_n \rangle$ with channel index {\it n} for localized waves (a) and diffusive waves (b). Sample lengths are $L$ = 23 (circle), 40 (square) and 61 (triangle) cm. The black dash lines are the fit to the data.} \label{Fig3.5}
\end{figure}   
We now consider the localization length for each of the transmission eigenchannels. The variation of $\ln \tau_n$ with respect to the channel index $n$ is presented in Fig. 3.5 (a) and (b) for localized and diffusive samples with different values of {\textsl g}, respectively. We find that for localized waves, $\langle \ln \tau_n\rangle$ falls linearly on a straight line, indicating equal spacing of $\langle \ln \tau_n\rangle$ as predicted by Dorokhov. We denote the inverse of the slope of the straight line by ${\textsl g}^{\prime\prime}$ and will identify it as the bare conductance for the localized samples. However, we find that for diffusive waves, $\ln \tau_n$ are also seen to be equally spaced.

The equal spacing of $\langle \ln \tau \rangle$ found for diffusive samples corresponds to a nearly uniform density of $\ln \tau$. The density of $\ln \tau_n$ for the most diffusive sample of $L$=23 cm is shown in Fig. 3.6. 
\begin{figure}[htc!]
\centering
\includegraphics[width=3in]{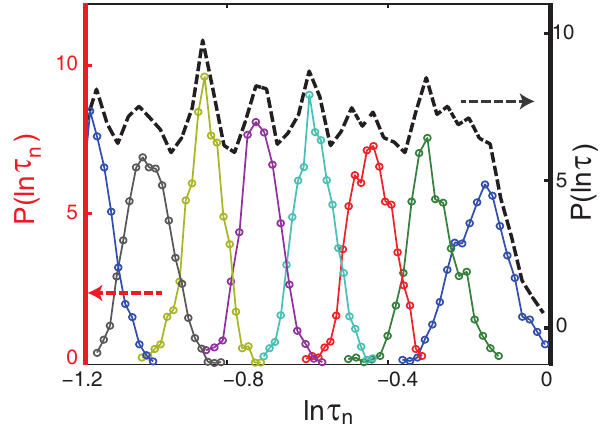}
\caption{Probability density of $\ln \tau_n$ and the density of $\ln \tau$, for diffusive samples with {\textsl g}=6.9.} \label{Fig3.6}
\end{figure}  

The probability density $P(\ln \tau)$ falls to 0 near $\ln \tau \sim 0$, which reflects the restriction $\tau_1 \le 1$. The nearly uniform density of $\ln \tau_n$ corresponds to the probability density of transmission eigenvalues, $P(\tau)=P(\ln \tau)\frac{d\ln \tau}{d\tau}={\textsl g}/\tau$. This distribution has a single peak at low values of $\tau$ in contrast to a later prediction of a bimodal distribution, which has a second peak near unity. This may be due to the difference between measurement of transmission matrix based on scattering between points on a grid as opposed to theoretical calculation based upon scattering between waveguide modes allowed in the sample leads. Since only a fraction of energy transmitted through the disordered medium is captured when the TM is measured on a grid of points, full information is not available and the measured distribution of transmission eigenvalues does not accurately represent the actual distribution in the medium. In particular, the bimodal distribution of transmission eigenvalues is not observed. In theoretical calculation in which scattering between waveguide modes is treated, all the transmitted energy can be captured. The impact of loss of information in the measurement of the transmission matrix upon the transmission eigenvalues has been calculated by Goetschy and Stone \cite{64}, which we will discuss in details in section 3.3. 

We expect that the bare conductance ${\textsl g}_0$ should be influenced by wave interactions at the sample interface, as is the case for transmission for a single incident channel \cite{5,65,66}. The impact of the interface on transmission can be found by considering the angular dependence of transmission. Measurement of total transmission in diffusive samples are well described by the expression,
\begin{equation}
T(\theta)=\frac{z_p\cos\theta +z_b}{L+2z_b}.
\end{equation}
in which $\theta$ is the angle between the normal and the wave as it penetrates into the sample. This expression is obtained from a model in which the incident wave is replaced by an isotropic source at a distance of travel from the interface, $z_p$, at a depth into the sample of  $z_p\cos\theta$, as illustrated in Fig. 3.7 \cite{5}. 
\begin{figure}[htc!]
\centering
\includegraphics[width=4in]{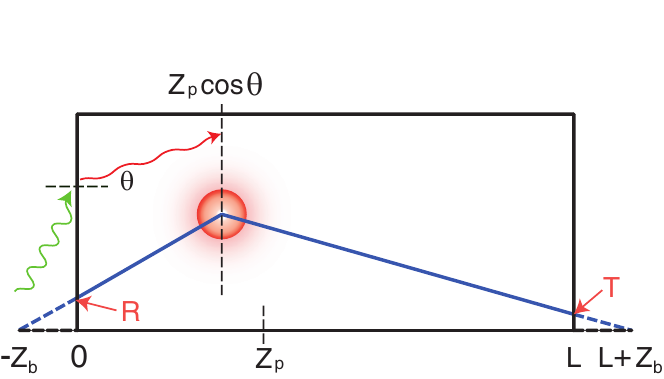}
\caption{Illustration of the photon diffusion model for the spatial variation of intensity inside the random sample, $I(z)$.} \label{Fig3.7}
\end{figure}

Diffusion within the sample gives a constant gradient of intensity to the right and left of the effective randomization depth $z_p \cos \theta$, which extrapolates to zero at a length $z_b$ beyond the sample boundary. The gradients of intensity at the input and output surfaces give the reflection and transmission coefficients, which together with the condition $T+R=1$, gives the expression for $T(\theta)$ above. The linear falloff of intensity near the sample surface boundaries should hold even for localized waves since the diffusion coefficient varies with depth into the sample but is hardly renormalized near the sample boundaries \cite{67,68,69}. Since $z_b$ and $z_p$ are proportional to the mean free path $\ell$, averaging over all incident angles gives, 
\begin{equation}
{\textsl g}_0=\frac{\eta N\ell}{L+2z_b}=\frac{\xi}{L_{eff}}.
\end{equation}
Here, $\eta$ is independent of $L$ and includes the effects of reduced flux into the sample due to external reflection of the incident waves at the interface and enhanced internal reflection. These effects tend to cancel in transmission so that $\eta \sim 1$. The values of $z_b$ of 6 cm for localized samples are obtained by fitting the diffusion model to the measured time of flight distribution \cite{40}.
\begin{figure}[htc!]
\centering
\includegraphics[width=4in]{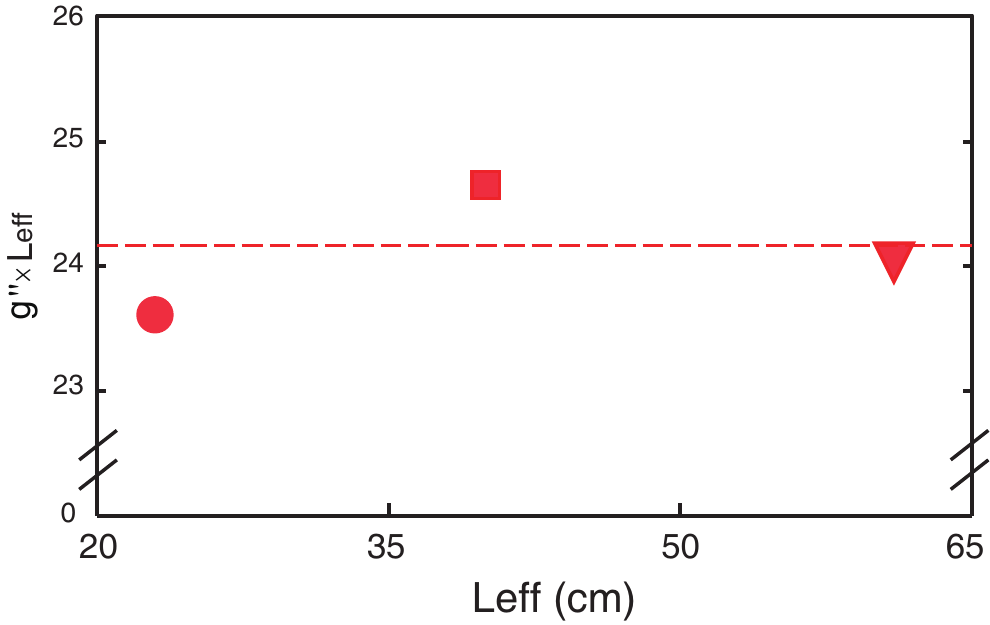}
\caption{Identification of ${\textsl g}^{\prime\prime}$ with the bare conductance ${\textsl g}_0$. The constant products of ${\textsl g}^{\prime\prime}L_{eff}$ for three different lengths for the localized samples give the localization length $\xi$ in the corresponding frequency range.} \label{Fig3.8}
\end{figure}
After taking boundary effects into account, we found that the product of ${\textsl g}^{\prime\prime}$ and the effective sample length, $L_{eff}$, gives constant values of 24 cm for localized samples. We therefore identified this product with the localization length $\xi$ and ${\textsl g}^{\prime\prime}$ with ${\textsl g}_0$, which is demonstrated in Fig. 3.8.

The relative values of transmittance $T$ for samples of different length and waves at different frequency are given by the ratio of the ensemble average of the sum of intensity for all pairs of incident and output points and the same measured in a tube emptied of scatterers. The absolute value of the transmittance $T$ and of the underlying transmission eigenvalues $\tau_n$ are obtained by equating ${\textsl g}=C\langle T\rangle$ for the most diffusive sample of length $L$ = 23 cm, in which {\textsl g} is taken to be the inverse of the spacing between neighboring $\langle \ln \tau_n \rangle$. The normalization factor $C$ is used to determine the values of {\textsl g} in other samples. 

\section{Imperfect control of transmission channels}
The impact of loss of information in the measurement of the transmission matrix upon the transmission eigenvalues is called imperfect control of transmission channels (ICC). Goetschy and Stone \cite{64} considered the case in which only $M_1$($M_2$) channels can be determined on the input (output) surfaces in a random system consisting of $N$ channels. The full transmission matrix $t$ is then mapped to $t^\prime=P_2tP_1$, where $P_1$ and $P_2$ are $N\times M_1$ and $M_2\times N$ matrices. This eliminates $N-M_1$ columns and $N-M_2$ rows of the original random matrix $t$. The density of transmission eigenvalues of $t^\prime$ changes from a bimodal distribution to the distribution characteristic of uncorrelated Gaussian random matrices as the degree of control $M_1/N$($M_2/N$) is reduced \cite{35}. 

This has been confirmed in simulation for scalar wave propagation through 2D random systems with perfectly reflecting transverse boundaries and semi-infinite leads attached to the random scattering region. The wave equation $\triangledown^2E(x,y)+k_0^2\epsilon(x,y)E(x,y)=0$ is discretized using a 2D tight-binding model on a square grid.  The variation of the dielectric function used in the simulation is sketched in Fig. 3.9.
\begin{figure}[htc!]
\centering
\includegraphics[width=4in]{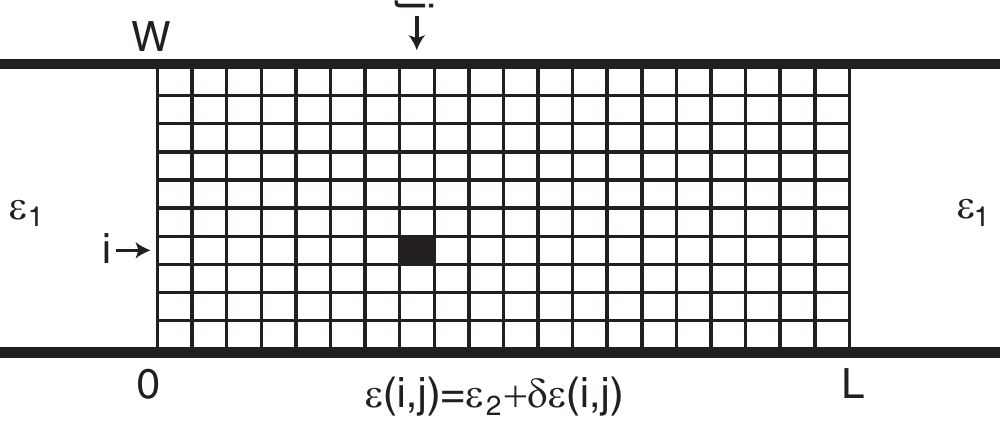}
\caption{Sketch of the simulation scheme for scalar wave propagation through a random waveguide. The dielectric constant at position $(i,j)$ is given by $\epsilon(i,j)=\epsilon_2+\delta(\epsilon)$.} \label{Fig3.9}
\end{figure}

The disordered medium is modeled by the position dependent dielectric constant $\epsilon(i,j)=\epsilon_2+\delta\epsilon(i,j)$, where $\epsilon_2$ is a constant background and $\delta\epsilon(i,j)$ is chosen from a random distribution. $\epsilon_1$ represents the dielectric constant of the leads attached to the scattering sample. The Green's function between grid points on the input and output surfaces, $G(y,y')$, is computed via the recursive Green's function method \cite{70,71}. The transmission coefficient between incoming transverse mode $m$ and outgoing transverse mode $n$ is obtained by projecting the Green's function onto the wavefunction of the transverse mode,
\begin{equation}
t_{m,n}=\sqrt{\nu_m/\nu_n}\int_0^W dy \int_0^W dy^\prime \phi_m(y)\phi_n(y^\prime)G(y,y^\prime).
\end{equation}
where $\nu_n$ is the group velocity of the mode $n$ and $\phi_n(y)=\sqrt{\frac{2}{W}}\sin(\frac{n\pi y}{W})$ is the wavefunction of the $n_{th}$ transverse mode in the lead. The number of propagating transverse modes $N$ depends upon the wave vector of the incident wave in the leads and the width of the sample.

Here, we employ this simulation method to study the impact of loss of full control in our measurements of the TM \footnote{The program for calculating transmission matrix through the 2D random system was kindly sent to us by Arthur Goetschy and A. Douglas Stone.}. In the simulation, we have chosen $\epsilon_1=n_1^2=1$, $\epsilon_2=n_2^2=1.5^2$. The dimensions of the sample and the wavelength are measured in units of the grid size, which is set to unity. The wavelength is $2\pi$ so that the wave vector in the leads is unity. The sample length is three times of its width of 200. $\delta \epsilon$ is uniformly distributed between [-0.55 0.55]. In this simulation, the number of propagation waveguide modes $N$ is equal to 66 and the dimensionless conductance of the random sample is {\textsl g}=6.67, which is close the value of {\textsl g} for the diffusive sample with $L=23$ cm in our experiment. Simulations are made for 10,000 random samples. We also consider the transmission matrix determined by the transmission coefficient between points at the input and output surfaces in the simulation, which is similar to the protocol in the experimental measurement. The field-field correlation function, $C_E(\Delta r)=\langle E(r) E^*(r+\Delta r)\rangle/\sqrt{\langle I(r)\rangle \langle I(r+\Delta r)\rangle}$, is presented in Fig. 3.10. The first zero of the real part of the field correlation function is at $\Delta r=2.7$ grid spacings and therefore the number of nearly independent points on the output surface in the simulation is approximately 66.
\begin{figure}[htc!]
\centering
\includegraphics[width=5in]{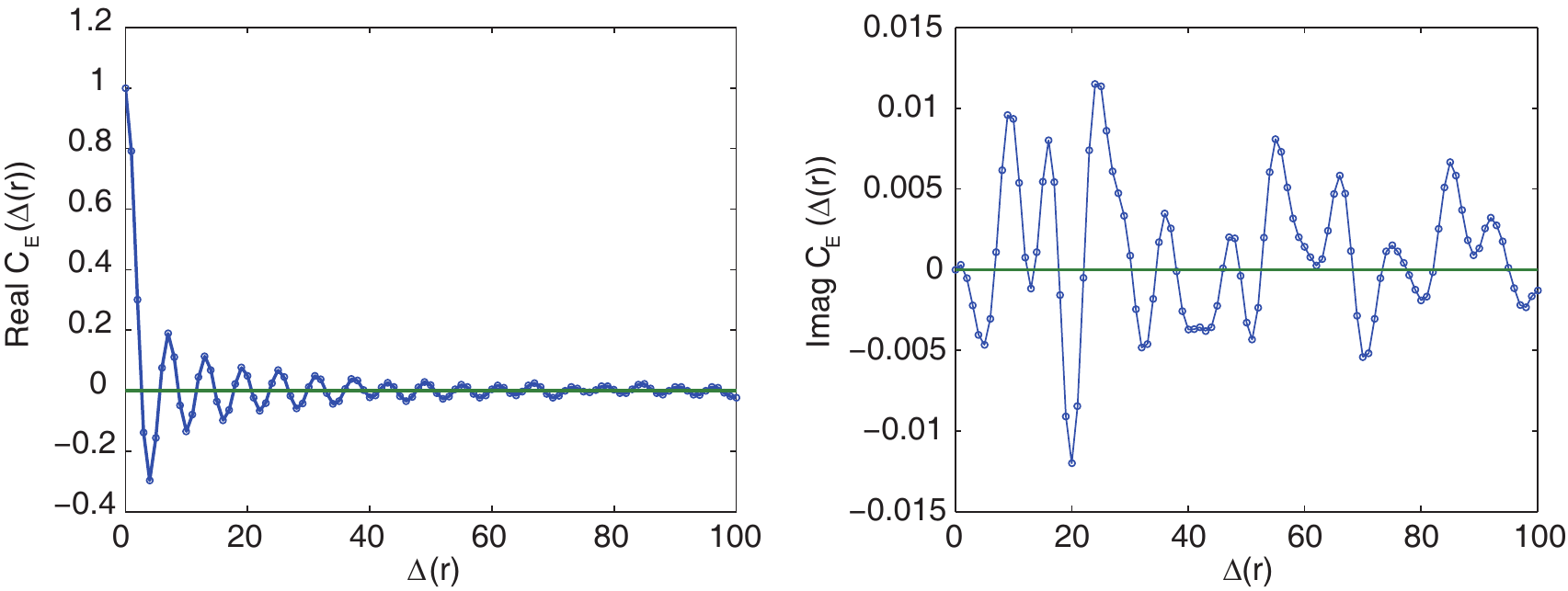}
\caption{Spatial field-field correlation of the transmitted field at the output surface. The first zero of the real part of the field-field correlation is $\sim \lambda/2$. Beyond the first zero, the correlation function oscillates around zero. The imaginary part of the field correlation function is almost zero.} \label{Fig3.10}
\end{figure}

\begin{figure}[htc!]
\centering
\includegraphics[width=3in]{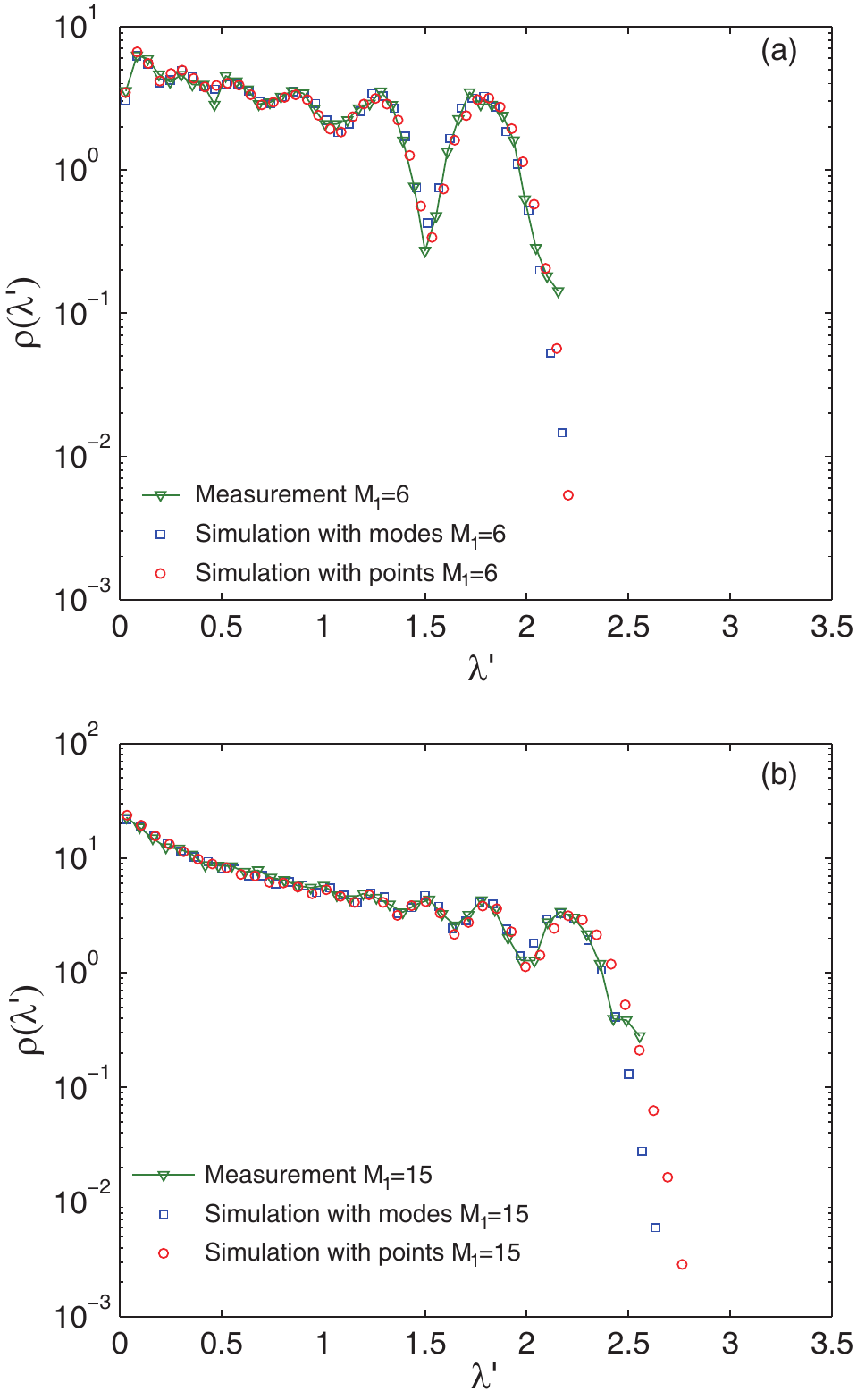}
\caption{The density of normalized $\rho(\lambda^\prime)$ for $M_1=6$ and 15 channels are measured in a system consisting of $N=66$ channels. In the simulations, singular values of transmission matrix based on scattering between orthogonal waveguide modes in the lead and based on scattering between arrays of points on the input and output surfaces are determined. The measurements are performed for diffusive samples of $L=23$ cm.} \label{Fig3.11}
\end{figure}
We now consider the changing of density of normalized singular values when the degree of control $M_1/N$ is gradually reduced, $\rho(\lambda_n^\prime)=\lambda_n/\sqrt{(\sum_{n=1}^{M_1} \lambda_n^2)/M_1}$. When the number of channels under control $M_1$ is much smaller than the dimensionless conductance {\textsl g}, mesoscopic correlation is not manifest in the measured transmission matrix and $\rho(\lambda^\prime)$ follows a quarter-circle law, $\rho(\lambda^\prime)=\frac{\sqrt{4-\lambda^{\prime2}}}{\pi}$, which is the characteristic distribution for uncorrelated Gaussian random matrices. This has been observed in in reflection in acoustics and in transmission in optics. We now only consider the case in which the number of channels $M_1$ controlled on the input and output sides of the sample is the same. In Fig. 3.11, we show for $M_1=6$ and 15 and find excellent agreement between measurements and simulations.

\begin{figure}[htc!]
\centering
\includegraphics[width=3in]{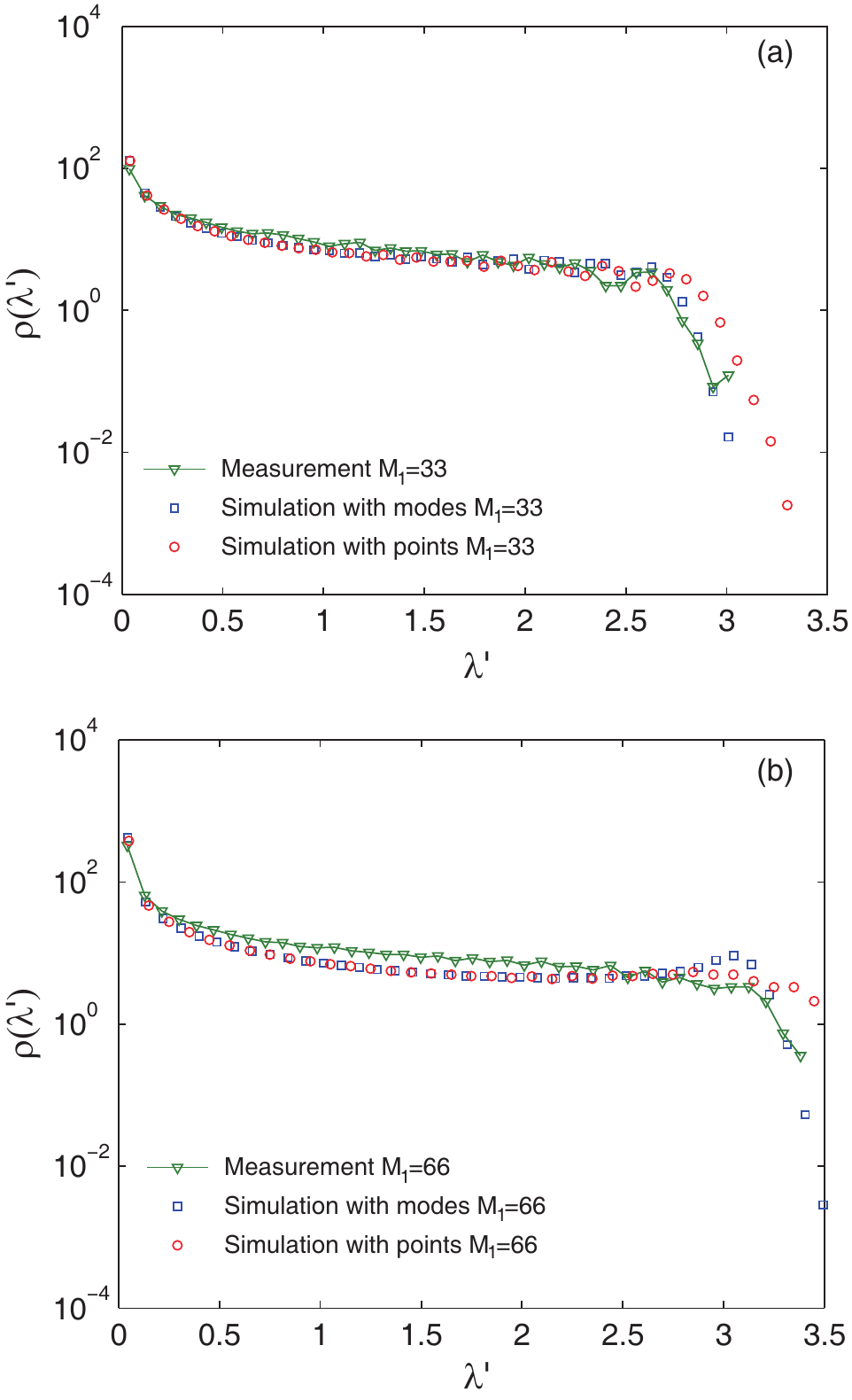}
\caption{The density of normalized $\rho(\lambda^\prime)$ for $M_1=33$ and 66 channels are measured in a system consisting of $N$ = 66 channels. The configuration setup is the same as in Fig. 3.11.} \label{Fig3.12}
\end{figure}
When the number of measured points $M_1$ approaches the total number of channels $N$, the measured density $\rho(\lambda^\prime)$ deviates from the density of $\lambda^\prime$ obtained from the transmission matrix based upon scattering between modes in the simulation. We also found that the density of obtained from the two protocols in the simulation also differ. This discrepancy, seen in Fig. 3.12, reflects the fundamental difference between measurement of transmission matrix based upon on points and the theoretical calculations of the transmission matrix which is based upon the scattering between the waveguide modes in the sample leads. Fields measured at points separated beyond the first zero of the field correlation function are still weakly correlated since the field correlation oscillates as seen in Fig. 3.10 and does not vanish. In contrast, transmission coefficient between different channels are independent. The impact of this correlation is more pronounced when more points are included in the measured transmission matrix and this leads to the difference observed between measurements and simulation results. We notice that when $M_1$ is close to $N$, the measured $\rho(\lambda^\prime)$ deviates from the transmission matrix based on points. This may be a consequence of correlation between measurements made in two polarizations, which is around 30\% when measured at the same point as discussed in Chapter 2.

\begin{figure}[htc!]
\centering
\includegraphics[width=4in]{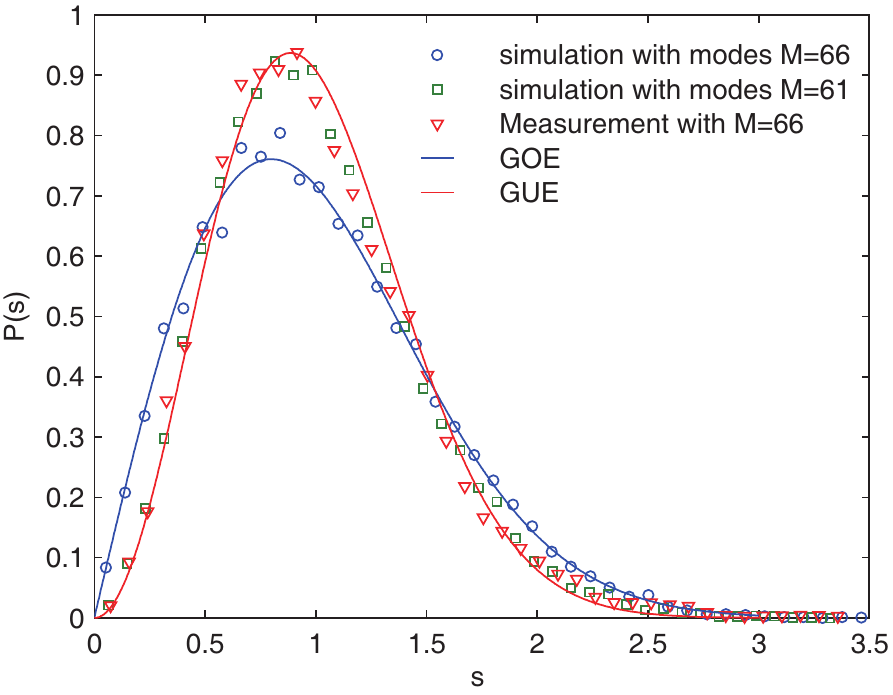}
\caption{Probability distribution of the nearest spacing normalized to its ensemble average, $P(s)$, compared with Wigner surmise for GOE and GUE. } \label{Fig3.13}
\end{figure}
The imperfect control of transmission channels also changes the repulsion between the measured transmission eigenvalues, in which a quadratic repulsion expected for the random systems without time-reversal symmetry is observed for systems with time-reversal symmetry \cite{64}. The change of interaction between transmission eigenvalues can be readily seen in the statistics of the nearest spacing between $x_n$, $s=(x_{n+1}-x_n)/(\langle x_{n+1}-x_n\rangle)$, where $x_n$ is related to the transmission eigenvalues via the relation, $\tau_n=1/\cosh^2(x_n)$. The probability distribution of $s$, $P(s)$, follows Wigner surmise for GOE or GUE, in the presence or absence of time-reversal symmetry, respectively \cite{35}. The results are given in Fig. 3.13. In the measurement, the transmission eigenvalues are normalized to the maximum value in the ensemble. When the degree of control of the transmission matrix is not perfect, the time-reversal condition for the scattering matrix is not satisfied and this leads to the change of the repulsion between the measured transmission eigenvalues.

The dependence of the probability density of the transmission eigenvalues on the degree of control of the transmission matrix reflects the changing degree of intensity correlation within the measured transmission matrix. The correlation in the measured transmission matrix can be easily examined by considering the relation between the fluctuations of the normalized total transmission $s_a$ and the number of measured channels $M_1$. var$(s_a)$ can be expressed by the intensity correlation function,
\begin{eqnarray}
&& C_{ab,a^\prime b^\prime}=\langle T_{ab}\rangle \langle T_{a^\prime b^\prime}\rangle [\delta_{aa^\prime}\delta_{bb^\prime}+\frac{2}{3{\textsl g}}(\delta_{aa^\prime}+\delta_{bb^\prime})+\frac{2}{15{\textsl g}^2}(1+\delta_{aa^\prime}\delta_{bb^\prime}+\delta_{aa^\prime}+\delta_{bb^\prime})]     \\
&& {\text var}(T_a)=\sum_{b\neq b^\prime=1}^{M_1} C_{ab,a^\prime b^\prime}=M_1\langle T_{ab}^2\rangle+\frac{2M_1^2}{3\textsl{g}}\langle T_{ab}\rangle^2+ \dots  \\
&& {\text var}(s_a)=\frac{{\text var}(T_a)}{\langle T_a\rangle^2}=\frac{{\text var}(T_a)}{M^2\langle T_{ab}\rangle^2}=\frac{1}{M_1}+\frac{2}{3\textsl{g}}+\dots
\end{eqnarray}

When the full transmission matrix is obtained, $M_1=N\gg {\textsl g}$, the bimodal distribution of transmission eigenvalues yields var($s_a$)=2/3{\textsl g}. In our experiment, we found $\rho(\tau)={\textsl g}/\tau$, which will give var($s_a$)=1/2{\textsl g}. When the number of measured channels $M_1$ is much smaller than {\textsl g}, var($s_a$)=1/$M_1$, indicating the mesoscopic intensity correlation between output channels $b$ and $b^\prime$ is not captured in the measurement. Therefore, the variance of normalized total transmission is proportional to $1/M_1$, given by the central limit theorem for the sum of $M_1$ independent random variable. This expression may as well be exploited to extract the value of the dimensionless conductance {\textsl g} in optical measurement. In optics, the number of channels $N$ is enormous and cannot be completely measured due to the finite numerical aperture. By fitting the relation between var($s_a$) and 1/$M_1$ to a straight line for small value of $M_1$, the value of {\textsl g} can be estimated from the interception of the straight line. This is demonstrated in Fig. 3. 14 for the diffusive sample with $L=23$ cm. By extrapolating the line, we find the value of var($s_a$)=0.074, which corresponds to the value of {\textsl g}=2/3var($s_a$) of approximately 9. 
\begin{figure}[htc!]
\centering
\includegraphics[width=3in]{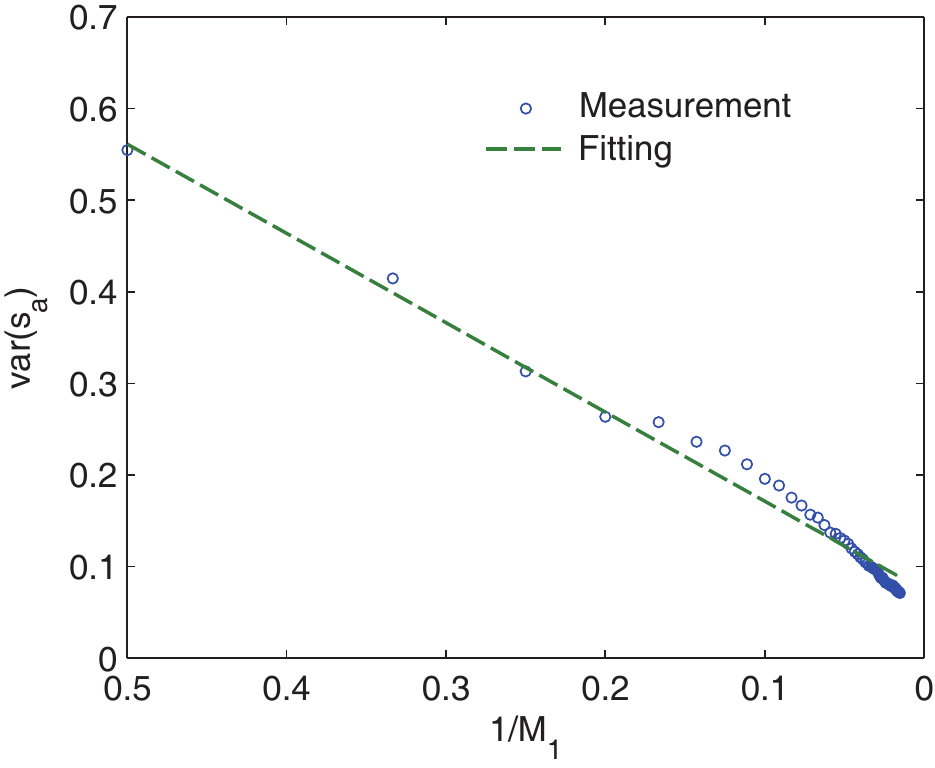}
\caption{Variation of the fluctuation of normalized total transmission vs. the inverse of the number of measured channels $M_1$ in the transmission matrix. The dashed black line is a fit to the data for points with $M_1^{-1}\ge 0.2$.} \label{Fig3.14}
\end{figure}

When wave localization occurs, the value of {\textsl g} falls below unity. Mesoscopic correlation is stronger for localized waves compared with diffusive waves. Therefore, we expect the degree of control in the measured transmission matrix for localized waves will be high. To explore this, we have performed numerical simulation for a localized sample with $N$=16 and a value of {\textsl g} which is equal to 0.35. In Fig. 3.15, we show $\rho(\lambda^\prime)$ for two simulation protocols of determining transmission matrix on points and waveguide modes for this ensemble and they are nearly indistinguishable.
\begin{figure}[htc!]
\centering
\includegraphics[width=3in]{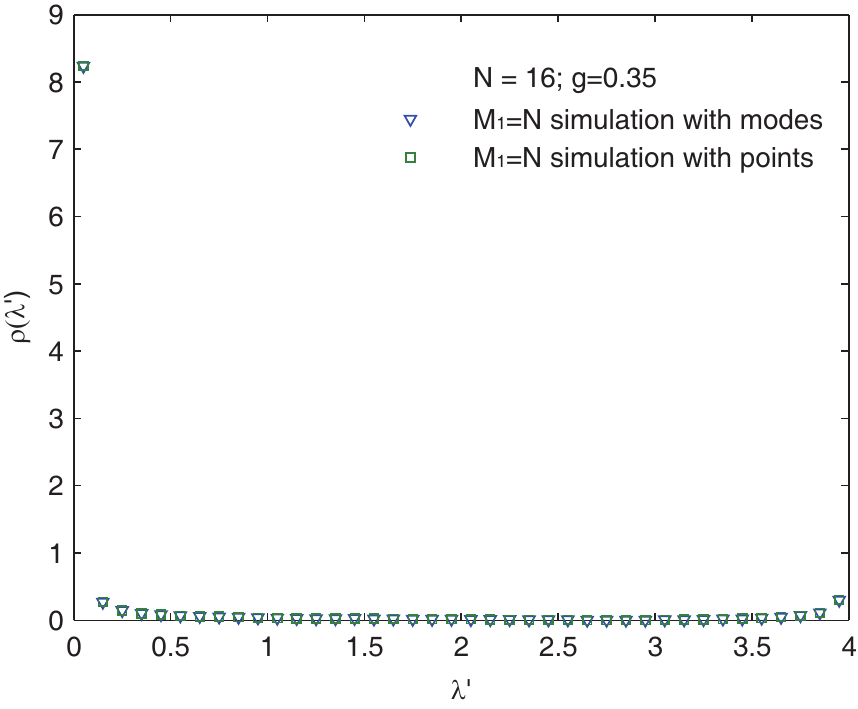}
\caption{Probability distribution of $\lambda^\prime$ for a localized sample with {\textsl g}=0.35. Number of channels $N$ in the simulation is 16.} \label{Fig3.15}
\end{figure}

This confirms that the degree of control is high for localized wave. In addition, the nearest spacing, $P(s)$, is found to follow a Wigner surmise for GOE, which also supports the factor that we have a good control over the measured transmission matrix. This is also observed in experiments for localized waves of {\textsl g}=0.37, which is shown in Fig. 3.16.
\begin{figure}[htc!]
\centering
\includegraphics[width=3in]{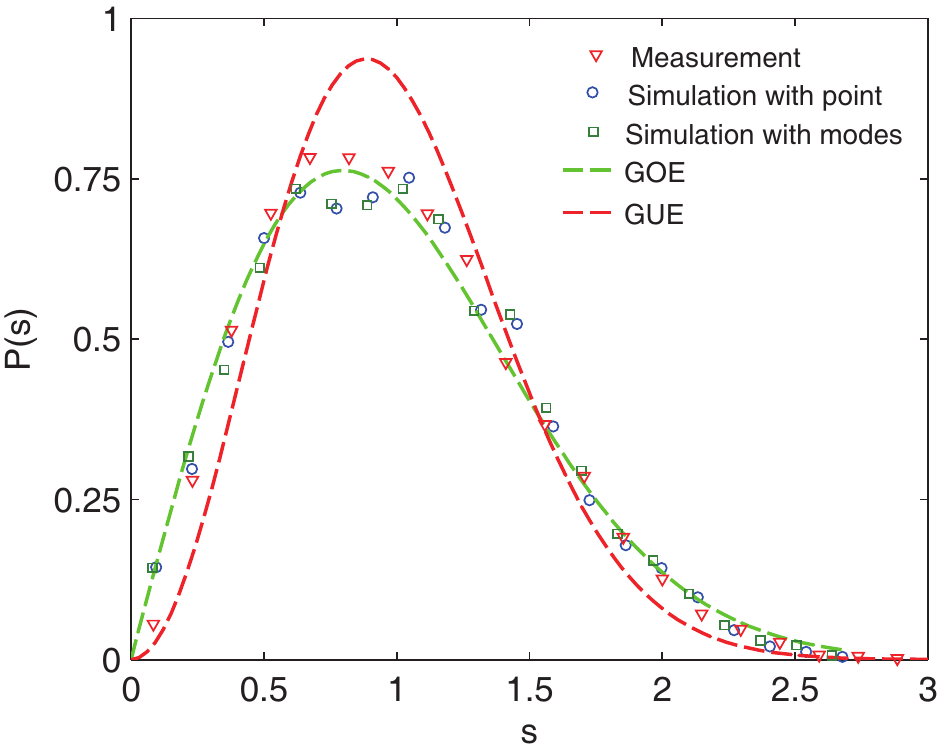}
\caption{Probability distribution of the nearest spacing normalized to its ensemble average, $P(s)$, compared with Wigner surmise for GOE and GUE, for localized waves. In the simulation, {\textsl g}=0.35 and $N$=16 and in the measurements, the value of {\textsl g} is 0.37 and $N$ = 30.} \label{Fig3.16}
\end{figure}

\section{Summary}
In this chapter, we have studied the eigenvalues of the transmission matrix in the Anderson localization transition. For localized waves, the transmittance is dominated by the highest transmission eigenvalue while for diffusive waves, approximately {\textsl g} transmission channels contribute appreciably to the transmittance $T$. The inverse of the measured localization length is found to be equally spaced for localized waves as predicted by Dorokhov. Once taking the wave interaction at the sample surface into account, the spacing is found to be the inverse of the bare conductance for localized waves. For diffusive waves, the density of the transmission eigenvalues is found to have a single peak at the low values in contrast to the theoretical prediction of a bimodal distribution with a second peak at high transmission close to unity. We explored the use of ICC to explain the disagreement between experiment and theory. Good agreement is found with the theory when the size of the measured transmission matrix is small relative to the total number of transmission channels. When $M_1$ is close to $N$, the measured density of transmission matrix deviates from the calculation. The discrepancy is due to the fundamental difference between measurement protocol and theoretical calculation. We also observe the transition of interaction between eigenvalues due to the loss of control of transmission channels and showed explicitly that the mesoscopic correlation vanishes when $M_1\ll {\textsl g}$. Due to the relative strong mesoscopic correlation for localized waves, the degree of control of transmission matrix for localized wave is high.

\chapter{Transmission statistics in single disordered samples}
\section{Introduction}
Enhanced fluctuations of conductance and transmission are the prominent feature of transport of wave in mesoscopic samples, where the wave is temporary coherent throughout the sample \cite{4,14,15}. Because of these large sample-to-sample fluctuations, measurement in individual samples has been considered to be of accidental rather than fundamental significance. Conductance and transmission fluctuations relative to the corresponding average over a random ensemble increase exponentially with sample length for localized waves \cite{10,11}. For diffusive waves, the lack of self-averaging in mesoscopic samples is observed in the universal conductance fluctuation \cite{29,30,31,32}. In the diffusive limit, ${\textsl g}\gg 1$, probability distribution of conductance is a Gaussian with the variance approaching a universal value of order unity. Thus, despite wide-ranging applications in communications, imaging and focusing in single samples or environments, studies of disordered systems have centered on the statistics in hypothesized ensembles of statistically equivalent samples. 

Studies of transport have shown that the statistics of propagation over a random ensemble of mesoscopic samples may be characterized in terms of a single parameter, {\textsl g}, the ensemble average of the conductance in units of the quantum of conductance $e^2/h$ \cite{9,12}. The similarity of key aspects of quantum and classical wave transport is seen in the equivalence of the dimensionless conductance and the transmittance {\it T}, which is the sum of flux transmission coefficients between the {\it N} incident and transmitted channels, $a$ and $b$, respectively, ${\textsl g}=\langle T\rangle=\langle \sum_{a,b=1}^N T_{ba}\rangle$. The threshold of the Anderson transition between freely diffusing and spatially localized waves in disordered media lies at {\textsl g}=1 \cite{12}. 

In contrast to in-depth studies of random ensembles, critical aspects of the statistics of transmission in single samples have not yet been explored. In this chapter, we will discuss the transmission statistics in single disordered samples \cite{72}. We treat the quasi-1D geometry for which the length of reflecting sides greatly exceeds the sample width, $L\gg W$. Examples of quasi-1D samples are disordered wires and random waveguides, for which measurement are presented here. It has been shown that when normalized by the total transmission in a single speckle pattern, the long-range correlation vanished and the probability distribution of relative intensity $T_{ba}/(\sum_{b}^N T_{ba}/N)$ is a negative exponential, $P(NT_{ba}/T_a)=\exp(-NT_{ba}/T_a)$. Since the statistics of relative intensity are universal, the statistics of transmission in a sample with transmittance $T$ would be completely specified by the statistics of total transmission $T_a$ relative to its average $T/N$ within the sample. 

We report here the essential statistics of transmission in single transmission matrices. We find the statistics of relative total transmission $NT_a /T$ and show it is determined by a single parameter, the participation number of the eigenvalues $\tau_n$, $M\equiv (\sum_{n}^N \tau_n)^2/\sum_{n}^N \tau_n^2$. We find, in the limit of large $N$, $M^{-1}$ is equal to the variance of $NT_a /T$. The distribution of relative total transmission changes from Gaussian to negative exponential over the range in which $M^{-1}$ changes from 0 to 1. 
\begin{figure}[htc!]
\centering
\includegraphics[width=3.8in]{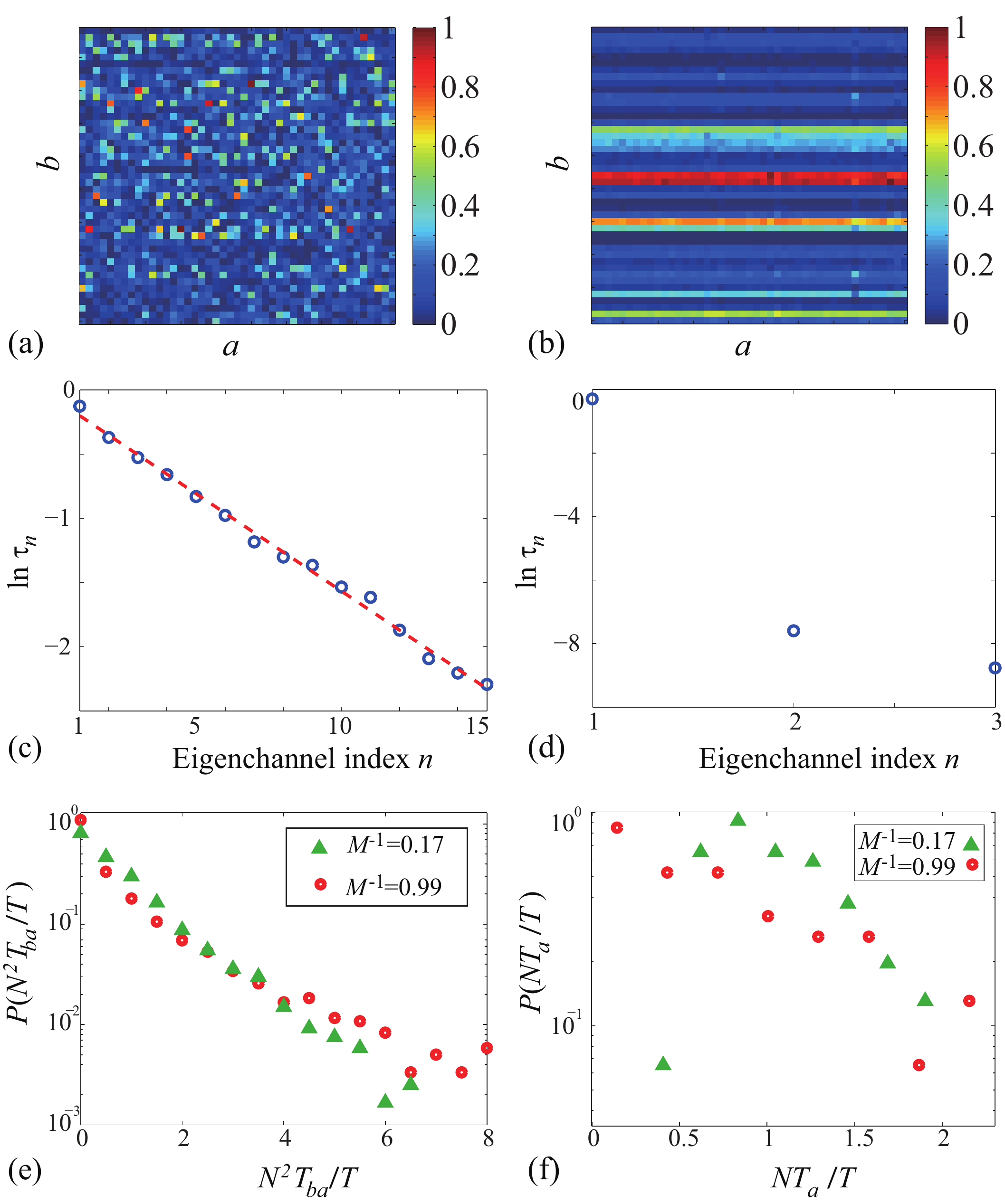}
\caption{Intensity normalized to the peak value in each speckle pattern generated by sources at positions {\it a} are represented in the columns with index of detector position {\it b} for all polarizations for (a) diffusive and (b) localized waves. (c,d) The transmission eigenvalues are plotted under the corresponding intensity patterns. For localized waves (d), the determination of the third eigenvalue and higher eigenvalues are influenced by the noise level of the measurements. Correlation between speckle patterns for different source positions are clearly seen in (b) due to the small numbers of eigenchannels {\it M} contributing appreciably to transmission. (e,f) Distributions of relative intensity $P(N^2T_{ba}/T)$ and relative total transmission $P(NT_a/T)$ for the two transmission matrices selected in this figure with $M^{-1}$=0.17 (green triangles) and $M^{-1}$=0.99 (red circles).}\label{fig4.1}
\end{figure}

\section{Statistics of single transmission matrices}
We utilize the measurement of microwave transmission matrix and random matrix theory calculation to study the statistics within the single transmission matrices. In Fig. 4.1, we present the transmission statistics in the transmission matrix from one random configuration drawn from ensemble with {\textsl g}=6.9 and 0.17 respectively. In Figs. 4.1(a) and 1(b), we show intensity pattern within a single transmission matrix drawn from ensembles with {\textsl g}=6.9 and 0.17, respectively. Plots of the transmission eigenvalues determined from the transmission matrices for the samples whose intensity patterns are shown in Figs. 4.1(a) and 1(b) are presented below the corresponding patterns. Values of  $\tau_n$ are seen to be substantial for a number of channels and to fall nearly exponentially for the sample in which the wave is diffusive. Since the transmitted speckle pattern for a source at any position $a$ is the sum of many orthogonal transmission eigenchannels, speckle patterns for different source positions are weakly correlated yielding the motley intensity pattern for the transmission matrix depicted in Fig. 4.1(a). In contrast, the first transmission eigenchannel for localized waves dominates transmission in Fig. 4.1(d) so that the normalized speckle patterns for each input are highly correlated and horizontal stripes appear in the Fig. 4.1(b). The probability distributions of relative intensity, $N^2T_{ba}/T$, and total transmission, $NT_a/T$, for the two transmission matrices depicted in Fig. 4.1(a) and 1(b) are shown in Figs. 4.1(e) and 1(f).

It has been shown that propagation in random ensembles can be characterized via the variance of the total transmission relative to its ensemble average, var($NT_a/\langle T\rangle$) \cite{21,22,23}. This suggests that the statistics of single samples may be characterized via the variance of relative total transmission in a single transmission matrix, var$NT_a/T$. This can be calculated for a single instance of the transmission matrix with $N\gg1$ using the singular value decomposition of the transmission matrix, $t=U\Lambda V^\dagger$. Here, $U$ and $V$ are unitary matrices with elements $u_{nb}$ and $v_{na}$, where $n$ is the index of the eigenchannel and index $a$ and $b$ indicate the input and output channels, respectively. The real and imaginary parts of $u_{nb}$ and $v_{na}$ are Gaussian random variables with zero mean and variance of 1/2$N$ . $\Lambda$ is a diagonal matrix with elements $\sqrt{\tau_n}$ along its diagonal.  

The total transmission from incident channel $a$, $T_a=\sum_{b}^N T_{ba}$, can be written as, $T_a=\sum_{n}^N \tau_n|v_{na}|^2$, giving the relative total transmission, $T_a/(T/N)=\sum_{n}^N \tau_n(N|v_{na}|^2)/T$. The second moment of $NT_a/T$ is,
\begin{equation}
\sum_{n}(\tau_n/T)^2\langle N^2|v_{na}|^4\rangle _a+\sum_{n\neq n^\prime}^N \tau_n\tau_n^\prime/T^2 \langle N^2|v_{na}|^2|v_{n^\prime a}|^2\rangle_a.
\end{equation} 
Here, $\langle \dots \rangle_a$ is the average over all incident points $a$ within a single transmission matrix. In the limit $N\gg 1$, $\langle |v_{na}|^2\rangle=1/N$ and $\langle |v_{na}|^2\rangle=2/N^2$. This yields,
\begin{equation}
\langle (\frac{NT_a}{T})^2\rangle = \frac{\sum_{n}^2 \tau_n^2+(\sum_{n}^N \tau_n)^2}{(\sum_{n}^N \tau_n)^2}. 
\end{equation}
In a given transmission matrix, $\langle NT_a/T\rangle=1$. Thus, Eq. 4.2 gives, 
\begin{equation}
{\rm var}(NT_a/T)=\frac{\sum_{n}^N \tau_n^2}{(\sum_{n}^N \tau_n)^2}.
\end{equation}
This parameter is distinct from $N_{{\rm eff}}$, which had been introduced earlier by Imry \cite{32} to denote the number of eigenchannels with $\tau_n \ge 1/e$ for diffusive waves. Whereas $N_{{\rm neff}} \rightarrow 0$ in the localization limit, $M$ approaches 1, indicating the transport through single transmission eigenchannels. 

\begin{figure}[htc!]
\centering
\includegraphics[width=4in]{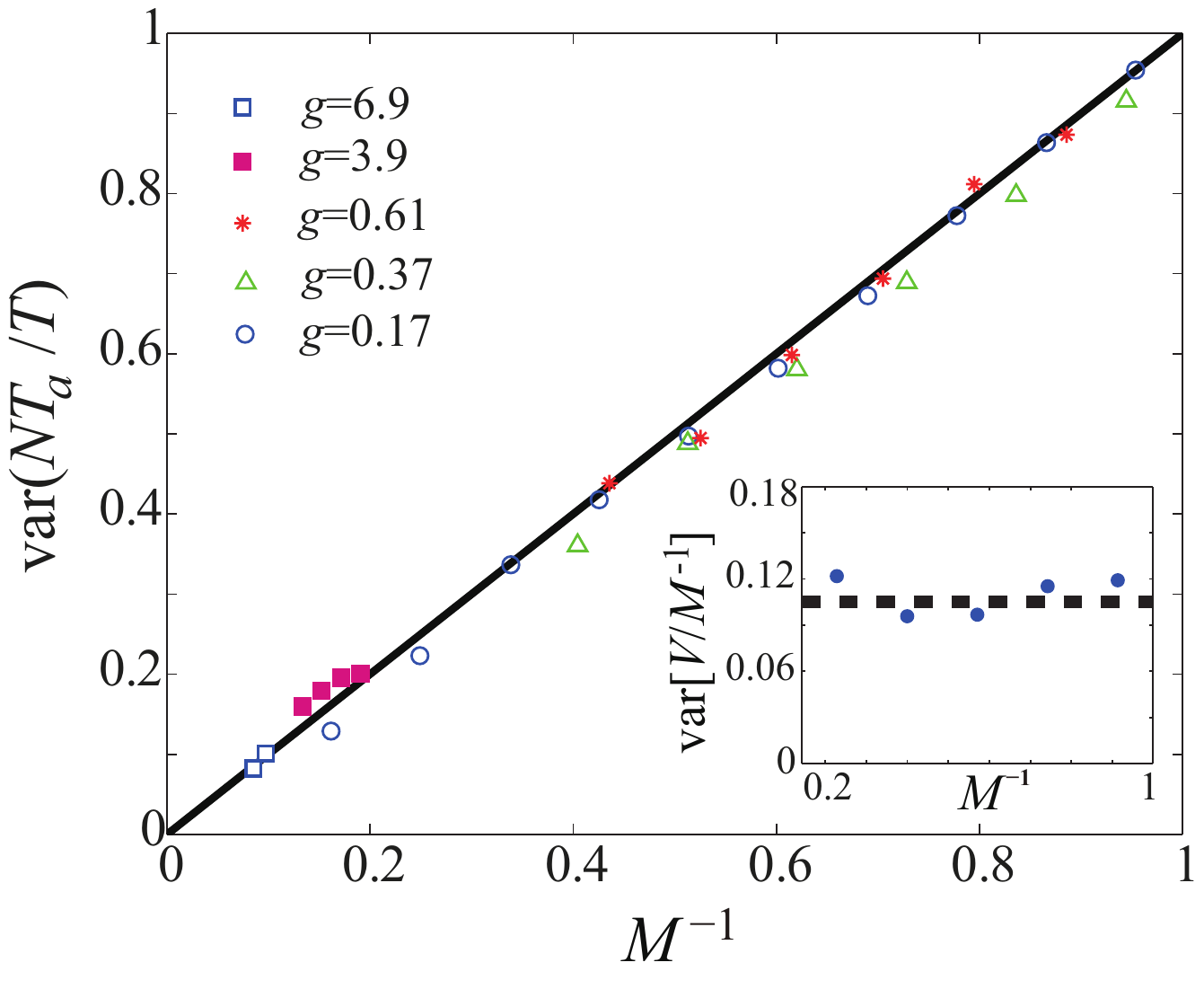}
\caption{Plot of the var$(NT_a/T)$ computed within single transmission matrices over a subset of transmission matrices drawn from random ensembles with different values of {\textsl g} with specified value of $M^{-1}$. The straight line is a plot of var($NT_a/T$)=$M^{-1}$. In the inset, the variance of $V/M^{-1}$ is plotted vs. $M^{-1}$ , where $V={\rm var}(NT_a/T)$.}\label{fig4.2}
\end{figure}
In order to compare the calculations with measurements in samples of small value of $N$, we have grouped the samples with same value of $M$ and compute the fluctuations of relative total transmission in the subsets of transmission matrices. This will suppress the Gaussian fluctuations due to limit statistics within single transmission matrix of small $N$. The results are presented in Fig. 4.2 and an excellent agreement between the calculation and experiment is seen and therefore confirms the validity of the Eq. (4.3). var[var$(NT_a/T)/M^{-1}$] is seen in the insert of Fig. 4.2 to be proportional to $1/N$ indicating that fluctuations in the variance over different subsets are Gaussian with a variance that vanishes as $N$ increases.

\begin{figure}[htc!]
\centering
\includegraphics[width=4.5in]{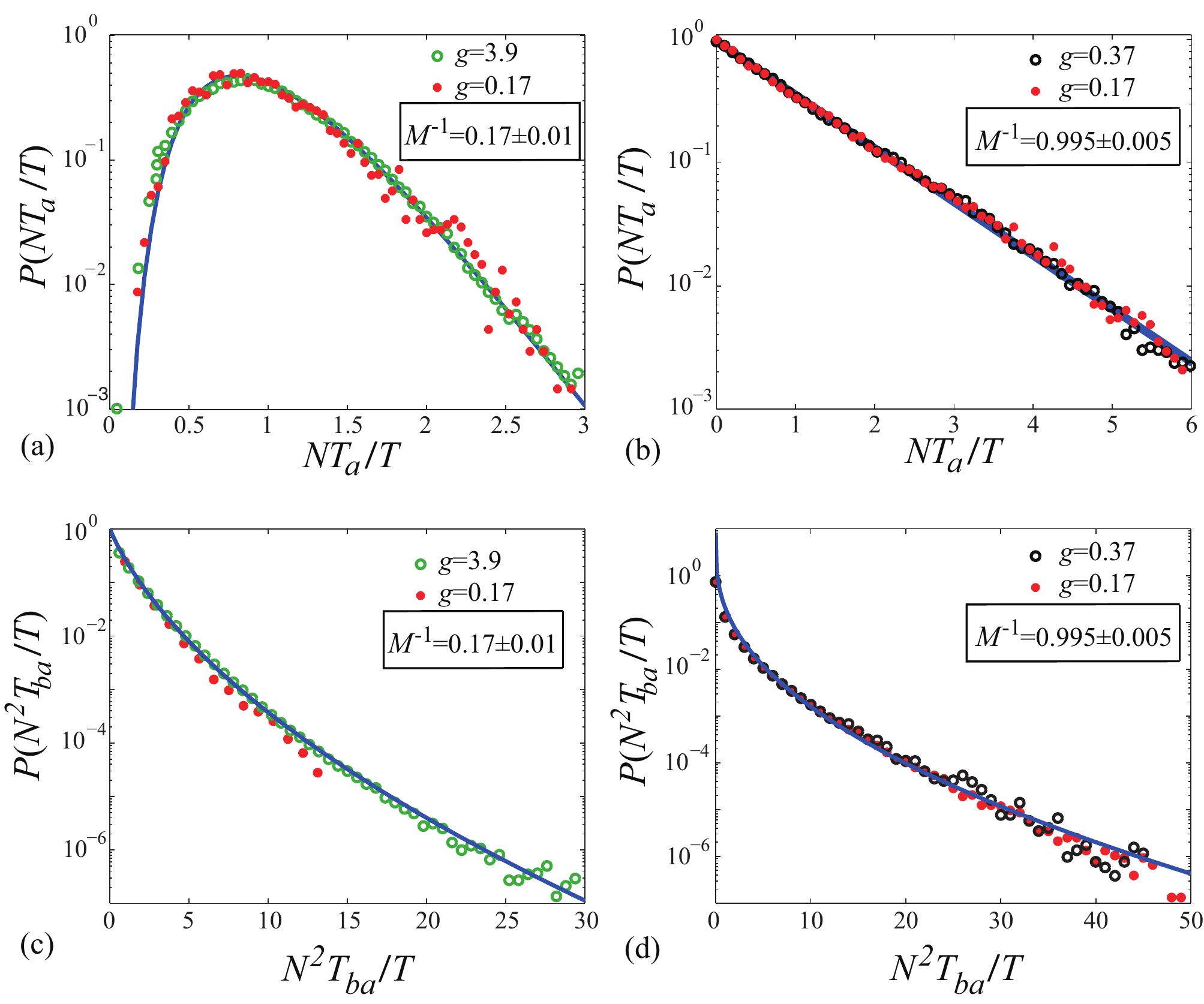}
\caption{ (a) $P(NT_a/T)$ for subsets of transmission matrices with $M^{-1}=0.17\pm 0.01$ drawn from ensembles of samples with {\it L}=61 cm in two frequency ranges in which the wave is diffusive (green circles) and localized (red filled circles). The curve is the theoretical probability distribution of $P(NT_a/\langle T\rangle)$ in which ${\rm var}(NT_a/T)$ is replaced by $M^{-1}$ in the expression for $P(NT_a/T)$ in Refs. 22 and 23. (b) $P(NT_a/T)$ for $M^{-1}$ in the range $0.995\pm 0.005$ computed for localized waves in samples of two lengths: {\it L}=40 cm (black circles) and {\it L}=61 cm (red filled circles). The straight line represents the exponential distribution, $\exp(-NT_a/T)$. (c,d) The corresponding intensity distributions $P(N^2T_{ba}/T)$ are plotted under (a) and (b). }\label{fig4.3}
\end{figure}
The central role played by $M$ can be appreciated from the plots shown in Fig. 4.3 of the statistics for subsets of samples with identical values of $M$ but drawn from ensembles with different values of {\textsl g}. The distributions $P(NT_a/T)$ obtained for samples with $M^{-1}$ in the range $0.17\pm0.01$ selected from ensembles with {\textsl g}=3.9 and 0.17 are seen to coincide in Fig. 4.3(a) and thus to depend only on $M^{-1}$. The curve in Fig 4.3(a) is obtained from an expression for $P(T_a/\langle T_a\rangle)$ for diffusive waves given in Ref. 22, in terms of the single parameter {\textsl g}, which equals 2/3var($T_a/\langle T_a\rangle$) in the limit of large {\textsl g}, in which {\textsl g} is replaced by $2/3M^{-1}$. The dependence of $P(NT_a/T)$ on $M^{-1}$ alone and its independence of $T$ is also demonstrated in Fig. 3(b) for $M^{-1}$ over the range 0.995$\pm$0.005 in measurements in different sample length with {\textsl g}=0.37 and 0.17. Since a single channel dominates transmission in the limit $M^{-1} \rightarrow 1$,we have, $NT_a/T=|v_{1a}|^2$, where $v_{1a}$ is the element of the unitary matrix $V$ which couples the incident channel $a$ to the highest transmission eigenchannel. The Gaussian distribution of the elements of $V$ leads to a negative exponential distribution for the square amplitude of these elements and similarly to $P(NT_a/T)=\exp(-NT_a/T)$, which is the curve plotted in Fig. 4.3(b). In Figs. 4.3(c) and 4.3(d), we plot the relative intensity distributions $P(N^2T_{ba}/T)$ corresponding to the same collection of samples as in Figs. 4.3(a) and 4.3(b), respectively. The curves plotted are the intensity distributions obtained by mixing the distributions $P(NT_a/T)$ shown in Figs. 4.3(a) and 4.3(b) with the universal negative exponential function for $P(NT_{ba}/T_a)$.

The departure of intensity and total transmission distributions within a transmission matrix from negative exponential and Gaussian distributions, respectively, is a consequence of mesoscopic intensity correlation \cite{4}. The results above for the statistics over a subset of samples with given $M$ suggest an expression for the cumulant correlation function of transmitted intensity relative to its average value for a single transmission matrix in the limit $N\gg 1$ or for a subset of transmission matrices with a specified value of $M$,
\begin{equation}
C^M_{ba,b^\prime a^\prime}=\langle [T_{ba}T_{b^\prime a^\prime}-(T/N^2)^2]/(T/N^2)^2\rangle_M.
\end{equation}
Because of the normalization by the $T/N^2$, infinite-range correlation of relative intensity between arbitrary incident and outgoing channels for transmission matrices with given $M$ vanishes. Such correlation in a random ensemble is known as $C_3$ and is due to fluctuations in $T$ \cite{47,48,49}. The values of unity and $M^{-1}$ for the variances of relative intensity and total transmission determines the sizes of the residual $C_1$ and $C_2$ terms, representing short- and long-range intensity correlation within the matrix, respectively and gives,
\begin{equation}
C^M_{ba,b^\prime a^\prime}=\delta_{a a^\prime}\delta_{b b^\prime}+M^{-1}(\delta_{a a^\prime}+\delta_{b b^\prime}).
\end{equation}
This gives ${\rm var}(N^2T_{ba}/T)=1+2M^{-1}$. This is confirmed by the close correspondence between the measured variances of relative intensity for the two values of $M^{-1}$ of 0.17 and 0.995 in Fig. 4.3 of 1.38 and 3.04 with the calculated values of 1.34 and 3. 
\section{Summary}
In this chapter, we explored the transmission statistics within single instance of transmission matrix and showed that the full statistics of transmission in single samples are determined by the eigenchannel participation number $M$. In particular, the variance of the total transmission relative to the average value is equal to the inverse of $M$, var($NT_a/T)=M^{-1}$. The probability distribution of $NT_a/T$, $P(NT_a/T)$, changes from a Gaussian distribution to a negative exponential function as $M^{-1}$ ranges from 0 to 1. We show the distinct roles of $T$ and $M$ in the statistics of transmission within single transmission matrix. While $M$ governs the internal statistics in single transmission matrix, the transmittance $T$ serves as an overall normalization factor, yielding the average total transmission. In order to obtain the full statistics in ensemble of random samples, the joint probability distribution of $T$ and $M$ has to be considered. In the diffusive limit, $\langle M\rangle$ is proportional to the $\langle T\rangle ={\textsl g}$.

\chapter{Fluctuations of ``optical'' conductance}
\section{Introduction}
Correlation between flux in disordered mesoscopic conductors leads to large fluctuations of conductance \cite{14,15}. The importance of fluctuations of temporally coherent waves in disordered samples was first recognized in calculations of electronic conduction mediated by localized states \cite{73,74,76}. Subsequently, universal conductance fluctuations were observed in diffusive mesoscopic resistors \cite{30,31,32}. The non-local intensity correlation has also been observed in reflection and transmission for classical waves, such as light, microwave and ultrasound, which gives enhanced sample to sample fluctuations of transmission \cite{47,48,49,50,51,52,53,54,55,56,57}. Such fluctuations make it insufficient to describe the transport through random systems by the mean value of conductance \cite{73}. However, probability distributions of conductance over ensembles of statistically equivalent random configurations could provide a basis for characterizing transport. 

Based on a maximum entropy model for the random transfer matrices, Stone, Mello, Muttalib and Pichard \cite{62} developed a unified treatment for quantum transport through disordered Q1D conductors, in which they obtained UCF for diffusive samples and a log-normal distribution of conductance for localized samples \cite{62}. The dimensionless conductance {\textsl g} is equal to the sum of the eigenvalues $\tau_n$ of the transmission matrix, which are closely related to the eigenparameters of the transfer matrix $\lambda_n$ or equivalent parameters $x_n$, $\tau_n=1/(1+\lambda_n)=1/\cosh^2(x_n)$. An intuitive Coulomb gas model originally proposed by Dyson \cite{34} to visualize the repulsion between eigenvalues of random Hamiltonian was extended to treat the interaction between $x_n$. Transmission eigenvalues $\tau_n$ are associated with positions of parallel line charges at $x_n$ and their images at $-x_n$ of the same sign embedded in a compensating continuous charge distribution. The logarithmic repulsion between two parallel lines of charges with same sign, $\ln |x_i-x_j|$, mimics the interaction between eigenvalues of the random matrix, while the oppositely charged background provides an overall attractive potential that holds the structure together. The repulsion between the first charge $x_1$ associated with the highest transmission eigenvalues $\tau_1$ and its image at -$x_1$ naturally provides a ceiling for $\tau_1$ of unity. The average spacing between neighboring lines of charge is $L/\xi$ while the background charge provides a screening length of $\sqrt{L/\xi}$, where $\xi$ is the localization length. In the diffusive limit $\xi /L \gg 1$, the screening length $\sqrt{L/\xi}$ is much larger than the average spacing $L/\xi$ between lines of charges. The repulsion between transmission eigenvalues are strong, which leads to the phenomenon of universal conductance fluctuations as conjectured by Imry \cite{32}. On the other hand, for localized waves $L\gg \xi$, the average spacing is much greater than the screening length and the repulsion between transmission eigenvalues is weak. In addition, when the spacing is large, the first transmission eigenvalue $\tau_1$ will be exponentially greater than the rest and $\textsl g$ will be dominated by $\tau_1$. The distribution of $x_1$ follows a Gaussian distribution and this leads to a log-normal distribution of conductance, ${\textsl g}\sim \exp(-2x_1)$ \cite{62,73}.

Measurements of fluctuations of conductance have been carried out in single specimens of ohmic samples as the applied voltage or magnetic fields are scanned \cite{29,77}. The most extensive measurements of the statistics of propagation have been made for classical waves. Probability distributions of transmitted intensity for a single coherent source, $T_{ba}$ and its sum over the output surface, $T_{a}= \sum_{b=1}^N T_{ba}$, the total transmission, as well as their relation to correlation between channels have been measured for diffusive and localized waves. However, the most spatially averaged quantity, the transmittance $T$, known as the ``optical" conductance, has not been measured for classical waves. Optical measurements of the transmission matrix have been carried out recently and exploited to focus light through strongly scattering media \cite{78}. But the degree of control of the measured matrix is too small for the impact of mesoscopic correlation to be manifest on the probability distributions of transmittance \cite{64,78}. In contrast, the degree of control of the measured microwave transmission matrix is relatively high and thus holds the promise for observing the mesoscopic fluctuation of transmittance $T$. The distribution of conductance in the crossover regime of ${\textsl g}\sim 1$ was first calculated by Muttalib and W\"{o}ffle \cite{79}, in which a highly asymmetrical distribution of $\ln {\textsl g}$ was found. The low values of $P({\textsl g})$ is well fit by a log-normal distribution and a Gaussian cutoff for the high values of {\textsl g} was found. It was later realized that the tail of $P({\textsl g})$ is essentially an exponential function for high values of {\textsl g} \cite{80}. This anomalous distribution of conductance has also been found in simulations \cite{81,82,83,84,85}. 

In this chapter, we will discuss the correlation between the transmittance $T$ and the underlying transmission eigenvalues $\tau_n$ in the crossover to strong localization. Just beyond the localization threshold of {\textsl g}=1, we find a one-sided log-normal probability distribution of {\it T} for {\textsl g}=0.37 and show that it is a consequence of the repulsion between charges and their images in the charge model. For low values of {\textsl g}, distributions of $T$ become log-normal with variances approaching -$\langle \ln T\rangle$ in accord with the single parameter scaling (SPS) hypothesis of localization for one dimensional random systems. We demonstrated that significant insight into the character of fluctuations of conductance and transmission can be gained by considering the joint distribution of {\it T} and the participation number of transmission eigenvalues, $M$, $P(T,M)$, rather than the more complex joint distribution of $\{\tau_n\}$. We argue that the distributions of intensity and total transmission may be calculated from $P(T,M)$ and demonstrate this for strongly localized waves in which transmission is dominated by single transmission eigenchannel. We also report universal fluctuations of $M$ for diffusive waves, in which var($M$) is found to be $\sim 0.3$ independent of the mean value of $\langle M\rangle$ in analog with UCF. We observe universal transmittance fluctuation for classical wave, but the universal value is considerably larger than the predicted value of 2/15 for Q1D geometry with time-reversal symmetry, in which $\ell \ll L \ll \xi=N\ell$ and ${\textsl g}=N\ell/L\gg 1$. The statistical measure of the rigidity of the energy levels in a random Hamiltonian developed by Dyson and Mehta, $\Delta_3$ statistics, is employed to describe the rigidity of the spectrum of $\ln \tau_n$ and to uncover the origin of UCF. The rigidity of the spectrum of $\ln \tau_n$ is weakened as the wave becomes localized in the random sample. As a result of the imperfect control of the measured TM, the rigidity of $\ln\tau_n$ for the diffusive samples deviates from the prediction for GOE. Instead, it is found to follow the prediction for GUE, which is associated with a system with broken time-reversal symmetry. The underlying repulsion between transmission eigenvalues which is the cause of the rigidity is also explored via the distribution of nearest spacing between neighboring $\ln \tau_n$.

\section{Probability distribution of ``optical'' conductance}
\begin{figure}[htc!]
\centering
\includegraphics[width=3.2in]{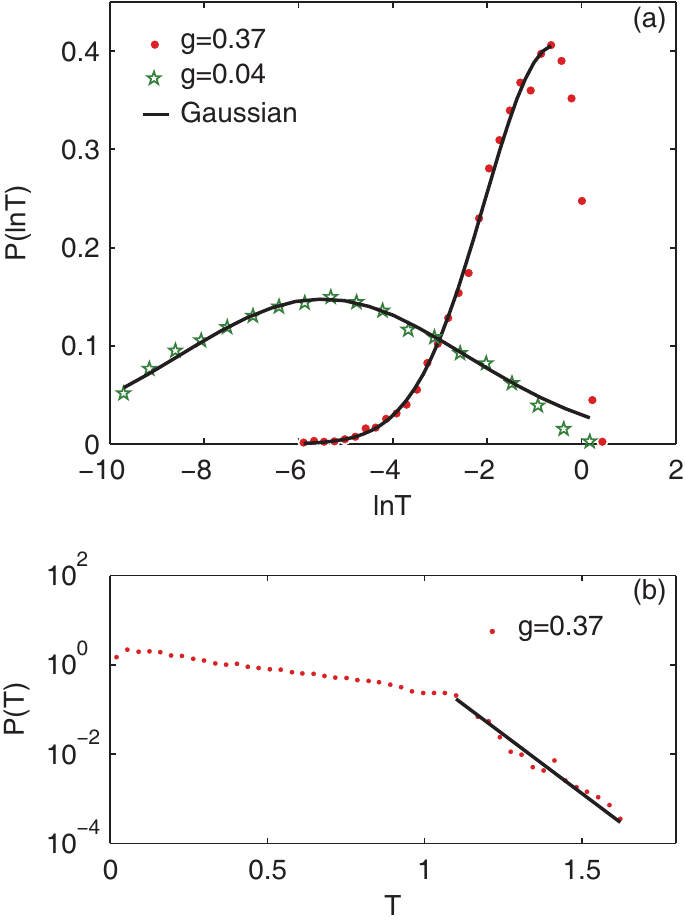}
\caption{ (a) Probability distribution of optical conductance for two random ensembles with values of $\textsl{g}=0.37$ and 0.045, respectively. $P(\ln T)$ for $\textsl{g}=0.37$ (red dots) and 0.045 (green asterisk). The solid black line is a Gaussian fit to the data. For $\textsl{g}=0.045$, all the data points are included, while for $\textsl{g}=0.37$, only data to the left of the peak are used in the fit. (b) $P(T)$ for the ensemble with {\textsl g}=0.37 in a semi-log plot. For high values of $T>1.1$, $P(T)$ falls exponentially.}\label{Fig1}
\end{figure}
Fig. 5.1 shows the measurements of the probability distribution of $\ln T$, $P(\ln T)$, for random ensembles with $\textsl{g}=0.37$ and 0.045. $P(\ln T)$ is seen to be Gaussian for $\textsl{g}=0.045$ and highly asymmetrical for $\textsl{g}=0.37$. The low transmission side of $P(\ln T)$ is well fit by a Gaussian distribution, but $P(\ln T)$ falls sharply for high transmission. For $T>1.1$, $P(T)$ falls exponentially with $e^{-11T}$. 

\begin{figure}[htc]
\centering
\includegraphics[width=5in]{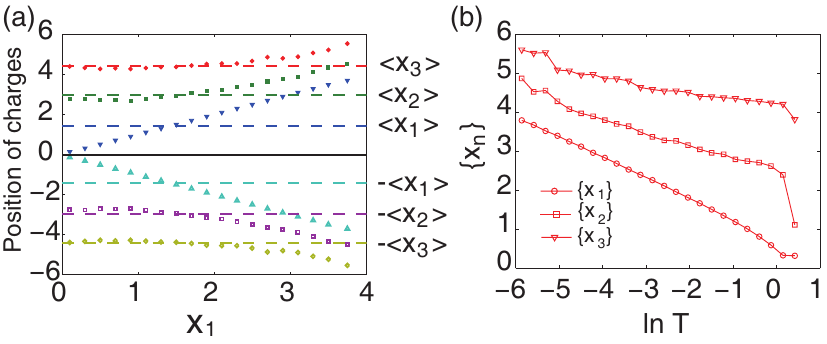}
\caption{ Charge model of transmission eigenvalues and conductance. (a) Average positions of charges and their images with respect to different positions of the first charge $x_1$ in the random ensemble with $\textsl{g}=0.37$. The dashed lines show the average positions of the charges for this ensemble. (b) Average positions of charges vs. $\ln T$ in the same ensemble.}\label{Fig2}
\end{figure}
The variation of the average positions of the charges as a function of the position of the first charge $x_1$ in the ensemble with $\textsl{g}=0.37$ is plotted in Fig. 5.2a. The average spacing between $x_1$ and $x_2$ increases as the value of $x_1$ decreases. This is due to the repulsion of the charge at $x_2$ by the charge at $x_1$ and the nearby images. At the same time, the spacing between $x_2$ and $x_3$ and their average positions hardly change as $x_1$ moves towards the origin. This reflects the tendency to heal large fluctuations in charge positions for more remote charges.

\begin{figure*}[!htbp]
\begin{center}
 \centerline{\includegraphics[width=5in]{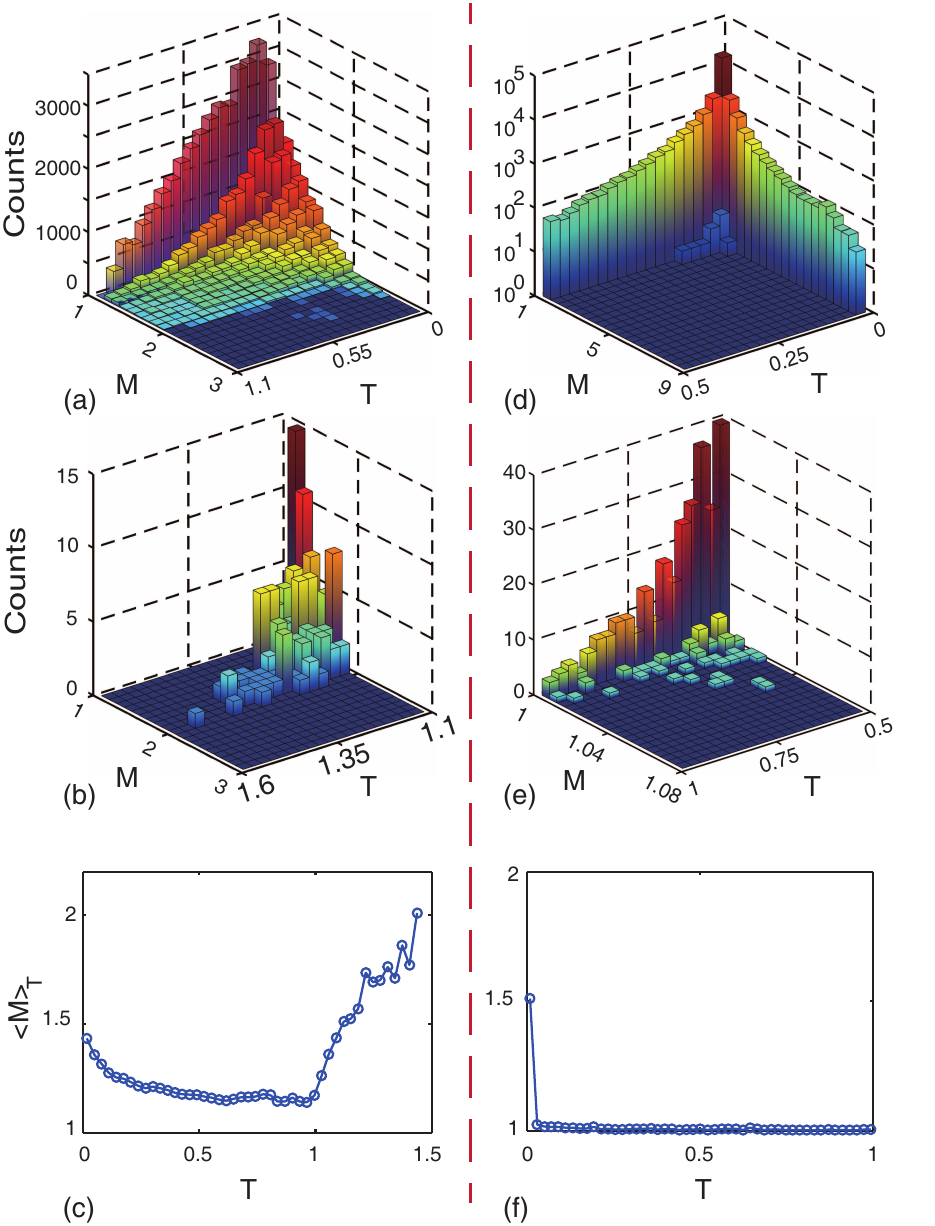}}
 \end{center}
\caption{ The joint probability distribution of $T$ and $M$, $P(T,M)$, for two random ensembles of {\textsl g}=0.37 (left column) and 0.045 (right column). The average value of $M$ vs. $T$ for the two ensemble are shown in 5.3(c) and (f).}\label{fig3}
\end{figure*}
The source of the sharp cutoff in $P(\ln T)$ can be seen by examining the spacing of the averages of $x_n$ for given value of $\ln T$ shown in Fig. 5.2b. A relatively high value of $T$ can only be achieved when the first charge is near the origin. This is an unlikely event because this charge is strongly repelled by its image. $P(T)$ would be expected to fall especially rapidly for values of $T$ above unity since this would require two charges along with their images to be close to the origin. This would correspond to large values for both $\tau_1$ and $\tau_2$. The number of transmission eigenvalues contributing substantially to transmission for large values of $T$, $\langle M\rangle_{T>1}$ would approach 2. This is seen in the joint distribution of $T$ and $M$, $P(T,M)$ in Figs. 5.3a and b and in the plot of $\langle M\rangle_{T}$ for {\textsl g}=0.37 in Fig. 5.3e.

The unusual nature of $P(T,M)$ for {\textsl g}=0.37 can be appreciated in a comparison in Fig. 5.3 of the statistics of $T$ and $M$ for {\textsl g}=0.045. For both these ensembles, $\langle M\rangle_{T}$ is distinctly larger than unity for {\it T}$<${\textsl g}. This tendency is more striking in the more strongly localized sample since $\langle M\rangle_{T}$ falls sharply and remains close to unity for high values of T. In contrast, $\langle M\rangle_{T}$ falls gradually at first for {\textsl g}=0.37 but then rises above $T \sim 1$ towards a value of 2. The increase of $\langle M\rangle_{T}$ above unity for small $T$ are associated with transmission at frequencies falling between the central frequency of the electromagnetic modes of the medium, where several weakly overlapping modes contribute to transmission. Spectra of $T$ do not show a succession of distinct Lorentzian lines even for {\textsl g}=0.045 for which $L$=102 cm $\sim 4\xi$ so that a number of electromagnetic modes make some contribution to transmission even when $T$ approaches unity. Thus, the superposition of modes produces a single channel that dominates other channels with $\langle M\rangle_{T>0.4} \sim 1.004$. The reason for this is that spatially separated resonances in deeply localized samples are hybridized into modes, which are linear combinations of these resonances \cite{85a}. Such coupled modes in 1D are predicted to exhibit a succession of peaks along the length of the sample, which are called necklace states \cite{85b}. The speckle patterns of these modes on the output surface should be similar since the field at the output will be dominated by the resonance closest to the output. The speckle pattern of the dominating transmission eigenchannel will be the same as for these quasi-normal modes.

\begin{figure}[htc]
\centering
\includegraphics[width=3in]{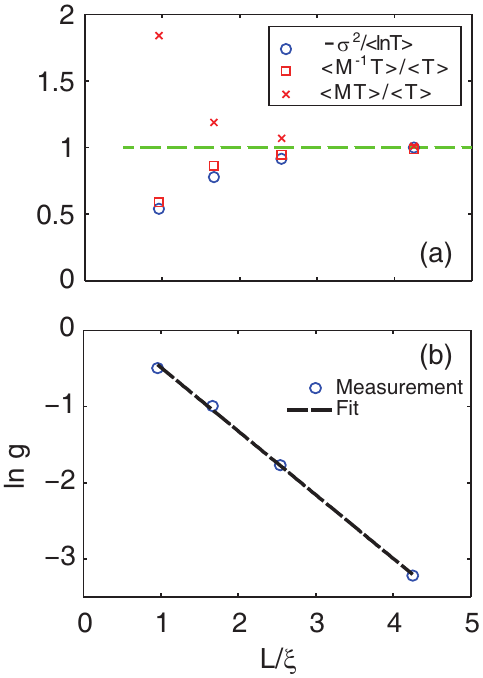}
\caption{(a) The ratio between $\sigma^2$ and $-\langle \ln T\rangle$, $\langle M T\rangle/\langle T\rangle$ and $\langle M^{-1} T\rangle/\langle T\rangle$ with respect to $L/\xi$. The dashed line is the prediction of SPS for large $L/\xi$. (b). Exponential decay of $\langle \ln T\rangle$ vs. sample length length $L$ for localized wave. The dashed black line is a linear fit to the data.}\label{Fig4}
\end{figure}
In the localization limit, $T\sim\tau_1\sim 4\exp(-2x_1)$ and $x_1$ is hypothesized to follow a Gaussian distribution. This leads to the log-normal distribution for {\it T} seen in Fig. 5.1a for $\textsl{g}=0.045$, as predicted by SPS theory of localization. SPS further predicts that in the limit of small $\textsl{g}$, var$(\ln T)\equiv \sigma^2=-\langle \ln T\rangle$, so that $P(\ln T)$ depends only upon the single parameter $-\langle \ln T\rangle=L/\xi$. The ratio of $\sigma^2$ and $-\langle \ln T\rangle$ as well as the average of $M$ weighted by $T$, ${\langle M T \rangle}/\langle T \rangle $, are plotted with respect to $L/\xi$ in Fig. 5.4. These functions are seen to approach unity for $L\gg\xi\sim24$ cm with the distribution of $T$ being log-normal, as predicted by SPS hypothesis, precisely when ${\langle M T \rangle}/\langle T \rangle $ approaches unity. At this point, the statistics of $T$ for the quasi-1D sample become one dimensional. We see in Fig. 5.4 that the weighted average of $M^{-1}$,  ${\langle M^{-1} T \rangle}/\langle T \rangle $, follows the ratio of SPS. In a single configuration, $M^{-1}$  is equal to the variance of the total transmission relative to its average in that configuration, $M^{-1}$=var$(NT_a/T)$. Thus SPS theory for fluctuations of $T$  within a random ensemble of 1D samples applies to quasi-1D samples when the fluctuations of total transmission within single samples arise from transmission through a single eigenchannel.

\begin{figure}[htc]
\centering
\includegraphics[width=3in]{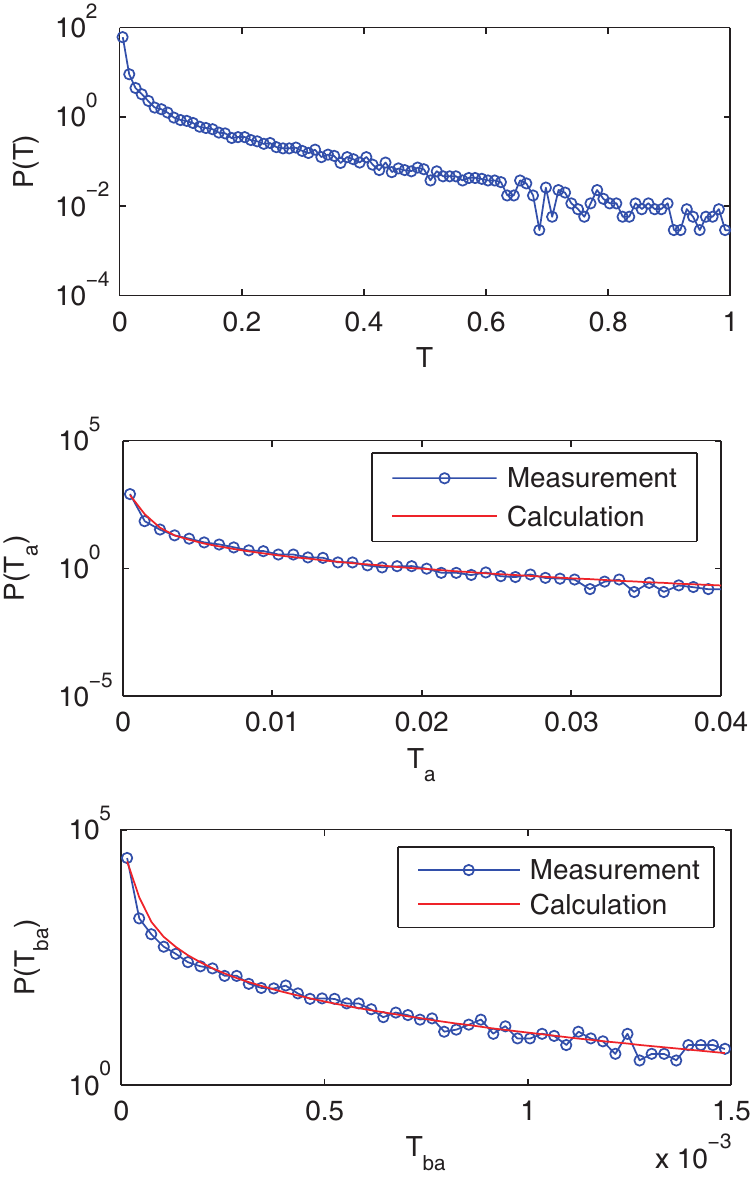}
\caption{ Statistics of transmittance, total transmission and intensity for the ensemble with {\textsl g}=0.045. The calculation is made based upon the assumption of $M=1$ in this ensemble.}\label{Fig5}
\end{figure}
A full account of the statistics of transmission is given by the joint probability distribution of all the transmission eigenvalues, $P(\{\tau_n\})$. We have seen that two functions of the set of transmission eigenvalues $\{\tau_n\}$, $T$ and $M$, are key localization parameters and that their joint distribution $P(T,M)$ give insight into the approach to localization. Clearly, integrating this distribution over $M$ gives $P(T)$, but more significantly, $P(T,M)$ also yields the second order distributions of statistical optics. We have shown that the statistics of relative transmission in a single transmission matrix depends only upon $M$, while the transmittance $T$ provides an overall normalization factor. The total transmission relative to its average in the transmission matrix, $NT_{a}/T$, has a variance of $M^{-1}$ and a distribution which changes from Gaussian to negative exponential over the range in which $M^{-1}$ changes from 0 to 1. An analytical expression for $P(NT_{a}/T;M)$ has not been obtained yet, but these distributions can be found from measurements of the transmission matrix. The distribution of total transmission in an ensemble of given $\textsl{g}$ can thus be expressed as $\int P(NT_a/T;M)P(T,M;{\textsl g})\, \mathrm{d}T\mathrm{d}M$. Since the distribution of intensity in a single speckle pattern is a negative exponential, $\exp(-NT_{ba}/T_a)/T_a$, the distribution of $T_{ba}$ is obtained by mixing this function with $P(T_a)$. Thus the statistics of intensity, total transmission and transmittance can be obtained from $P(T,M;{\textsl g})$. These statistics are presented in Fig. 5.5 for the sample in which {\textsl g}=0.045 and $\langle M T\rangle/\langle T \rangle$=1.02. The degree to which transmission is via a single channel can be appreciated by comparing the measured $P(T_a)$ to the calculation based on the assumption of {\it M}=1 for all values of {\it T}, where $P(NT_a/T)=\exp(-NT_a/T)/(T/N)$. The tendency towards higher values of $M$ for small $T$ seen in Fig. 5.3f leads to a small discrepancy at low values of $T_a$ between the measurement and the calculation. This is readily repaired by more accurately representing the full distribution of $M$ for small values of $T$. The intensity distribution calculated under the same assumption of $M$=1 is shown in Fig. 5.5c and seen to be in excellent agreement with measurements.

We now investigate the statistics of $M$, since it plays a central role in the statistics of transmission. Fig. 5.6 shows $P(M^{-1})$ for both diffusive and localized waves. 
\begin{figure}[htc]
\centering
\includegraphics[width=3in]{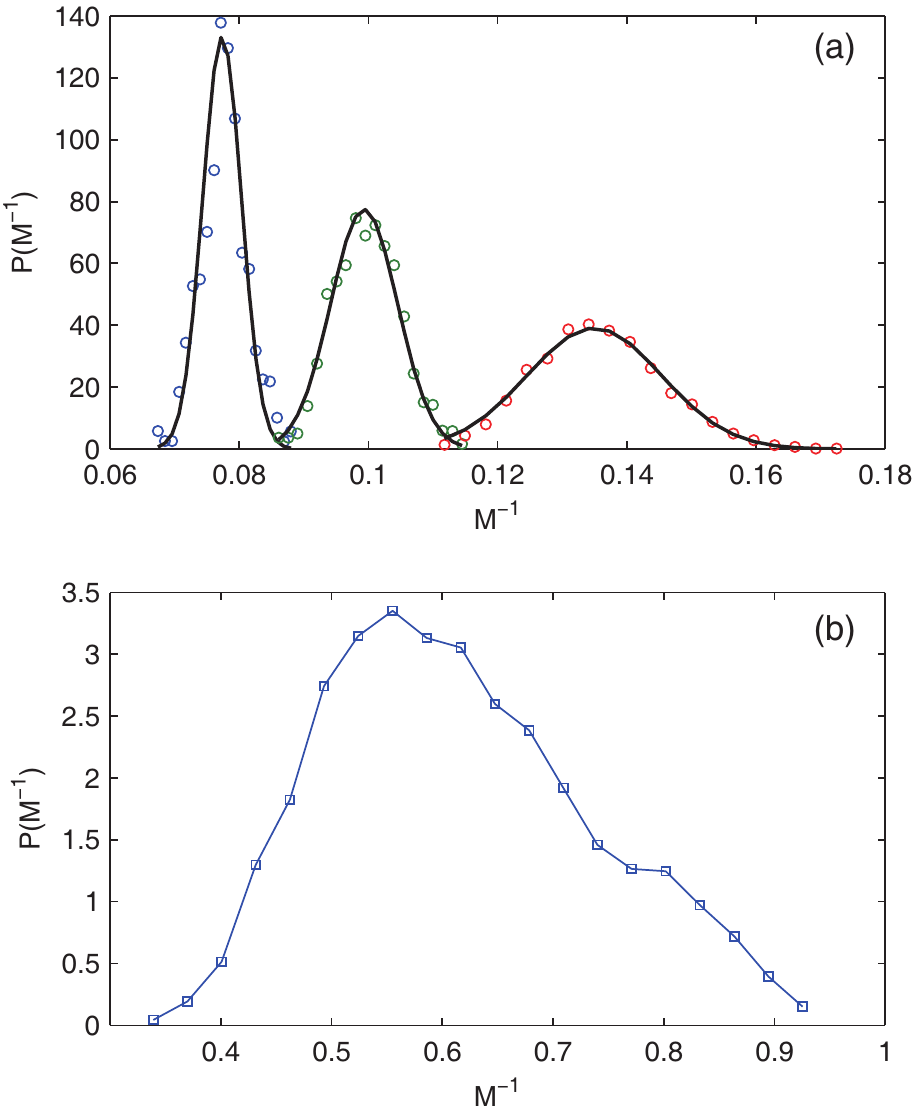}
\caption{Probability distribution of $M^{-1}$ for (a) three diffusive ensembles with $L$= 23, 40 and 61 cm and (b) localized waves with $L$=23 cm.}\label{Fig5.6}
\end{figure}

\begin{figure}[htc!]
\centering
\includegraphics[width=3in]{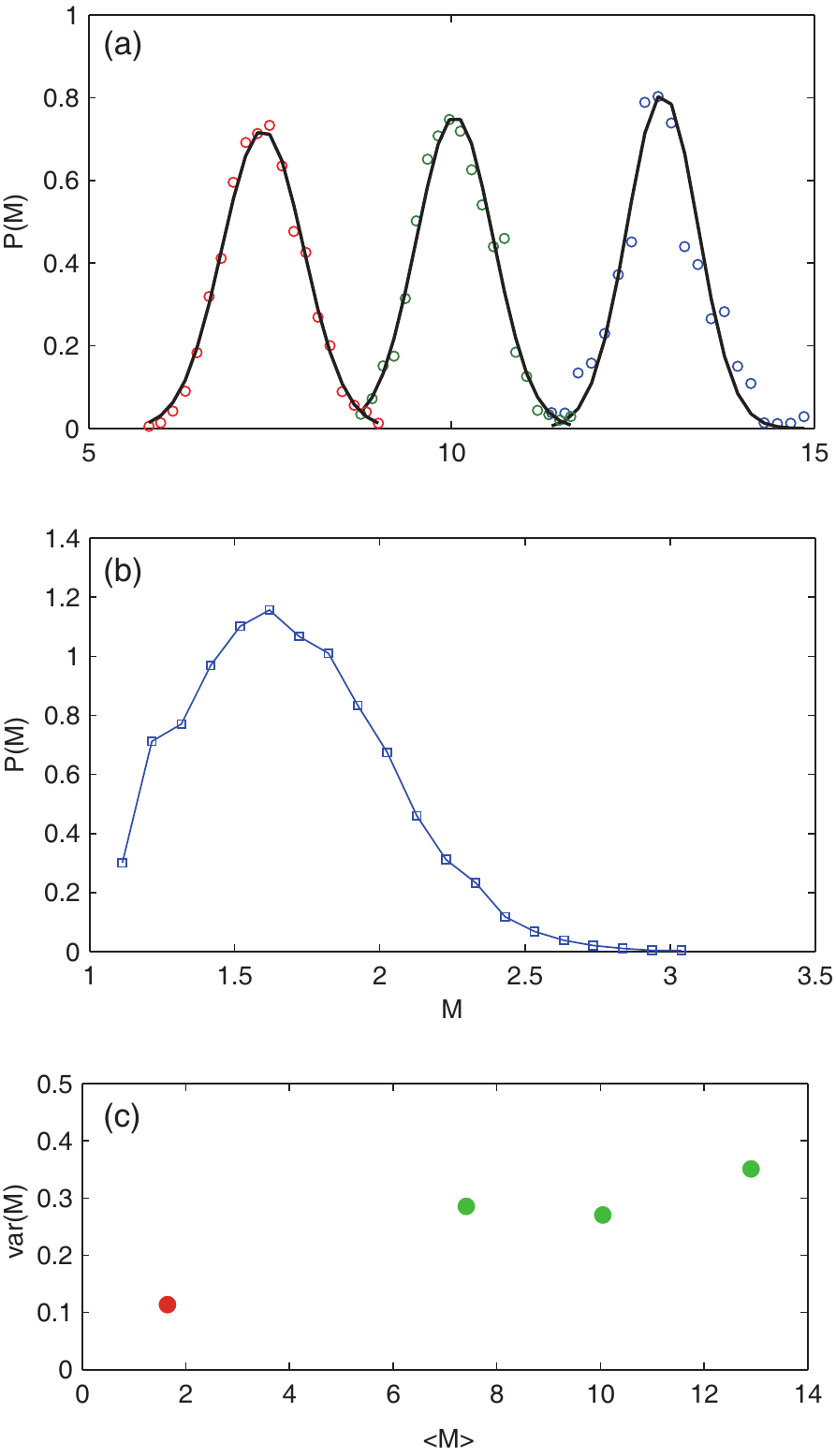}
\caption{Universal fluctuation of $M$ for diffusive waves. (a) Probability distribution of $M$ for three diffusive samples of lengths $L$= 23, 40 and 61 cm. (b) $P(M)$ for the localized sample with $L=23$ cm. (c) Var($M$) vs. $\langle M\rangle$ for the four samples studied in (a) and (b).}\label{Fig5.7}
\end{figure}
In order to improve statistics, we increase the number of matrices analyzed by picking 5 different random combinations of $N$/2 from the measurements to construct the transmission matrix at each frequency in a given sample configuration. For three diffusive samples, $P(M^{-1})$ is well fit by a Gaussian distribution with a decreasing value of variance as the mean value of $M^{-1}$ falls. A rather surprising feature is seen when we plot the distribution of $M$ instead, which is shown in Fig. 5.7. $P(M)$ for three diffusive samples with different lengths follows a Gaussian distribution with nearly the same variance. This is reminiscent of the phenomenon of universal conductance fluctuations. The fact that fluctuation of $M^{-1}$ for diffusive samples is not universal is similar to earlier finding that the fluctuation of resistance of mesoscopic conductors is not universal whereas its inverse, conductance, has a universal variance. The values of $\langle M\rangle$ and var($M$) for corresponding samples are presented in Fig. 5.7c. The scaling of $\langle M\rangle$ vs. 1/$L$ for the three diffusive samples is explored and demonstrated in Fig. 5.8a and we find that $\langle M\rangle$ is approximately inversely related to the sample length $L$. The scaling is seen to be in accord with diffusive transport when $\langle M\rangle$ is plotted with respect to the effective sample length $L_{eff}=L+2z_b$, in which $z_b$=13 cm is the extrapolation length found from the fit to the time of flight distribution. This is presented in Fig. 5.8b where three data points more closely falls on a line. 
\begin{figure}[htc]
\centering
\includegraphics[width=3in]{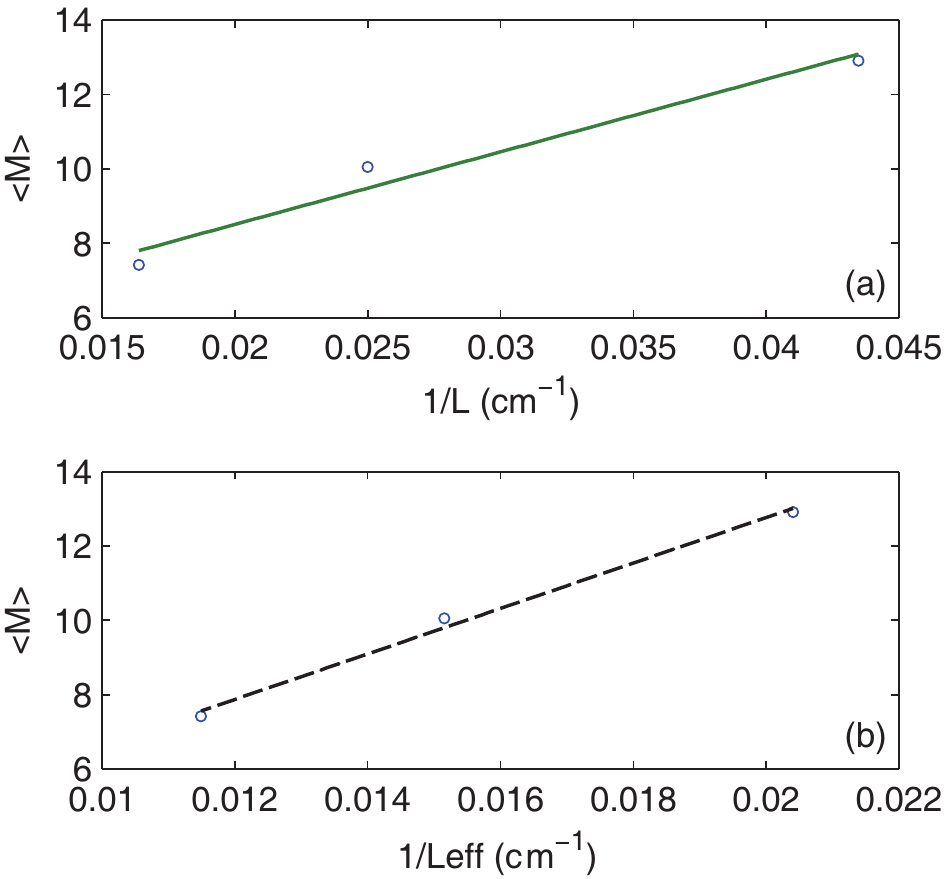}
\caption{Scaling of $\langle M\rangle$ for the three diffusive samples as in Fig. 5.7c. with respect to (a) the sample length $L$ and (b) the effective sample length $L_{eff}$.}\label{Fig5.8}
\end{figure}

Assuming a bimodal distribution of the transmission eigenvalues in a large transmission matrix $N\gg 1$, $M$ is given by, $\langle M\rangle=\frac{{\textsl g}^2}{\int_0^1 \tau^2\rho (\tau) \rm{d}\tau}=\frac{2}{3{\textsl g}}$. In contrast, $\rho(\tau)={\textsl g}/\tau$ will yield, $\langle M\rangle=1/2{\textsl{g}}$. The value of dimensionless conductance and the variance of {\it T} is presented in Table 5.1, based on the relation of $\langle T\rangle={\textsl g}=1/2\langle M\rangle$. 
\begin{center}
\begin{tabu} to 0.8\textwidth{|X[3,c]|X[1,c]|X[1,c]|}
\hline
Sample length $L$ (cm)  & {\textsl g}=$\langle T\rangle$  & var($T$) \\
\hline
23    & 6.5 & 0.26\\ \hline
40    & 5.0 & 0.42\\ \hline
61    &  3.7 & 0.23\\ \hline
\end{tabu}
\end{center}
\begin{table}[htc]
  \caption{The mean and variance of transmittance $T$ for three diffusive sample of lengths 23, 40 and 61 cm.}
\end{table}

The variance of $T$ is found to be $\sim 0.3$ and independent of the sample length $L$ or the mean value $\langle T\rangle=${\textsl g}. We believe this is the first direct demonstration of universal conductance fluctuation for classical waves. UCF was first observed in small metals rings in which the fluctuation of conductance is of order unity independent of the size of the sample. The $C_3$ correlation which is linked to UCF has been observed for classical waves. However, the value of var({\it T}) is considerably larger than the theoretical prediction for Q1D disordered random media of 2/15. This could be a consequence of the wave interaction at the sample interface, which gives the large value of the extrapolation length $z_b$. The strong interaction at the sample interface alters the value of fluctuation of transmittance, which is analogous to the impact of tunnel barrier on the universal conductance fluctuations in disordered electronic systems \cite{85c}. Some electrons will be instantaneously reflected by the tunnel barrier and therefore cannot make any contribution to the measured conductance. This effect disappears when the sample length $L$ is much greater than $\ell$, as does the impact of $z_b$ on transmission when $L\gg \ell$ for classical waves. Recent supersymmetry calculations \cite{85d} have shown that the interaction at the sample surfaces can substantially reduce the high transmission eigenvalues and therefore suppress the bimodal distribution. In addition, since the transmittance is obtained by summing the transmitted intensity on arrays of discrete points on the input and output surface, the lack of spatial averaging may give a larger fluctuation of $T$. The non-negligible correlation between the two orthogonal polarization measurement also enhances the fluctuation. The value of var($T$)$\sim 0.3$ reported is obtained by equating ${\textsl g}=\langle M\rangle/2$. We note that, when we take ${\textsl g}=2\langle M\rangle/3$, the variance of $T$ is $\sim 0.5$, but is still independent of the sample size. 

It is seen that the fluctuation of both $M$ and $T$ are universal for the diffusive samples and $\langle M\rangle$ is proportional to $\langle T\rangle$ in the limit of large {\textsl g}. However, $M$ and $T$ are two distinct parameters in the statistics of transmission. To see this, the average value of $M$ as a function of the normalized transmittance $s=T/\langle T\rangle$ for the diffusive sample of length $L=23$ cm is plotted and no clear correlation between $M$ and $s$ is seen.
\begin{figure}[htc]
\centering
\includegraphics[width=3in]{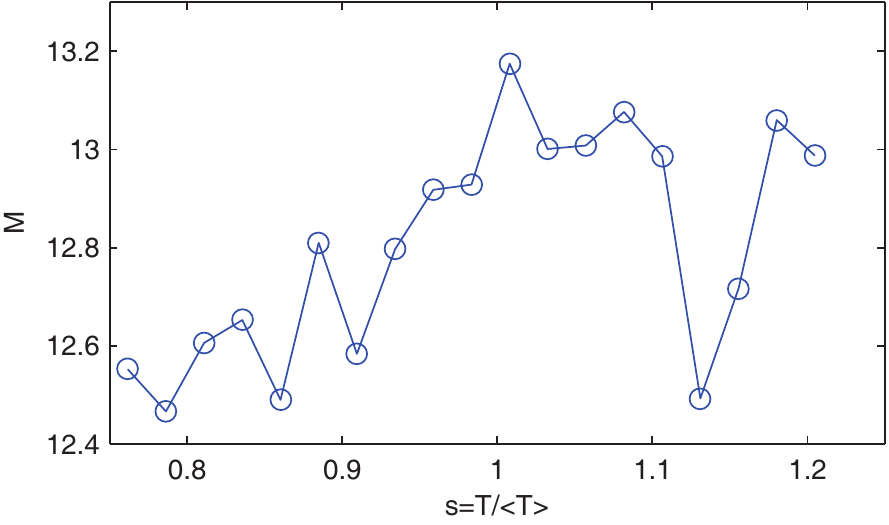}
\caption{The average value of $M$ for different value of $s$ in the ensemble of diffusive samples of length $L=23$ cm.}\label{Fig5.9}
\end{figure}
Furthermore, if they were highly correlated, var($M$) will be much greater than var($T$), since $\langle M\rangle$ is larger than $\langle T\rangle$. We have also performed numerical simulations based on Recursive Green's function method and found that the variance of $M$ and $T$ is about the same for a diffusive sample with ${\textsl g}\sim 6.7$. In Fig. 5.10, we show the correlation function of $T$ and $M$ with frequency shift $\Delta \nu$, defined as $C(\Delta \nu)=\langle x(\nu) x(\nu+\Delta \nu)\rangle-\langle x(\nu)\rangle \langle x(\nu+\Delta \nu)\rangle$. The value of correlation function at $\Delta \nu=0$ is equal to the variance. 
\begin{figure}[htc]
\centering
\includegraphics[width=3in]{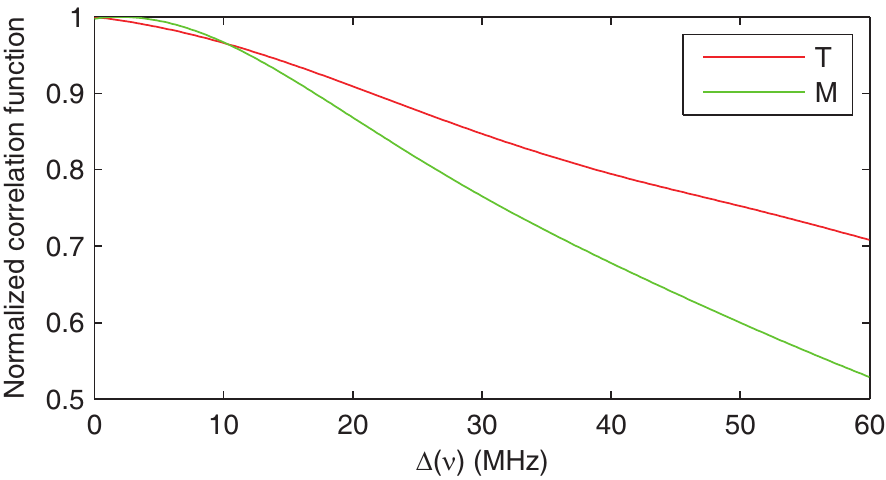}
\caption{The correlation function of $T$ and $M$ with frequency shift $\Delta \nu$. The curves are normalized by its maximum value in the figure to give the better comparison between the two correlation function.}\label{Fig5.10}
\end{figure}

The charge model provides an intuitive explanation of UCF \cite{62}. For diffusive samples, there are approximately ${\textsl g}$ charges packed in the interval of [0,1] with average spacing of $1/{\textsl g}$. The position of the charges varies at different sample realization and this results in the fluctuation of the measured conductance at different configurations. Due to the Coulomb interaction, the position of the the charges are rigid and the vibration around their equilibrium position is reduced. The distribution of the positions is a Gaussian for all the charges except the first one $x_1$. The distribution of $x_1$ is different from the rest is because the repulsion by its image. It is therefore expected to follow the Wigner surmise for the GOE \cite{62}. Because of the rigidity of the positions of the charges, the fluctuation of number of charges between 0 and 1 is small and therefore, the fluctuation of conductance is always of order of unity irrespective of the number of charges within [0,1], which is the mean value of the conductance in the ensemble. The 
fluctuation is also independent of the total number of charges $N$, in which $N$ depends on the transverse size of the sample. 

\begin{figure}[htc]
\centering
\includegraphics[width=3in]{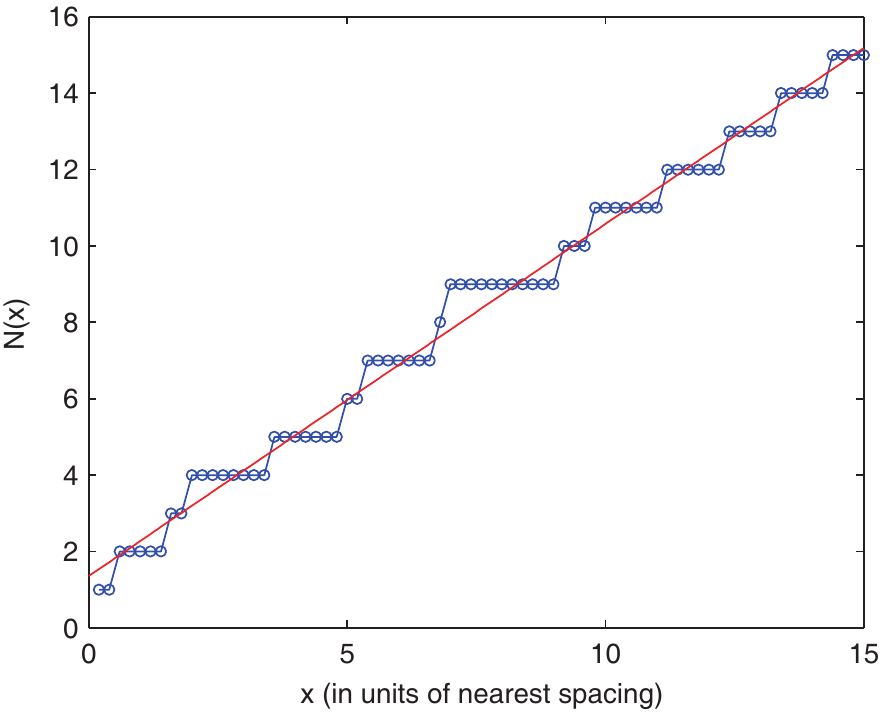}
\caption{The staircase counting function $N(x)$ for one set of $\ln \tau_n$ for diffusive sample with $L$=23 cm. The step for $x$ is 1/5 of the average nearest spacing between $\ln \tau_n$ in the random ensemble.}\label{Fig5.11}
\end{figure}
The strong repulsion between the charges tends to keep the charges at their equilibrium positions and therefore main a long-range order of the charge positions. A measure of the rigidity of the spectrum is called the $\Delta_3$ statistics, which was first proposed by Dyson and Metha \cite{85e} to explain the spectra of energy levels obtained in the slow neutron scattering experiments. The $\Delta_3$ statistics measures the deviation of a given energy sequence from a perfect uniform sequence. For a given sequence of length $S$, in which $S$ is measured in terms of the nearest average spacing, $\Delta_3(S)$ is defined as the least-squares deviation of the staircase function $N(x)$ from the best linear fit over the range of $S$,
\begin{equation}
\Delta_3(S)=\frac{1}{S}{\rm min}(\int_\alpha^{S+\alpha} (N(x)-Ax-B)^2 \rm{dx}).
\end{equation} 
Here, $N(x)$ is the counting function, which increases by one when it moves across an energy level. We have found that spectrum of $\ln \tau_n$ on average is rigid for both diffusive and localized waves. We can now consider the rigidity of spectrum of $\ln \tau_n$ using the $\Delta_3$ statistics. For large $n$, logarithm of $\tau_n$ is close to the value of -2$x_n$. One example of the counting function for diffusive sample with length of 23 cm is given in Fig. 5.11. The straight line in the figure is the best linear fit to the $N(x)$. 

\begin{figure}[htc]
\centering
\includegraphics[width=3in]{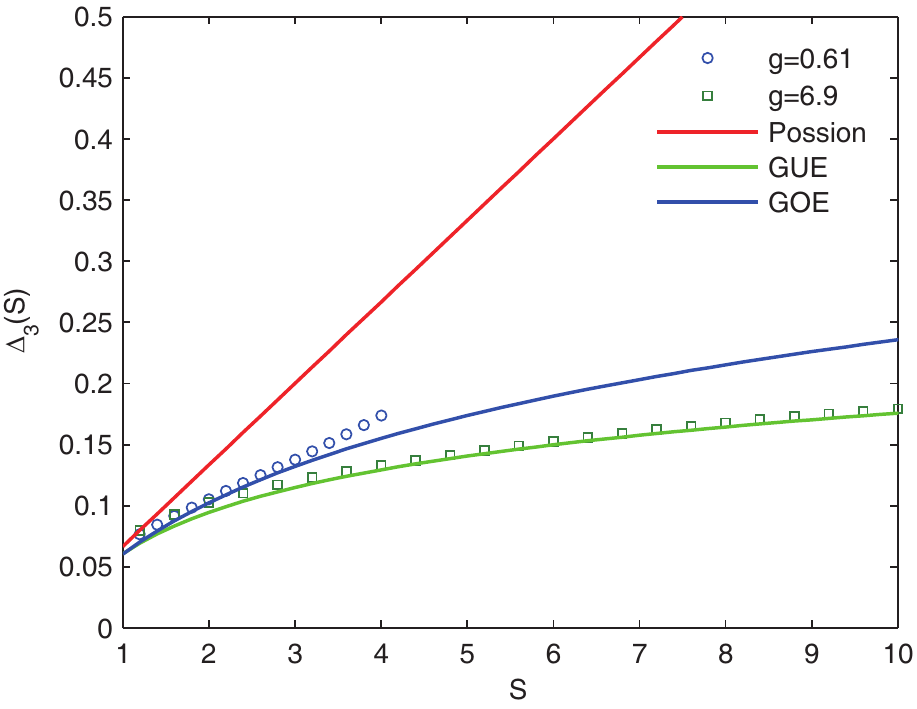}
\caption{Measure of level rigidity, $\Delta_3(S)$ statistics for diffusive (green squares) and localized (blue dots) sample with length $L=23$ cm. The red line is for the localization limit in which there is not interaction between $\ln\tau_n$.}\label{Fig5.12}
\end{figure}
When averaging over all the sample configurations and different starting values of $\alpha$, the $\Delta_3(S)$ for the diffusive and localized sample with $L=23$ cm is presented in Fig. 5.12. We see that for large value of $S$, $\Delta_3(S)$ approaches the prediction for the GUE systems in the diffusive limit. This suggests that the spectrum of $\ln \tau_n$ is long-range ordered. The agreement with GUE instead of GOE may be attributed to the degree of the control of the measured transmission matrix discussed previously \cite{64}. In contrast, the rigidity of $\ln \tau_n$ is substantially weakened for localized waves. In the localized limit, one would expect the rigidity of $\ln \tau_n$ follows the $S$/15 for a Poisson process, since there is no interaction between neighboring $\ln \tau_n$ and the position of $\ln \tau_n$ is independent of the positions of other charges. It is worth noting that for localized waves, for small value of $S$, $\Delta_3(S)$ agrees with prediction for GOE systems. This could be due to the fact that we have better control of the measured TM for localized waves.

The measure of the interaction between neighboring $\ln\tau_n$ that leads to the rigidity of $\ln \tau_n$ will be the nearest spacing distribution of $\ln\tau_n$. The results are presented in Fig. 5.13.
\begin{figure}[htc]
\centering
\includegraphics[width=3in]{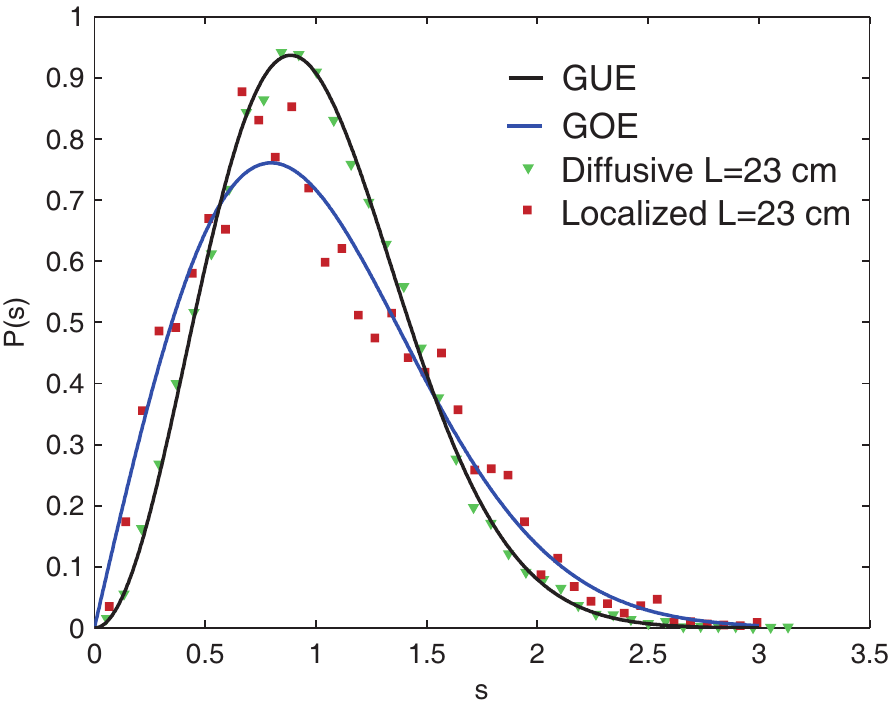}
\caption{Probability distribution of the nearest spacing of $\ln\tau_n$, $s=\frac{\ln\tau_n-\ln\tau_{n+1}}{\langle \ln\tau_n-\ln\tau_{n+1}\rangle}$, $P(s)$ for diffusive and localized sample of length $L=23$ cm.}\label{Fig5.13}
\end{figure} 
It is seen that the nearest spacing distribution for diffusive waves agrees with the Wigner surmise for GUE systems and is close to the Wigner surmise for GOE system for moderately localized sample. It could be the weak interaction that gives to the short-range order of the spectrum logarithm of the transmission eigenvalues for localized waves as seen in Fig. 5.12. It is indicative that the $\Delta_3$ statistics gives a better measure of closeness to localization than does the statistics of the nearest spacing distribution. 

\section{Summary}
In this chapter, we present measurements of the distribution of the ``optical" conductance in the Anderson localization transition. The distribution of $T$ changes from a Gaussian to a log-normal distribution as the number of $\tau_n$ contributing to conductance appreciably decreases. We have observed a one-sided log-normal distribution of transmittance in the crossover to Anderson localization at $L/\xi \sim$ 1.63 and a log-normal distribution for deeply localized waves at $L/\xi \sim$ 4. The impact of the joint distribution of transmission eigenvalues upon these distributions is interpreted both in terms of a charge model including image charges and in terms of $P(T,M)$. We show that SPS is approached in a quasi-1D sample as $\langle M\rangle \rightarrow 1$ with the statistics becoming one-dimensional in this limit. We show explicitly that $P(T,M)$ yields the statistics of intensity and total transmission for deeply localized waves, but, since $P(NT_{a}/T)$ in a single transmission matrix depends only on $M$, $P(T,M)$ determines the statistics of intensity, total transmission, and transmittance in any random ensemble. These results illustrate the power of a unified approach to mesoscopic physics that can treat integrated and local flux as arise in studies of electronic conductance and statistical optics. We find that the fluctuation of $M$ is universal for diffusive waves and report the first direct observation of UCF for classical waves. The weakening of the rigidity of spectrum of $\ln \tau_n$ is observed when approaching wave localization.  

\chapter{Focusing through random media}
\section{Introduction}
Waves, such as light, microwave, ultrasound and acoustic, are the ultimate tools for non-invasive and contact free imaging. Tremendous efforts have been made to control wave propagation in diverse media. For instance, optical elements, such as lens and mirrors, are used to guide light propagation. A revolutionary invention in optical imaging is the optical microscope, which allows us to resolve images of size around few hundred nanometers \cite{86}. It was realized by Abbe \cite{87} that there exists a fundamental limit for the lens based optical microscope, which is known as the diffraction limit. This fundamental limit reflects the inability of capturing the evanescent waves in the far field imaging process. Near field microscope can reach a very high resolution since the information of the evanescent waves are preserved in the near field measurement \cite{89}. Some engineered materials, called Metamaterials, can also be used to perform subwavelength imaging as demonstrated in superlens \cite{88}, in which is the evanescent waves are amplified in the far field and in hyperlens \cite{91,90}, where the evanescent waves are supported inside the medium with hyperbolic dispersion. 

All of these imaging techniques are vulnerable to scattering, since the phase of the multiply scattered wave is completely randomized. Recently, Vellekoop and Mosk \cite{92,93,94} have developed a wavefront shaping method to correct the spatial distortion due to scattering of light in a turbid medium. The phase pattern of the incident wavefront is adjusted by a spatial light modulator. Employing a learning algorithm with an intensity feedback from a target spot on the back of the sample, they were able to focus light through the opaque sample and enhance the intensity at the focal spot by 3 orders in magnitude \cite{92}. They were also able to increase the total transmitted light by 44\% \cite{93}. The ability of focusing wave through inhomogeneous media was first demonstrated in acoustics by means of time-reversal \cite{95}. The signal transmitted through a scattering medium from a source is recorded in time by arrays of transducers. The recorded time signals are played back in time and a pulse emerges at the location of the source. Assume $f(t)$ is the time-response to an incident pulse from the source, the system response to the time-reversed pulse $f^*(-t)$, $h(t)$ is given by the convolution of $f(t)$ and $f^*(-t)$,
\begin{eqnarray}
&&f(t) = \int_{-\infty}^{\infty} a(\omega)e^{-i\omega t}d\omega. \\
&&h(t) = \rm{Conv}(f(t),f^*(-t)) =\int_{-\infty}^\infty e^{-i\omega t}|a(\omega)|^2d\omega.
\end{eqnarray} 
Here, $\ast$ is the complex conjugation. Since all the frequency components of $f(t)$ at $t=0$ add in phase, a focused pulse then emerges at $t=0$. 

Analogous to the method of time-reversal, wavefront shaping exploits the spatial degrees of freedom of the incident wave to focus monochromatic radiation through random systems. In order to focus at the output channel $\beta$, the phase of the incident wavefront has to be adjusted so that the transmitted electric field at $\beta$ from different input channels $a$ arrive in phase and interfere constructively. Thus, to obtain the optimal focusing, the correct wavefront will be the phase conjugation of the transmitted electric field $t_{\beta a}$, $t^*_{\beta a}/\sqrt{\sum_a |t_{\beta a}|^2}$ \cite{93}, where $t_{\beta a}$ is the transmission coefficient from input channel $a$ to output channel $\beta$ and $\sqrt{\sum_a |t_{\beta a}|^2}$ is the normalization to set the incident power to be unity. With the information of the transmission matrix of the disordered medium, it is possible to focus radiation at any desired spot or simultaneously at several spots by phase conjugating the measured transmission matrix. This has been demonstrated in microwave and optical measurements of transmission matrices \cite{78,72,96}.  

For a pulse impinging on a scattering medium, the temporal profile of the pulse is distorted as is its spatial profile \cite{40}. Because of the multiple scattering, the residence time of wave inside the sample varies and this gives the so-called time flight distribution when the time-varying intensity $I(t)$ is averaged over a random ensemble. The width of the time of flight distribution is significantly broadened as compared to the incident pulse. Recently, the wavefront shaping technique has been extended to control the pulse transmission through random media. Employing a generic learning algorithm with the intensity at a point at a given time delay as a feedback or via feedback from nonlinear process such as two-photon fluorescence, optical pulses have been focused through random media \cite{98,99}. 

In this chapter, we demonstrate focusing of monochromatic radiation through random media via phase conjugation of microwave transmission matrix and relate the focusing to the statistical parameters that characterizes the random medium. The phase conjugation was applied numerically, since in our experiment, we cannot modify the incident wavefront to achieve focusing experimentally. The spatial contrast between average intensity at the focus and the background is found to be equal to $1/(1/M-1/N)$, in which $M$ is the eigenchannel participation number of the measured transmission matrix of size $N$ \cite{72}. We found that, the contrast is given by the formula, even when $N$ is smaller than the number of independent channels in the medium. For a random medium with $N \gg 1$ and in the diffusive limit ${\textsl g}\gg 1$, the focusing contrast is the same as the inverse of the long-range correlation $\kappa$. The spatial profile of the focus can be expressed in terms of the square of the field-field correlation in space and the long-range correlation $\kappa$ of the random media. A time dependent transmission matrix has also been exploited to focus a pulse transmitted through the random waveguide at a point for a given time delay and the spatial contrast is given by $1/(1/M(t^\prime)-1/N)$, where $M(t^\prime)$ is the eigenchannel participation number at time $t^\prime$. The profile of the focused pulse is equal to the square of the field-field correlation function in time and is the same as the incident Gaussian pulse. The decreasing of $M(t^\prime)$ in time is closely linked to the number of quasi-normal modes contributing appreciably to transmission at the time $t^\prime$. 

\section{Focusing monochromatic radiation through random waveguides}
\begin{figure}[htc]
\centering
\includegraphics[width=3in]{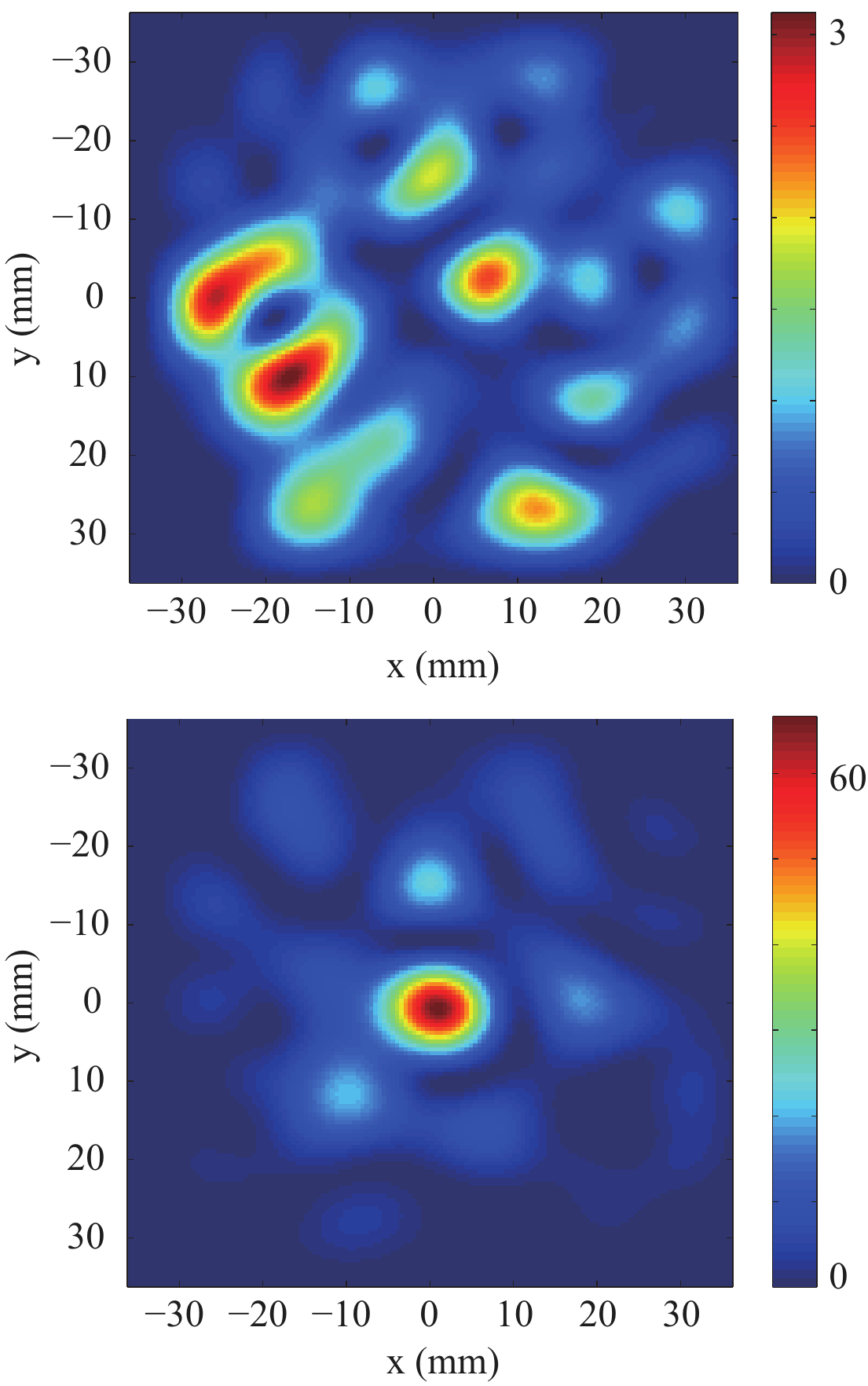}
\caption{Intensity speckle pattern generated for {\it L}=23 cm for diffusive waves. The speckle pattern is normalized to the average intensity within the speckle pattern. Focusing at the center point via phase conjugation is shown in 6.1(b).}\label{Fig5.12}
\end{figure}
In Fig. 6.1, a speckle pattern produced by radiation from point {\it a} on the center of the input surface for a diffusive sample of length $L=23$ cm. The speckle pattern is measured for single polarization and the intensity is normalized to the average value of the pattern. The Whittaker-Shannon sampling theorem is used to obtain high-resolution patterns. To focus the radiation at the center of the output surface $\beta=(0,0)$, the incident wavefront should be $t^*_{\beta a}/\sqrt{\sum_a |t_{\beta a}|^2}$. The corresponding speckle pattern is given in Fig. 6.1(b) and a focal spot emerge at target spot $\beta=(0,0)$. The intensity at the focal spot $I_\beta$ is equal to,
\begin{equation}
I_\beta = |\frac{t_{\beta a}t^*_{\beta a}}{\sqrt{\sum_a |t_{\beta a}|^2}}|^2=\sum_a |t_{\beta a}|^2 = T_\beta.
\end{equation}  
The background intensity when phasing conjugating the transmission matrix is given by, 
\begin{equation}
I_{b\ne \beta}=\frac{|\sum_n \tau_nu_{nb}u^*_{n\beta}|^2}{T_\beta}=\frac{\sum_{n,n^\prime} \tau_n\tau_{n^\prime}u_{nb}u^*_{n\beta}u^*_{n^\prime b}u_{n^\prime \beta}}{\sum_n \tau_n|u_{n\beta}|^2}.
\end{equation}
The $u$ matrix is obtained by the singular decomposition of the transmission matrix $t$, $t=U\Lambda V^\dagger$. We consider the ratio within a single sample realization between the average intensity at the focus and the average intensity at the background as the focusing contrast,
\begin{equation}
\mu=\frac{\langle I_\beta \rangle_\beta}{\langle I_{b\ne \beta}\rangle_\beta}.
\end{equation}
in which $\langle \dots \rangle$ indicates averaging over all possible focusing points. The average intensity at the focus is $\langle T_\beta\rangle_\beta=\sum_n \tau_n/N$. The background intensity averaged over all possible $b\ne \beta$ and $\beta$ is given by,
\begin{equation}
\langle I_b\rangle_{b\ne \beta, \beta} = \frac{\langle \sum_n \tau_n^2|u^2_{n\beta}|/T_\beta\rangle_\beta}{N-1}-\frac{\langle T_\beta \rangle_\beta}{N-1}.
\end{equation}
We found that $\langle \sum_n \tau_n^2|u^2_{n\beta}|/T_\beta\rangle_\beta$ can be well approximated as the ratio between the average of the numerator and of the denominator, $\langle \sum_n \tau_n^2|u^2_{n\beta}|\rangle_\beta/\langle T_\beta\rangle_\beta\sim\frac{T}{M N}$. This yields the average background intensity when focusing is achieved,
\begin{equation}
\langle I_b\rangle_{b\ne \beta, \beta}\sim\frac{T}{M N}-\frac{T}{N^2}.
\end{equation}
Here, $M=\frac{(\sum_n \tau_n)^2}{\sum_n \tau^2_n}$ is the eigenchannel participation number and {\it N} is the number of channels in the measured transmission matrix. The average intensity is equal to $T/N^2$ and it is clearly seen that the background intensity is enhanced by a factor of $N/M$ when focusing is achieved via phase conjugation. This is due to that more weight is given to high transmission eigenchannels when phase conjugation of the TM is applied. The contrast for optimal focusing is,
\begin{equation}
\mu=\frac{\langle I_\beta\rangle_\beta}{\langle I_b\rangle_{b\ne \beta, \beta}}=\frac{T/N}{\frac{T}{MN}-\frac{T}{N^2}}=\frac{1}{(1/M-1/N)}.
\end{equation}
In the limit of a large transmission matrix $N\gg 1$, the contrast is simply the eigenchannel participation number {\it M}. Since for localized waves, the transmission is dominated by the first eigenchannel and therefore the value of $M$ approaches 1. We then expect that focusing for localized wave cannot be achieved by phase conjugation, since the background intensity is also enhanced by a factor of $N$. This is confirmed in Fig. 6.2. 
\begin{figure}[htc]
\centering
\includegraphics[width=3in]{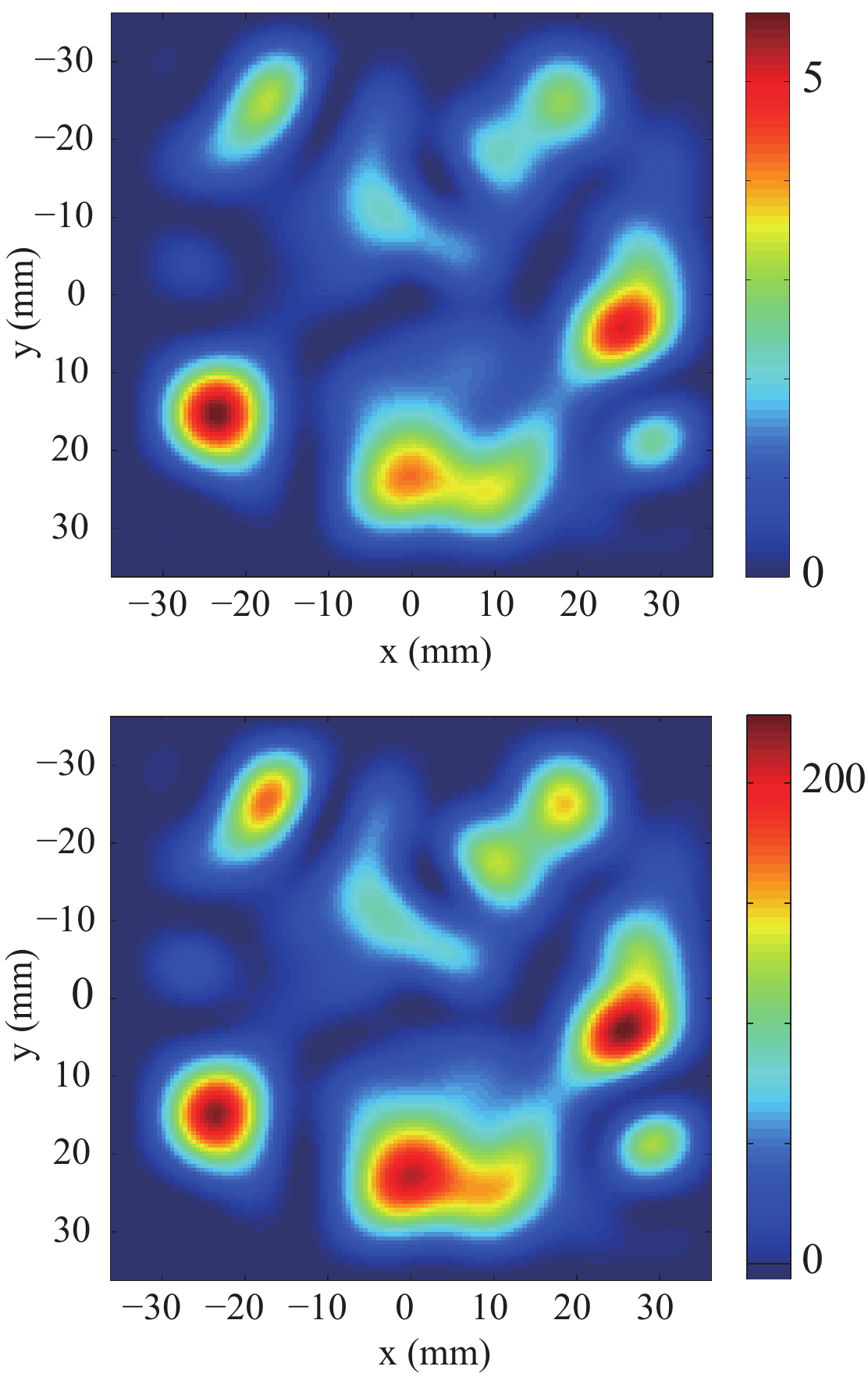}
\caption{Intensity speckle pattern generated for {\it L}=61 cm for localized waves. The speckle pattern is normalized to the average intensity within the speckle pattern. Phase conjugation is applied to focus at the center point shown in 6.2(b) and no focusing is obtained.}\label{Fig5.12}
\end{figure}

We now consider the situation where only $N^\prime$ channels can be measured instead of the $N$ channels and calculate the optical focusing contrast defined earlier. We found that the formula is still valid when only a fraction of the TM can be determined experimentally, with $M$ and $N$ replaced by $M^\prime$ and $N^\prime$. Here, $M^\prime$ is the eigenchannel participation number for the reduced transmission matrix of size $N^\prime$.  This is demonstrated in Fig. 6.3. This can be applied to optical measurements of the transmission matrix in which $N^\prime$ is much smaller than $N$.  
\begin{figure}[htc]
\centering
\includegraphics[width=3in]{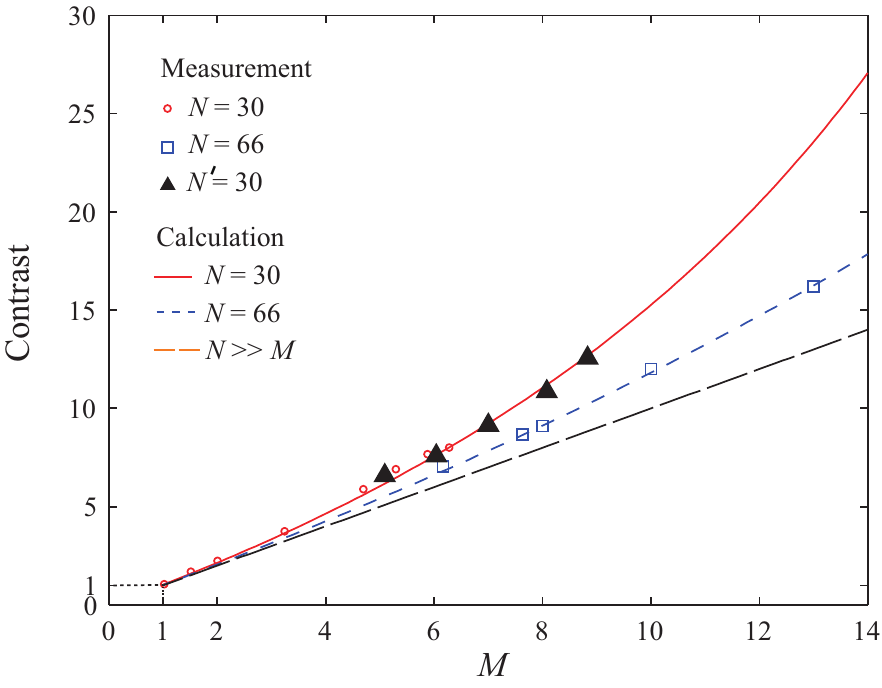}
\caption{Contrast in maximal focusing vs. eigenchannel participation number {\it M}. The open circles and squares represent measurements from transmission matrices $N=30$ and 66 channels, respectively. The filled triangles give results for $N^\prime \times N^\prime$ matrices with $N^\prime=30$ for points selected from a larger matrix with size $N=66$. Phase conjugation is applied within the reduced matrix to achieve optimal focusing. Eq. 6.8 is represented by the solid red and dashed blue curves for $N=30$ and 66, respectively. In the limit of $N\gg M$, the contrast is given by Eq. 6.8 is equal to $M$, which is shown in long-dashed black lines.}\label{Fig5.12}
\end{figure}

Since the variance of relative total transmission within large single transmission matrix is equal to the inverse of $M$, $\rm{var}(s_a/s)=1/$$M$. In the diffusive limit in which the fluctuation of $s$ is much smaller than fluctuation of $s_a$, var($s_a$) can be estimated to be $1/M$. At the same time, since the long range correlation is linked to the fluctuation of normalized total transmission, $\kappa=\rm{var}(s_a)$. $\kappa$ is also proportional to the inverse of the dimensionless conductance {\textsl g}. The focusing contrast can therefore be expressed in terms of the fundamental parameter which characterize the wave propagation through the disordered media. 

Significantly higher spatial resolution is achieved by focusing in disordered systems than in free space since the resolution is not limited by the aperture of the emitting array. Instead, the resolution is equal to the field correlation length, which is the inverse of the width of the $k$-vector distribution of the scattered waves. The spatial variation of intensity in the focused speckle pattern $\langle I_{foc}(\Delta r)\rangle$ reflects the decay of the ensemble average of the coherent sum of eigenchannels at the focus toward the average value of the incoherent sum. $\langle I_{foc}(\Delta r)\rangle$ can be expressed as a function of the degree of extended intensity correlation $\kappa$ and the square of the field-field correlation function $F(\Delta (r))=|\langle E(r)E^*(r+\Delta r)\rangle|^2/[\langle I(r)\rangle \langle I(r+\Delta r)\rangle]$. The average intensity in the focus pattern normalized to the average intensity at the focus is, 
\begin{equation}
\frac{I_{foc}(\Delta r)}{N\langle I\rangle}\sim\frac{\langle |\sum_n \tau_n u_{nb}u^*_{n\beta}|^2\rangle}{\langle |\sum_n \tau_n|u_{n\beta}|^2|^2 \rangle}.
\end{equation}
The numerator on the right-hand side of the Eq. 6.9 is also written as $\sum_n \langle \tau_n^2\rangle \langle u_{nb}u^*_{n\beta}u^*_{nb}u_{n\beta}\rangle+\sum_n \sum_{n\ne n^\prime} \langle \tau_n\tau_{n^\prime}\rangle \langle u_{nb}u^*_{n\beta}u^*_{nb}u_{n^\prime \beta}\rangle$. Since the components of the singular vectors $u_{nb}$ are circular Gaussian variables, the average product $\langle u_{nb}u^*_{n\beta}u^*_{n^\prime b}u_{n\beta}\rangle$ can be broken into the $|\langle u_{nb}u^*_{n\beta}|^2+\langle |u_{nb}|^2\rangle \langle |u_{n\beta}^2|\rangle$. In this expression, $|\langle u_{nb}u^*_{n\beta}\rangle|^2$ is the square of the field correlation function of the singular vectro $u_{nb}$, which can be approximated as $|\langle u_{nb}u^*_{n\beta}\rangle|^2=F(\Delta r)/N^2$. This gives $\langle u_{nb}u^*_{n\beta}u^*_{nb}u_{n\beta} \rangle=F(\Delta r)/N^2+1/N^2$.

For $n\ne n^\prime$, the singular vectors $u_{nb}$ and $u_{n^\prime b}$ are uncorrelated so that $\langle u_{nb}u^*_{n\beta}u^*_{nb}u_{n\beta} \rangle=\langle u_{nb}u^*_{n\beta}\rangle \langle u^*_{n^\prime b} u_{n^\prime \beta}\rangle = F(\Delta r)/N^2$. Eq. 6.9 can then be written as,
\begin{equation}
\frac{I_{foc}(\Delta r)}{N\langle I\rangle} = \frac{F(\Delta r)+\kappa}{1+\kappa}.
\end{equation}
The ensemble average of the intensity in the focused patterns is shown in Fig. 6.4 and seen to be in excellent agreement with Eq. 6.10 using measurement of the square of the field correlation function $F(\Delta r)$ and $\kappa$. For localized waves, Eq. 6.10 no longer holds, but good agreement with the focused pattern is obtained when $\kappa$ in Eq. 6.10 is replaced with $1/(\mu-1)$ with $\mu$ obtained experimentally.
\begin{figure}[htc]
\centering
\includegraphics[width=3in]{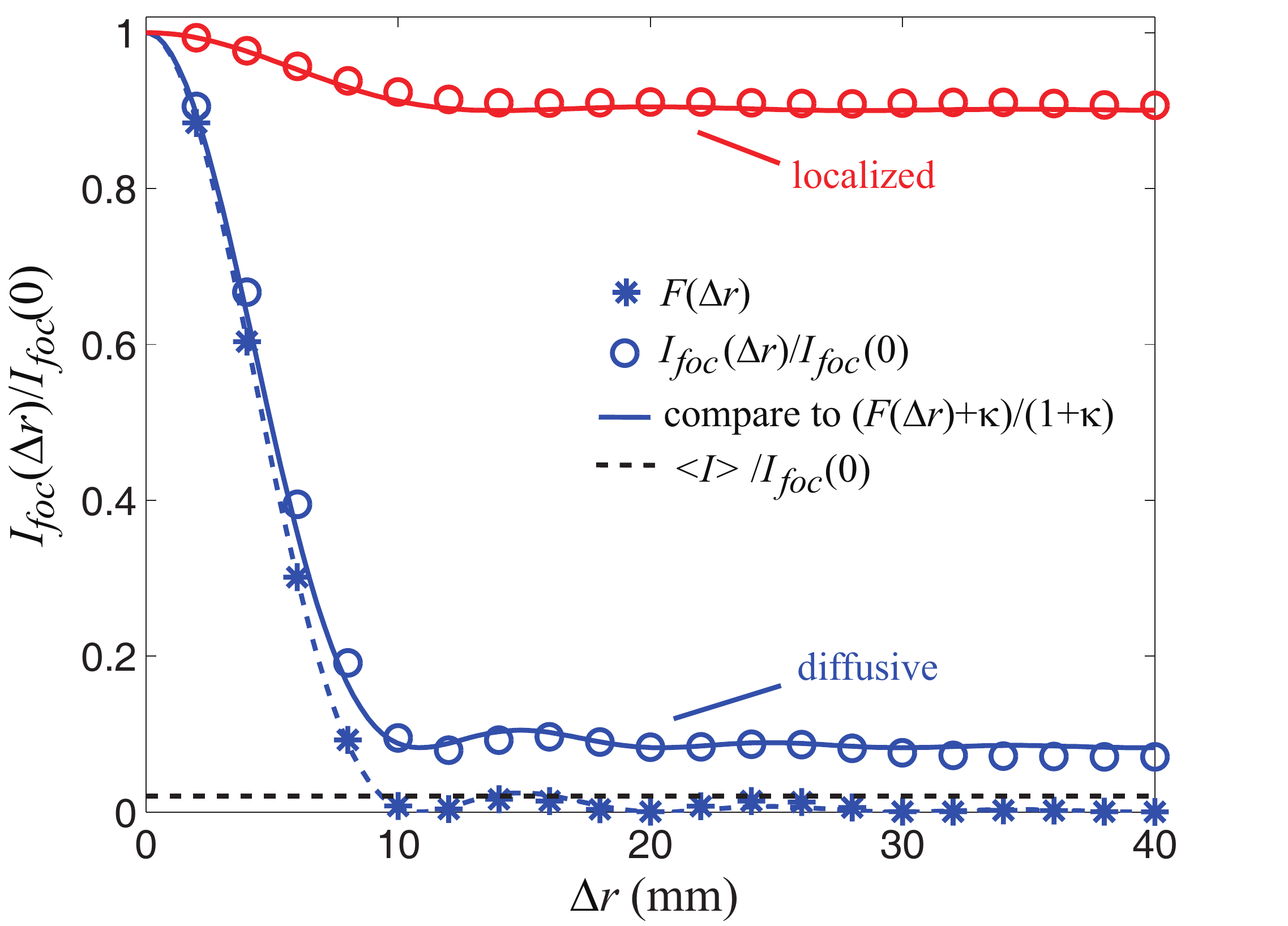}
\caption{The ensemble average of normalized intensity for focused radiation (blue circles) is compared to Eq. 6.10 (blue solid line) for $L=61 $ cm, $\kappa$ is replaced by $1/(\mu-1)$ in Eq. 6.10. $F(\Delta r)$ (blue dots) is fit with the theoretical expression obtained from the Fourier transform of specific intensity (dashed blue line). The field has been recorded along a line with a spacing of 2 mm for 49 input points for $L=61$ cm. The black dashed line is proportional to $\langle I\rangle/\langle I_{foc}(0)\rangle=1/N$}\label{Fig5.12}
\end{figure}

\section{Focusing pulse transmission through random waveguides}
In this section, we will discuss focusing pulse transmission through the random waveguides in space and time by phase conjugating a time-dependent transmission matrix at a given time delay $t^\prime$ \cite{100}. Spectra of transmission matrix for microwave propagation through random waveguide with length $L=61$ cm over the frequency range 14.7-15.7 GHz in 3200 steps. The transmission matrix is measured for $N^\prime=45$ points on the input and output surfaces for single polarization. 

We obtain the time dependent transmission matrix from spectra of the transmitted field between all points $a$ and $b$, $t_{ba}(\nu)$. These spectra are multiplied by a Gaussian pulse centered in the measured spectrum at $\nu_0=$15.2 GHz with bandwidth $\sigma_\nu=$150 MHz and then Fourier transformed into the time domain. This gives the time response at the detector to an incident Gaussian pulse launched by a source antenna with bandwidth $\sigma_t$=$1/2\pi/\sigma_\nu$. The time variation $I_{ba}(t^\prime)$ of an incident pulse launched at the center of the input surface and detected at the center of the output surface in a single realization of the random sample is shown in Fig. 6.5(a). Individual peaks in intensity have widths comparable to the width of the incident pulse. The average of the time of flight distribution $\langle I(t^\prime)\rangle$ over transmission coefficients for $N^{\prime 2}$ pairs of points and 8 random configurations is also presented and seen to be significantly broadened over the incident pulse. 

\begin{figure}[htc!]
\centering
\includegraphics[width=2.5in]{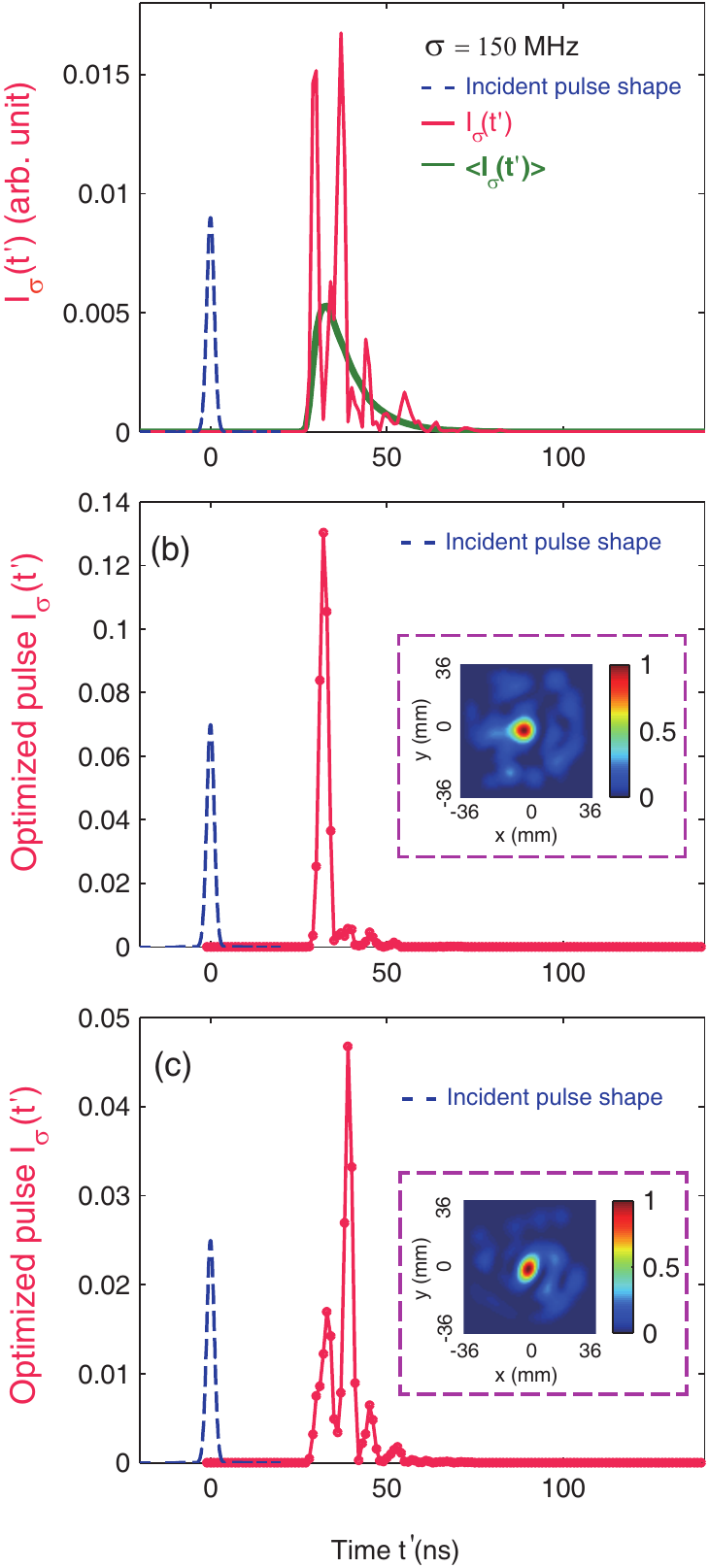}
\caption{ Spatiotemporal control of wave propagation through a random waveguide. (a) Typical response of $I_{ba}(t^\prime)$ and the time of flight distribution $\langle I(t^\prime)\rangle$ found by averaging over an ensemble of random samples. The incident pulse is sketched in the dashed blue curve. (b) and (c), phase conjugation is applied numerically to the same configuration as in (a) to focus at $t^\prime$= 33 ns and 40 ns at the center of the output surface in (b) and (c), respectively. The Whittaker-Shannon sampling theorem is used to obtain high-resolution spatial intensity patterns shown in the inset of (b) and (c).} \label{Fig1}
\end{figure}

The intensity that would be delivered to a point at the center of the output surface of the waveguide $\beta$=(0,0) at a selected time $t^\prime$ if the transmission matrix were phase conjugated at time $t^\prime$ is investigated. The results for $t^\prime$= 33 ns and 40 ns are shown in Figs. 6.5(b) and 6.5(c), respectively. In both cases, a sharp pulse emerges at the selected time delay with intensity peaked at $\beta$=(0,0).

The spatial profile of focusing for a monochromatic wave above an enhanced constant background is equal to the square of the field correlation function in space. Similarly, the temporal profile of the focused pulse is seen in Fig. 6.6 to correspond to the square of the field correlation function in time, $|F_E^\sigma(\Delta t)|^2$, where $F_E^\sigma\equiv \langle E_\sigma(t^\prime)E_\sigma^*(t^\prime+\Delta t)\rangle/(\langle I(t^\prime)\rangle \langle I(t^\prime+\Delta t)\rangle)^{1/2}$. For an incident Gaussian pulse, the square of the field correlation function is equal to the intensity profile of the incident pulse and is independent of delay time.
\begin{figure}[htc]
\centering
\includegraphics[width=3in]{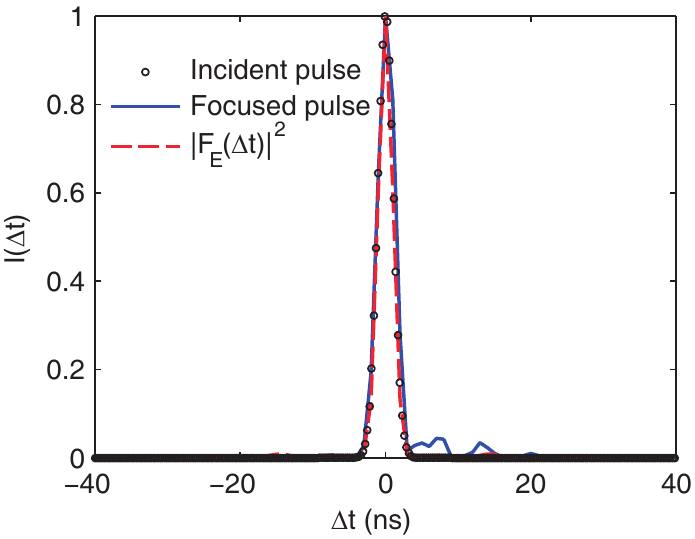}
\caption{ Profile of the focused pulse compared with the square modulus of the field correlation function in time and the profile of incident Gaussian pulse. The focused pulse in Fig. 6.5(b) has been plotted relative to the time of the peak. All curves are normalized to unity at $\Delta t$=0 ns.} \label{Fig2}
\end{figure}	

We have shown previously that in a large single transmission matrix in steady state, $\mu=1/(1/M-1/N)$. Here, $M$ is the eigenchannel participation number when the full transmission matrix of size {\it N} is measured. Because the full transmission matrix is not accessible in the experiment, the density of measured transmission eigenvalues differs from theoretical prediction. Nonetheless, we find in steady state measurements that when part of the transmission matrix is measured, the contrast in focusing via phase conjugation is given by, 
\begin{equation}
\mu=1/(1/M^\prime-1/N^\prime).
\end{equation}           
where $M^\prime$ is the eigenvalue participation number of the measured transmission matrix of size $N^\prime$. This is a property of random transmission matrices and therefore should apply as well to transmission matrices at different time delays provided the field within the transmission matrix is randomized. In Fig. 6.7, we present the time evolution of $\langle M^\prime\rangle$ and $\langle \mu\rangle$. Near the arrival time of the ballistic wave, the value of $M^\prime$ is close to unity and the contrast is not described by Eq. 6.11. This is because ballistic wave is associated with the propagating waveguides modes with the highest speed and therefore the transmitted field is not randomized. Once the transmitted waves at the output have been multiply scattered, a random speckle pattern develops and the measured contrast is in accord with Eq. 6.11. 
\begin{figure}[htc]
\centering
\includegraphics[width=3in]{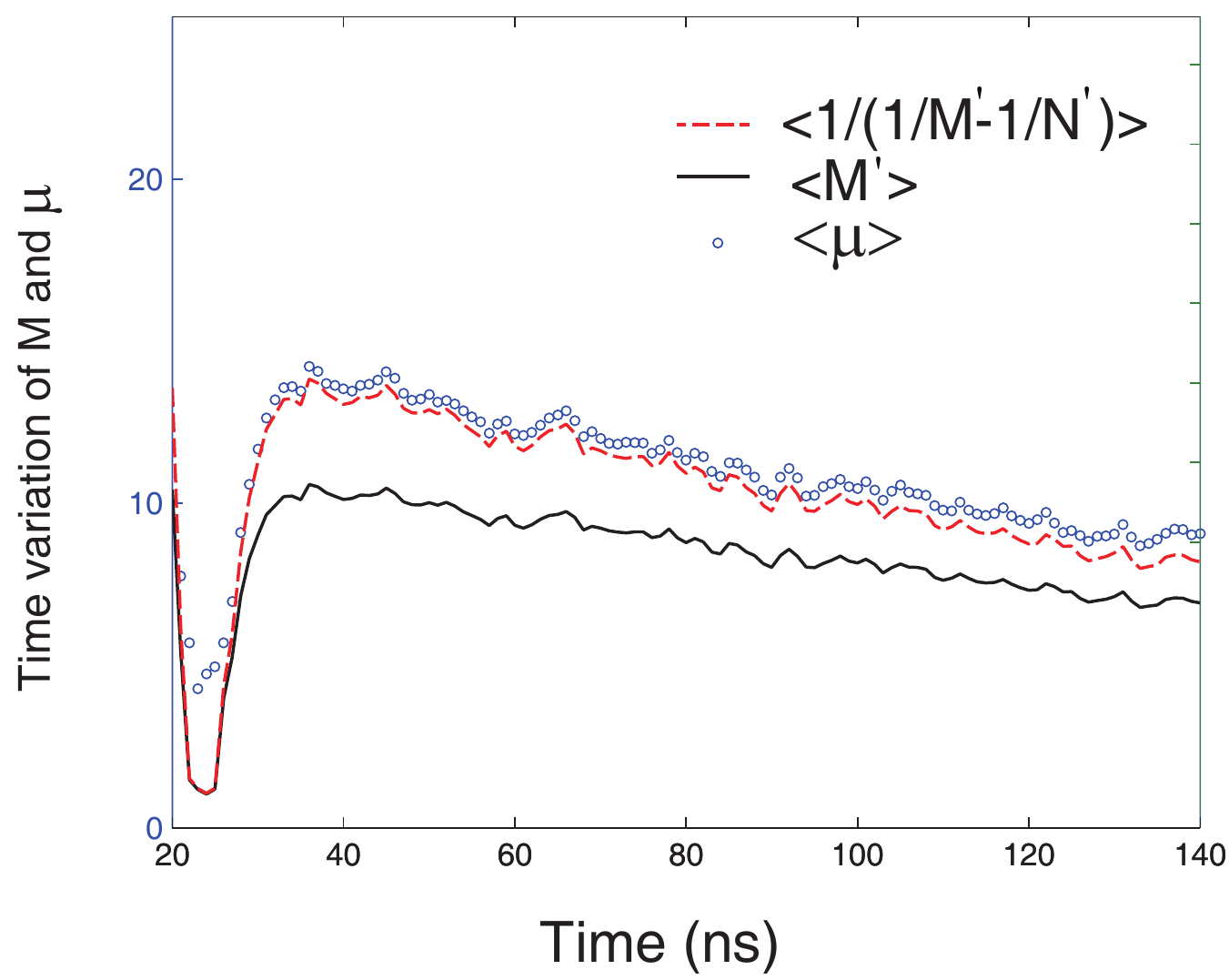}
\caption{Time evolution of $\langle M^\prime \rangle$ (lower solid curve) and the maximal focusing contrast $\langle \mu \rangle$. $\mu$ is well described by Eq. 6.11 after the time of the ballistic arrival, $t^\prime \sim 21$ ns. At early times, the signal to noise ratio is too low to analyze the transmission matrix.} \label{Fig3}
\end{figure}

After the arrival of ballistic waves, the value of $M^\prime$ is seen in Fig. 6.7 to increase rapidly before falling slowly. This reflects the distribution of lifetimes and the degree of correlation in the speckle patterns of quasi-normal modes \cite{44}. Just after the ballistic pulse, transmission is dominated by the shortest-lived modes. These modes are especially short lived and strongly transmitting because they are extended across the sample as a result of coupling between resonant centers. Sets of extended modes that are close in frequency could be expected to have similar speckle patterns in transmission, so that a number of such modes might then contribute to a single transmission channel. As a result, the number of independent eigenchannels of the transmission matrix contributing substantially to transmission would be relatively small at early times and $M^\prime$ would be low. At late times, only the long-lived modes contribute appreciably to the transmission. Thus for intermediate times, modes with wider distribution of lifetimes than at either early or late times contribute to transmission and these modes are less strongly correlated than at early times so that $M$ and the contrast are peaked. 

\section{Summary}
In this chapter, we have shown that it is possible to focus monochromatic and pulse transmission through disordered random waveguides. The contrast $\mu$ of the focusing via phase conjugation is related the eigenchannel participation number $M$ of the measured transmission matrix of size $N$, $\mu=1/(1/M-1/N)$. The spatial and temporal profile of the focused beam is the close to the square of the field correlation function in space and time for a diffusive waves. The dynamics of the $M$ in time is investigated and the initial rise and subsequent falloff of the $M$ is explained in terms of the decaying of the quasi-normal modes within the random media.  

\chapter{Densities of states and intensity profiles of transmission eigenchannels inside opaque media}
\section{Introduction}
A disordered medium fills with a fine-grained random interference pattern of intensity when illuminated by a monochromatic wave. When the intensity is averaged over many realizations of the disorder, however, all trace of the underlying wave is lost in the diffuse profile of intensity which corresponds to the density of randomly scattered particles of the wave such as photons or electrons. Nonetheless, coherence of the field is preserved in the transmission values $\tau$ of the eigenchannels of TM which link the incident and transmitted wave via a small number of strongly transmitting channels among mostly dark channels \cite{32,60,61,63,101}. This coherence should also modify the energy density inside the medium and the dynamics of transmission \cite{102,103}. But even the spatially averaged wave energy within the sample, known as the density of states (DOS), which controls spontaneous and stimulated emission \cite{104} and wave localization \cite{10}, has not been measured. In this chapter, we will show that microwave spectra of the TM yield the photon dwell time and DOS for each eigenchannel and, in conjunction with computer simulations, provide the average of the intensity profile inside the sample for each eigenchannel. The channel dwell time increases dramatically with transmission and grows quadratically with sample length. Measurements of the DOS from the sum over eigenchannels are in close correspondence with the sum of the DOS over electromagnetic modes of the medium. The control of the energy density within the sample demonstrated here may enable imaging and enhanced absorption within complex system for applications such as medical diagnosis and therapy, energy harvesting and low thresholds lasing in random media.

\section{Dynamics of eigenchannels of TM}
The power of the TM to mold the flow of waves through random samples has been demonstrated in sharp focusing of sound, light and microwave radiation \cite{78,96} and by enhanced transmission of specific eigenchannels \cite{45,105}. Similarly the excitation by eigenchannels with a diversity of intensity profiles and photon dwell times {\it inside} the sample might be controlled and exploited in numerous applications. For example, illuminating a sample with wavefronts which excite channels with diverse spatial and temporal characteristics inside the sample can be used to obtain a depth profile the optical absorption, emission or nonlinearity within a sample. The possibility of depositing energy well below the surface would lengthen the residence time of emitted photons in active random systems and so could dramatically lower the lasing threshold of amplifying diffusive media. In contrast, residence times of emitted photons are relatively short because of the shallow penetration of the pump laser due to multiple scattering so that lasing thresholds are high in traditional random lasers \cite{106,107,108}.

The DOS is the integral of the intensity $I_a(z)$ over the sample volume for unit flux summed over all 2N channels {\it a} of the incident wave on both sides of the sample, 
\begin{equation} 
\rho=\frac{1}{\pi}\sum_a^{2N}\int_0^L I_a(z)dz.
\end{equation}
The DOS is also equal to the sum of delay times weighted by $I_{ab}$ over all $(2N)^2$ pairs of incoming and outgoing propagation channels \cite{109},
\begin{equation} 
\rho(\omega)=\frac{1}{\pi}\sum_{b,a}^{2N} I_{ba}\frac{d\varphi_{ba}}{d\omega}.
\end{equation}
The phase derivative $d\varphi_{ba}/d\omega$ is the single channel delay time for a narrow band pulse propagating from channel $a$ to $b$ \cite{110}. Because this expression involves the full scattering matrix involving field transmission coefficient in both transmission and reflection for waves incident on both sides of the sample, the DOS has not been measured previously for classical waves. The TM may be expressed via the singular value decomposition as, $t=\sum_n^N \sqrt{\tau_n}\bf{u}_n\bf{v}^\dagger_n$. For a non-dissipative disordered system with time reversal symmetry, the DOS can alternatively be expressed in terms of the transmission eigenchannels, 
\begin{equation} 
\rho(\omega)=\frac{1}{\pi}\sum_n^N \frac{d\theta_n}{d\omega}.
\end{equation}
in which $\frac{d\theta_n}{d\omega}=\frac{1}{i}(\frac{d\bf{v}_n^*}{d\omega}.\bf{v}_n-\frac{d\bf{u}_n^*}{d\omega}.\bf{u}_n)$ is the residence time of the energy of the wave for the $n^{\it th}$ transmission eigenchannel. The contributions to the DOS from transmission and reflection are $\rho_t=\frac{1}{\pi}\sum_n^N \tau_n\frac{d\theta_n}{d\omega}$ and $\rho_r=\frac{1}{\pi}\sum_n^N (1-\tau_n)\frac{d\theta_n}{d\omega}$, respectively \cite{111}.  

\begin{figure}[htc]
\centering
\includegraphics[width=3.5in]{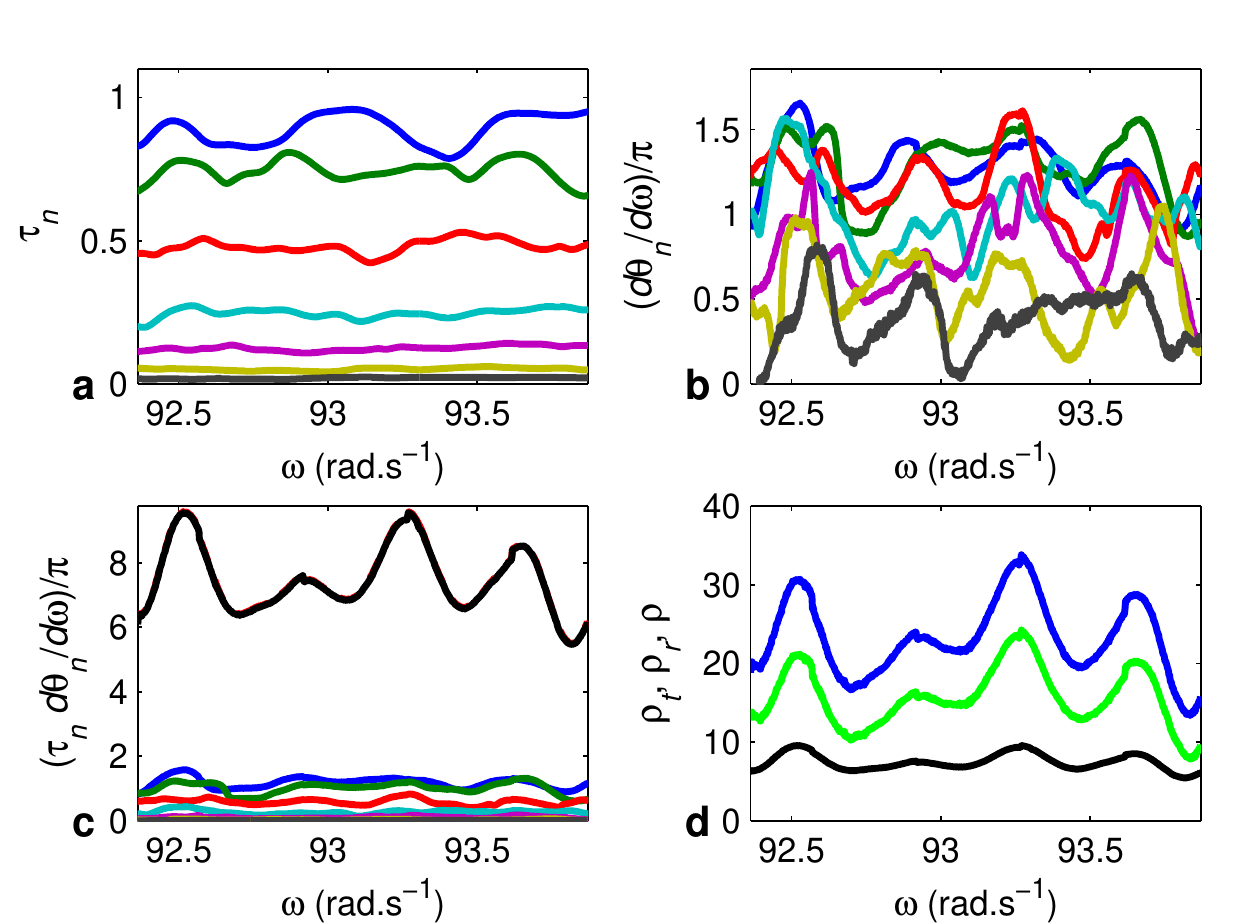}
\caption{Measured spectra of eigenchannel properties in a single realization of the random sample. (a) Spectra of eigenvalues $\tau_n$ for $n= 1,5,9,13,17,21,25$. (b) Spectra of the corresponding $d\theta_n/d\omega$. (c) Spectra of $\tau_nd\theta_n/d\omega$, and transmission component of the DOS, $\rho_t$, found from the singular value decomposition of the TM (black line), $1/\pi\sum_n d\theta_n/d\omega$, and from the sum of the weighted channel delay times (red dashed line), $1/\pi\sum_{a,b} I_{ab}d\varphi/d\omega$, (these last two curve overlap). (d) Spectra of the transmission DOS $\rho_t$ (black line) that was found in (c), reflection DOS $\rho_r$ (green line) and their sum, which gives the DOS $\rho=\rho_t+\rho_r$ (blue line).} \label{Fig7.1}
\end{figure}
Spectra for $\tau_n$ and $d\theta_n/d\omega$ for a single random configuration drawn from an ensemble of samples for which {\textsl g}=6.9  are shown in Figs. 7.1a and 7.1b. The contribution of each channel to $\rho_t$ as well as the sum over all channels is shown in Fig. 7.1c, while the sum of $\rho_t$ and $\rho_r$ to give $\rho$ is shown in Fig. 7.1d. The perfect overlap of $\rho_t$ obtained from the sum over channels with the spectrum of $\frac{1}{\pi}\sum_{b,a}^N I_{ba}d\varphi/d\omega$ confirms that the DOS can be apportioned into reflection and transmission contributions and can be written as the sum of the eigenchannel DOS (EDOS) $d\theta_n/d\omega$. 

\begin{figure}[htc]
\centering
\includegraphics[width=4.5in]{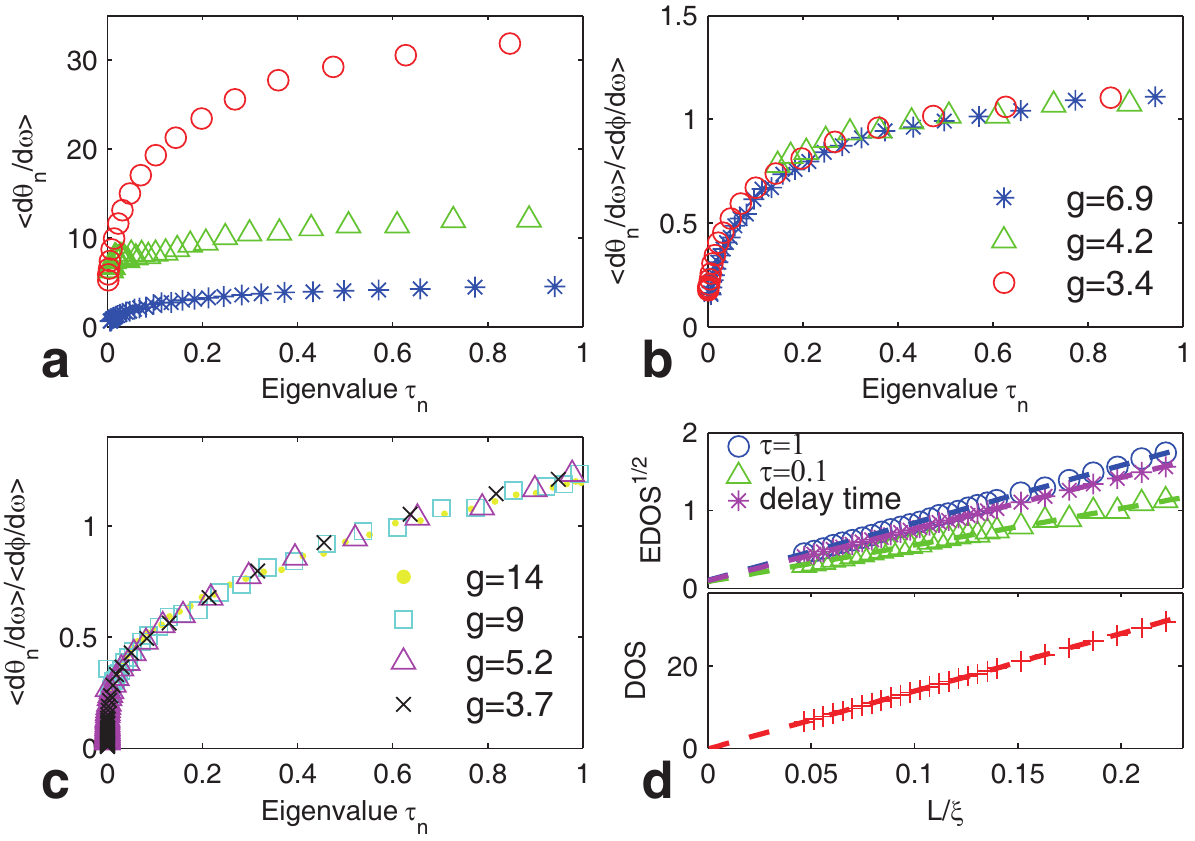}
\caption{Scaling of eigenchannel dwell time. The experimental residence time of the eigenchannels $\langle d\theta_n/d\omega\rangle$ (EDOS) (a) and residence time normalized by the average delay time, $\langle d\theta_n/d\omega\rangle/\langle d\varphi/d\omega\rangle$, (b) are plotted with respect of the eigenvalues for $L=61$ cm (red dots), $L=40$ cm (green triangles) and $L=23$ cm (blue stars). (c) Simulations of eigenchannel residence time and EDOS with different values of {\textsl g}. (d) Scaling of $\sqrt{\langle d\theta_n/d\omega\rangle}$ for $\tau_n=1$ (blue dots) and $\tau_n=0.1$ (blue triangles), of $\sqrt{\langle d\varphi/d\omega\rangle}$ (blue stars) showing that the EDOS increases quadratically with $L$, while the DOS (red crosses) increases linearly.} \label{Fig7.2}
\end{figure}
The single eigenchannel delay time $d\theta_n/d\omega$ is seen in Fig. 7.2a to increase with $\tau_n$ and {\it L} in measurements in three sample lengths. When normalized by the ensemble average of the delay time in each ensemble, denoted by $\langle \frac{d\varphi_{ba}}{d\omega}\rangle$, the curves collapse to a single curve, both for measurements in Fig. 7.2b and recursive Green’s function simulations used in Chapter 3 in Fig. 7.2c. These curves are not identical because internal reflection present in the experimental sample is minimized in the simulated sample by index matching the sample and background to create a sample in which universal features can more easily be discerned. Since $\langle d\varphi_{ba}/d\omega \rangle$ scales as $L^2$ for diffusive waves as seen in Fig. 7.2d, the constant ratio of the delay time $\langle d\theta_n/d\omega\rangle_{\tau_n}$ for a channel of fixed $\tau_n$, to the average delay time for different sample lengths indicates that the EDOS for a given value of $\tau_n$ will also scale as $L^2$. This is confirmed in plots of the scaling of the EDOS for $\tau=0.1$ and 1 shown in Fig. 7.2d. Though the EDOS scales as $L^2$, the DOS is seen in Fig 7.2d to scale as $L$, as expected. The DOS is linear in $L$ since the number of open channels is proportional to {\textsl g} which falls inversely with $L$, $\textsl{g}=\xi/L$.

To find an expression for the average of the longitudinal variation of the intensity inside the sample for a given eigenchannel $\langle I_n(z)\rangle$ or given transmission eigenvalue $\langle I(z;\tau_n)\rangle$, we carry out recursive Green's function simulations for scalar wave propagation in multichannel random samples \cite{112}. In the simulation, the sample leads are uniform with a refractive index $n_0=1$ while $\epsilon(x,y)$ in the scattering medium is given by, $\epsilon(x,y)=\epsilon_1+\delta\epsilon_1(x,y)$, in which $\epsilon_1$ is equal to 1 and $\delta\epsilon_1(x,y)$ is selected from a uniform distribution with a standard deviation equal to 0.4. The number of propagating waveguide modes in the leads $N$ is chosen to be 66 and the sample length is $L=0.07\xi$. Simulations are made for 250 sample configurations. We also carry out scattering matrix simulations in a system that presents perhaps the simplest statistics: an electromagnetic plane wave normally incident upon a layered system with fixed number of layers $L$ with alternating indices of refraction between $n_1$ and $n_2$ and with randomness of layer thickness drawn from a distribution that is much greater than the average layer thickness. This is a 1D system with fixed number of internal reflecting interfaces all with the same index mismatch, and hence with identical scattering strength. The phase shift for the wave propagating between interfaces is consequently totally random. The mean free path in this system is equal to the localization length $\xi$. Previous simulations carried out in the slab geometry noted that the peak in intensity moved towards the incident face of the sample in channels with lower values of $\tau_n$.

\begin{figure}[htc]
\centering
\includegraphics[width=4.2in]{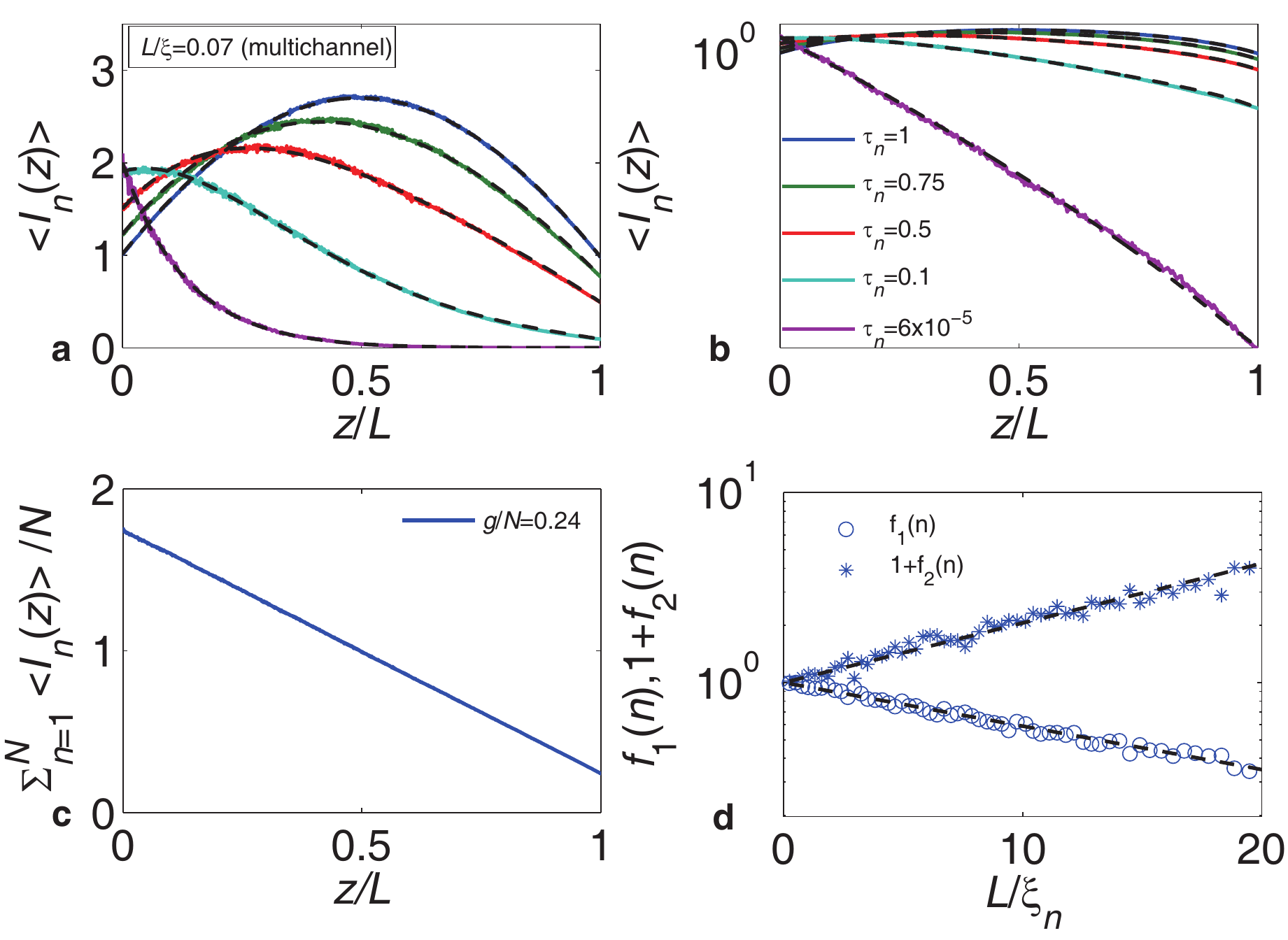}
\caption{Simulations of intensity inside sample. (a) and (b) Intensity inside the sample for eigenchannels with $\tau_n=1$ (blue line), $\tau_n$=0.75 (green line), $\tau_n$=0.5 (red line), $\tau_n$=0.1 (cyan line) and $\tau_n=6\times 10^{-6}$ (purple line) for $L/\xi$=0.07. The dashed black lines are the fit of these intensities using Eq. (7.4). (c) The sum of the eigenchannels $I_n(z/L)$ falls linearly from $2-{\textsl g}/N$ to ${\textsl g}/N$ as expected for diffusive waves. (d) Semilogarithmic plot of coefficients $f_1(n)$ and $1+f_2(n)$ giving the best fit of Eq.(7.4) to $\langle I_n(z/L)\rangle$.} \label{Fig7.3}
\end{figure}
An expression for $\langle I_n(z)\rangle$ that will correspond to these simulations must satisfy a number of conditions. Among these are that the expression for the transmitted flux must be the analytic continuation of the intensity inside the sample, $\langle I_n(z=L)\rangle=\tau_n$. We consider the case of an index matched sample so that additional nonuniversal parameters need not be introduced. Since the sample is index matched, the angular distribution of diffusing particles can be assumed to be the same inside and outside the sample for all channels. In addition, the local intensity is the sum of forward and backwards propagating particles of the wave.  As a result, the intensity at $z=0$ is the sum of unit flux flowing to the right plus the reflected flux $(1-\tau_n)$ so that, $\langle I_n(z=0)\rangle=2-\tau_n$. The average intensity distribution for channels with $\tau_n=1$ is seen in Figs. 7.3a to be $1+ F_1(z)$, where $F_1(z)$ is a symmetrical function peaked at $L/2$ and vanishes at the boundaries. These considerations together with the observed shift of the intensity peak to the front of the sample suggest the variation of $I_n(z) $with $\tau_n$ is given by,
\begin{equation}
 \begin{aligned} 
& \langle I_n(z)\rangle = (1+f_1(n)\exp(f_2(n)z)F_1(z))\tau_n\cosh((L-z)/\xi_n). \\
& F_1(z)=\langle I(z,\tau=1)\rangle-1 =M\frac{4z(L-z)}{L^2}.
\end{aligned}
\end{equation}
Here, $\xi_n$ is related to $\tau_n$ via the relation $\tau_n=1/\cosh^2(L/2\xi_n)$ and can be considered as the localization length of the $n^{th}$ eigenchannel. The sum of intensity profiles over all channels in Fig. 7.3c for the sample with $N=66$ and $L=0.07\xi$ gives the linear decay of intensity within the medium required by Fick's law for diffusing particles. Plots of $f_1(n)$ and $f_2(n)$ which give the best fit to $\langle I_n(z)\rangle$ as shown in Figs. 7.3d. $f_1(n)$ falls exponentially, $f_1(n)=\exp(-\alpha L/\xi_n)$, while $f_2(n)$ increases as an exponential function, $f_2(n)=\exp(\beta L/\xi_n)-1$. In addition, we found that, when normalized by its maximum value, the function $F_1(z/L)$ for 1D and multi-channel samples with different values of $L/\xi$ collapse to a single curve, as long as $L/\xi$ is much smaller than unity. This shows that the shape of $F_1(z/L)$ is the same for diffusive waves and this is demonstrated in Fig. 7.4.
\begin{figure}[htc]
\centering
\includegraphics[width=3.5in]{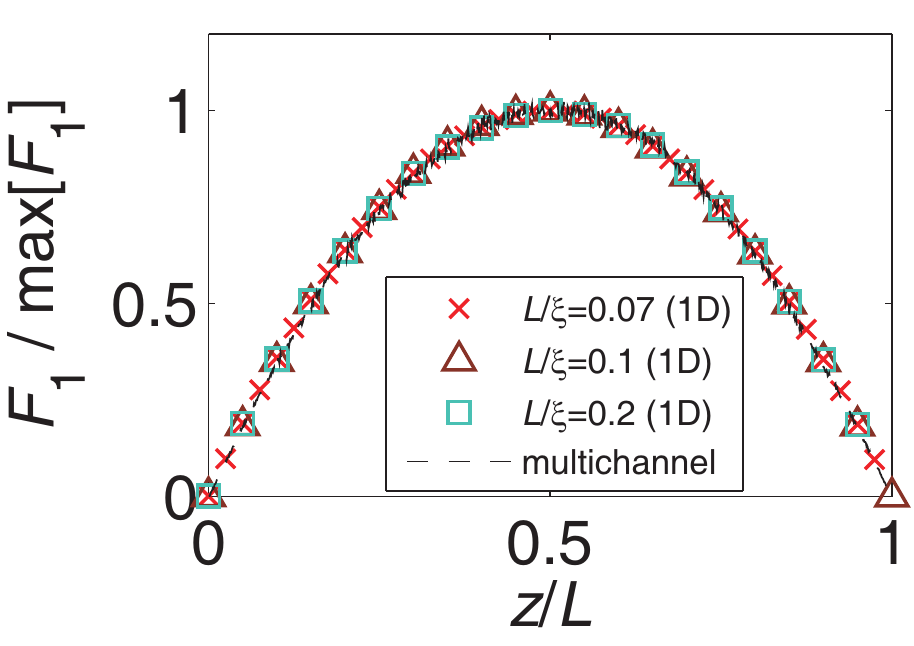}
\caption{Universality of $F_1(z/L)$. The normalized functions $F_1(z/L)$ for 1D sample with $L/\xi=0.07, 0.1,0.2$ and for multi-channel sample with $L/\xi=0.07$ are seen to overlap. } \label{Fig7.4}
\end{figure}

We confirm the measurement of the DOS by comparing the sum of the EDOS to the sum of the contributions to the DOS of electromagnetic modes of the medium. We carry out the comparison for waves near the Anderson localization transition for which the degree of modal overlap is appreciable but still not so large so that the full set of mode central frequencies $\omega_n$ and linewidths $\Gamma_n$ cannot be accurately determined. Normalizing the integral of the Lorentzian line for each mode to unity, we obtain the DOS,
\begin{equation}
\rho(\omega)=\sum_n \rho_n(\omega)=\frac{1}{\pi}\frac{\Gamma_n/2}{(\Gamma_n/2)^2+(\omega-\omega_n)^2}.
\end{equation}
The DOS obtained as the sum over channels and modes is shown in Figs. 7.5a and 7.5b, respectively. 
\begin{figure}[htc]
\centering
\includegraphics[width=5in]{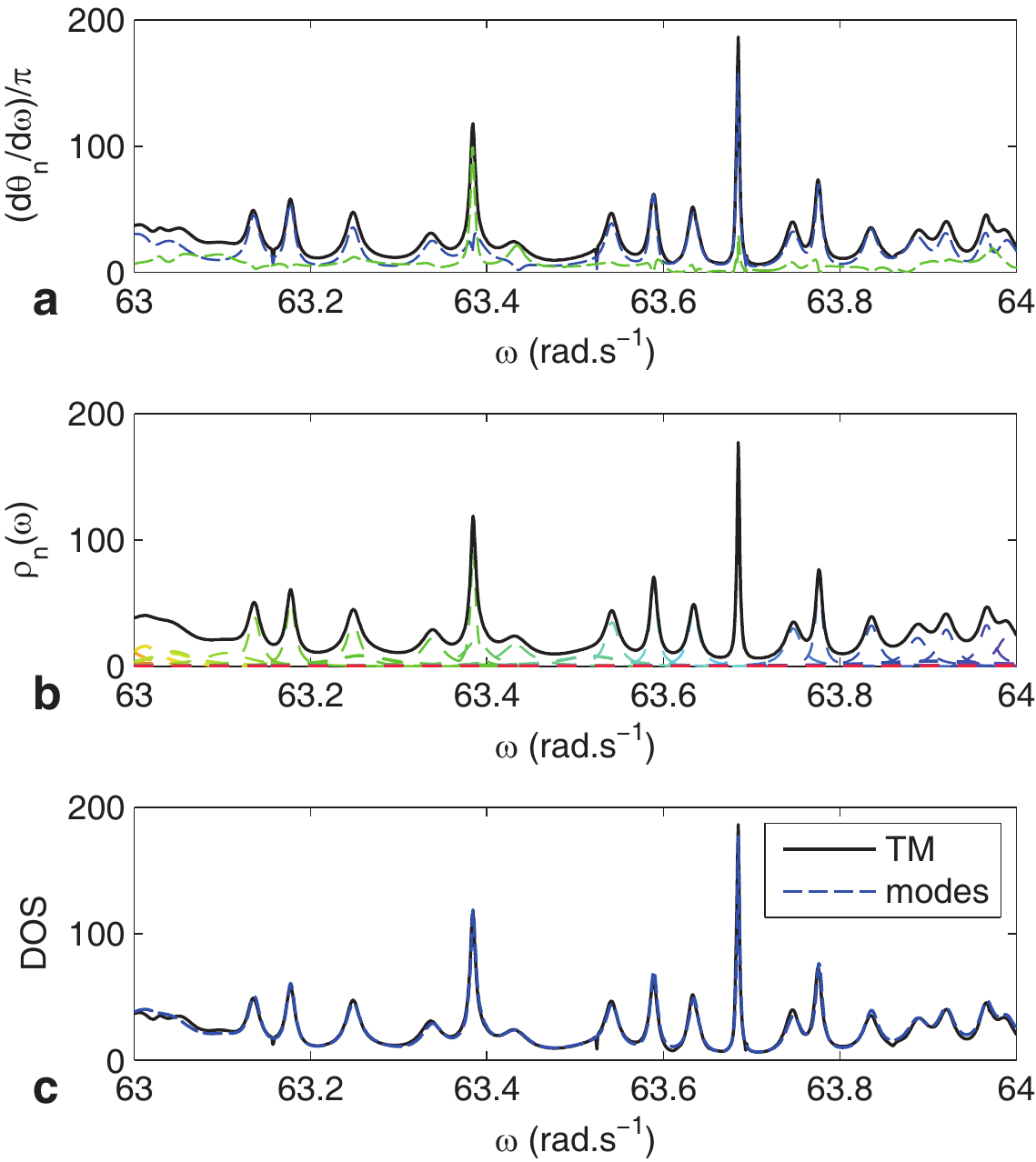}
\caption{DOS determined from channels and modes. (a) For localized waves ($L/\xi$=1.7) in which the eigenvalues decrease rapidly and only two eigenchannels (dashed lines) contribute appreciably to the DOS (black line). (b) Contributions of the modes (dashed lines) from Eq. (7.5) to the DOS (black line). (c) Comparison of the DOS retrieved from the TM (red line) and the decomposition into modes (blue dashed line).} \label{Fig7.5}
\end{figure}
The two approaches are seen in Fig. 7.4c to be in good agreement. Whereas the mode approach can only be performed for localized waves, the determination of the DOS using channel can also be applied for diffusive waves.

\section{Summary}
In this chapter, we explored the dynamics of transmission eigenchannels and relate the dynamics to the DOS of the disordered medium and the intensity distribution {\it inside} for different excitation of eigenchannels. The ability to sum the EDOS to obtain the DOS shows that wave propagation within random systems can be decomposed into the behavior of individual channels. This is the basis of the usefulness of channels as a description of transport and their great potential for controlling the flow of light inside random media. 

\chapter{Conclusions}
In this thesis, we have measured the microwave transmission matrix in the crossover to Anderson localization. The correlation between {\it T} and the underlying transmission eigenvalues $\tau_n$ is found and used to explain the fluctuations of transmittance over the random ensemble. Statistics of transmission relative to the average value of transmission in single TM are found and expressed in terms of the eigenchannel participation number $M$. The second order transmission statistics over a random ensemble is given by the joint probability distribution of $T$ and $M$. This represents a considerable simplification from the joint distribution of the full set of transmission eigenvalues $\tau_n$. We found that SPS is approached in a multi-channel random system when $M\rightarrow 1$. The power of transmission matrix to control wave transport is demonstrated by focusing in steady-state and pulsed transmission via phase conjugation of TM. The spatial contrast of optical focusing is shown to be $M$ for a large TMs and the profile of the focus is directly related to the spatial and temporal correlation function of the random media. The dynamics of transmission eigenchannel are seen to give the density states of a disordered medium and to be closely related to the intensity profile of individual eigenchannel within the medium.

\appendix{Appendix A \\ Optimal focusing via phase conjugation}
In Section 6.2, we gave the contrast of focusing monochromatic radiation through random waveguides in terms of the eigenchannel participation number {\it M} and the size {\it N} of the measured transmission matrix, $\mu=1/(1/M-1/N)$. In this Appendix, we provide detailed derivation of the formula. 

Via singular value decomposition, the field transmission matrix {\it t} could be expressed as, $t=U\Lambda V^\dagger$, where $U$ and $V$ are unitary matrices and $\Lambda$ is a diagonal matrix with $\sqrt{\tau_n}$ along the main diagonal. The elements of {\it U} and {\it V} are, respectively, $u_{nb}$ and $v_{na}$, in which {\it n} is the eigenchannel index and $a$ and $b$ are the input and output channels. The real and imaginary part of $u_{nb}$ and $v_{na}$ are Gaussian random variables with zero mean and variance of $1/2N$. The total transmission from incident channel {\it a}, $T_a=\sum_b T_{ba}$, can be written as, $T_a=\sum_n \tau_n|v_{na}|^2$. 

Maximal focusing through random media at a target spot $\beta$ can be achieved by phase conjugating the field transmission coefficients between the target and incident points in the input plane, so that all the fields from the incident points arrive in the target in phase and interfere constructively. Therefore, the correct wavefront to focus at $\beta$ is $t^*_{\beta a}/\sqrt{T_\beta}$, which $T_\beta=\sum_a |t_{\beta a}|^2$ is the normalization factor to normalize the incident power to be unity. In this way, the intensity at the focal spot will be $I_\beta=(\sum_a t^*_{\beta a}t_{\beta a})^2/T_\beta=T_\beta$. The focused intensity $T_\beta$ is $N$ times of the ensemble average transmission $T_{ba}$, $\langle T_{\beta}\rangle=N\langle T_{ba}\rangle$. 

However, the background intensity is enhanced relative to the ensemble average transmission when applying phase conjugation, since more weight is given to the high transmitting eigenchannels. Therefore, the focusing contrast is not equal to the number of channels {\it N}. When focusing is achieved via phase conjugation, the electric field at the background spot {\it b} is given as, 
\begin{align}
E_{ba}=\frac{\sum_a t^*_{\beta a}t_{ba}}{\sqrt{T_\beta}}=\frac{\sum_a (\sum_n \sqrt{\tau_n}u_{nb}v_{na})(\sum_{n^\prime} \sqrt{\tau_{n^\prime}}u^*_{n^\prime b}v^*_{n^\prime a})}{\sqrt{T_\beta}}.
\end{align}
Because $\sum_a v_{na}v^*_{n^\prime a}$ is equal to $\delta_{n n^\prime}$, $E_{ba}$ can be written as,
\begin{equation}
E_{ba}=\frac{\sum_n \tau_nu_{nb}u^*_{n \beta}}{\sqrt{T_\beta}}.  
\end{equation}
The intensity $I_{ba}$ is then given by,
\begin{equation}
I_{ba} = |E_{ba}|^2=\frac{\sum_{n,n^\prime} {\tau_n u_{nb}u^*_{n \beta} \tau_{n^\prime} u^*_{n^\prime b} u_{n^\prime \beta}}}{T_\beta}
\end{equation}
Since the variance of total transmission $T_a$ in a single transmission matrix is equal to $1/M$, in the diffusive limit, the fluctuation of $T_\beta$ can be neglected and this gives,
\begin{align}
\langle I_b\rangle 
& = \frac{\langle \sum_{n,n^\prime} {\tau_n u_{nb}u^*_{n \beta} \tau_{n^\prime} u^*_{n^\prime b} u_{n^\prime \beta}}\rangle_{\beta, b\ne\beta}}{\langle T_\beta\rangle} \\
& =\frac{\frac{\sum_{n,n^\prime,\beta,b} {\tau_n u_{nb}u^*_{n \beta} \tau_{n^\prime} u^*_{n^\prime b} u_{n^\prime \beta}}}{N(N-1)}-\frac{\sum_{n,n^\prime,\beta} \tau_n\tau_{n^\prime}|u_{n\beta}|^2|u_{n^\prime \beta}|^2}{N(N-1)}}{\langle T_\beta\rangle} \\
& = \frac{\frac{\sum_{n} \tau_n^2}{N(N-1)}-\frac{\langle T^2_\beta\rangle}{N-1}}{\langle T_\beta\rangle}
\end{align}
In a given TM, the average total transmission is $\langle T_\beta\rangle=T/N$ and $\langle T_\beta^2\rangle=\langle T_\beta\rangle^2(1+1/M)$. For $N\gg 1$, the contrast $\mu$ is then,
\begin{align}
\mu
& =\frac{\sum_n \tau_n/N}{\frac{1}{N-1}\frac{\sum_n \tau_n^2}{\sum_n \tau_n}-\frac{\sum_n \tau_n}{N(N-1)}-\frac{\sum_n \tau_n}{MN(N-1)}} \\
& = 1/(\frac{1}{M}-\frac{1}{N-1}) \\
& \sim 1/(1/M-1/N)
\end{align}

\end{document}